  \documentclass[amsmath,amssymb,apj]{emulateapj-rtx4}
\usepackage{color}
\begin{document}


\title{Diversity of kilonova light curves}


\author{Kyohei Kawaguchi}
\affil{Institute for Cosmic Ray Research, The University of Tokyo, 5-1-5 Kashiwanoha, Kashiwa, Chiba 277-8582, Japan}
\affiliation{Center for Gravitational Physics,
  Yukawa Institute for Theoretical Physics, 
Kyoto University, Kyoto, 606-8502, Japan} 
\author{Masaru Shibata}
\affil{Max Planck Institute for Gravitational Physics (Albert Einstein Institute), Am M\"{u}hlenberg 1, Potsdam-Golm, 14476, Germany}
\affiliation{Center for Gravitational Physics,
  Yukawa Institute for Theoretical Physics, 
Kyoto University, Kyoto, 606-8502, Japan} 

\and
\author{Masaomi Tanaka}
\affil{Astronomical Institute, Tohoku University, Aoba, Sendai 980-8578, Japan}
\affil{National Astronomical Observatory of Japan, Mitaka, Tokyo, Japan}

\newcommand{\angstrom}{\text{\normalfont\AA}}
\newcommand{\rednote}[1]{{\color{red} (#1)}}
\newcommand{\UTF}[1]{\rednote{UTF[#1]}}

\begin{abstract}
We perform radiative transfer simulations for kilonova in various situations, including the cases of prompt collapse to a black hole from neutron-star mergers, high-velocity ejecta possibly accelerated by magnetars, and a black hole-neutron star merger. Our calculations are done employing ejecta profiles predicted by numerical-relativity simulations and a new line list for all the r-process elements. We found that (i) the optical emission for binary neutron stars promptly collapsing to a black hole would be fainter by  $\gtrsim1$--$2\,{\rm mag}$ than that found in GW170817, while the infrared emission could be as bright as that in GW170817 if the post-merger ejecta is as massive as $\approx0.01\,M_\odot$; (ii) the kilonova would be brighter than that observed in GW170817 for the case that the ejecta is highly accelerated by the electromagnetic energy injection from the remnant, but it would decline rapidly and the magnitude would become fainter than in GW170817 within a few days; (iii) the optical emission from a black hole-neutron star merger ejecta could be as bright as that observed in GW170817 for the case that sufficiently large amount of matter is ejected ($\gtrsim0.02\,M_\odot$), while the infrared brightness would be brighter by $1$--$2\,{\rm mag}$ at the same time. We show that the difference in the ejecta properties would be imprinted in the differences in the peak brightness and time of peak. This indicates that we may be able to infer the type of the central engine for kilonovae by observation of the peak in the multiple band. 
\end{abstract}

\keywords{gravitational waves --- stars: neutron --- radiative transfer}

\section{Introduction}

The simultaneous detection of gravitational waves (GWs) from a neutron star-neutron star (NS-NS) merger~\citep{TheLIGOScientific:2017qsa} and its electromagnetic (EM) counterparts opened a new window of the multi-messenger astronomy. EM counterparts to GW170817 were observed over the entire wavelength range, from gamma-ray to radio wavelengths~\citep{GBM:2017lvd,Andreoni:2017ppd,Arcavi:2017xiz,Coulter:2017wya,Cowperthwaite:2017dyu,Diaz:2017uch,Drout:2017ijr,
Evans:2017mmy,Hu:2017tlb,Valenti:2017ngx,Kasliwal:2017ngb,Lipunov:2017dwd,Pian:2017gtc,Pozanenko:2017jrn,Smartt:2017fuw,Tanvir:2017pws,Troja:2017nqp,Utsumi:2017cti,Mooley:2017enz,Hallinan:2017woc}. In particular, a counterpart in optical and infrared wavelengths (named as SSS17a, AT 2017gfo or DLT17ck), is identified as the emission so-called a kilonova (also known as macronova). 

A kilonova is the emission which has been expected to be associated with NS mergers as a consequence of the mass ejection from the system~\cite[e.g.,][]{Rosswog:1998hy,Ruffert:2001gf,Hotokezaka:2012ze}. Since the ejected material is composed of neutron-rich matter, heavy radioactive nuclei can be synthesized in the ejecta by the so-called {\it r-process} nucleosynthesis~\citep{Lattimer:1974slx,Eichler:1989ve,Korobkin:2012uy,Wanajo:2014wha}, and EM emission in the optical and infrared wavelengths could be powered by radioactive decays of heavy elements~\citep{Li:1998bw,Kulkarni:2005jw,Metzger:2010sy,Kasen:2013xka,Tanaka:2013ana}. Previous studies~\citep{Li:1998bw,Kasen:2013xka,Kasen:2014toa,Barnes:2016umi,Wollaeger:2017ahm,Tanaka:2017qxj,Tanaka:2017lxb} show that light curves of kilonovae depend on the mass, velocity, and opacity of ejecta. The peak luminosity and the time at the peak luminosity are approximately estimated as
\begin{eqnarray}
	L_{\rm peak}&\approx& 1.3\times10^{42}\,{\rm erg s^{-1}}\left(\frac{f}{10^{-6}}\right)\left(\frac{M}{0.03\,M_\odot}\right)^{1/2}\nonumber\\
	&\times&\left(\frac{v}{0.2\,c}\right)^{1/2}\left(\frac{\kappa}{1\,{\rm cm^2g^{-1}}}\right)^{-1/2}\label{eq:Lpeak}
\end{eqnarray}
and
\begin{eqnarray}
	t_{\rm peak}&\approx& 4.9\,{\rm days}\nonumber\\
	&\times&\left(\frac{M}{0.03\,M_\odot}\right)^{1/2}\left(\frac{v}{0.2\,c}\right)^{-1/2}\left(\frac{\kappa}{1\,{\rm cm^2g^{-1}}}\right)^{1/2},\nonumber\\\label{eq:tpeak}
\end{eqnarray}
where $M$, $v$, $\kappa$, $f$, and $c$ are the ejecta mass, velocity, opacity, heating efficiency with respect to the restmass energy (see Ref.~\cite{Li:1998bw} for the detail), and the speed of light, respectively. In particular, previous studies~\citep{Kasen:2013xka,Tanaka:2013ana,Kasen:2014toa,Barnes:2016umi,Wollaeger:2017ahm,Tanaka:2017qxj,Tanaka:2017lxb} show that opacity of ejecta varies significantly ($\kappa=0.1$--$10\,{\rm cm^2g^{-1}}$) depending on the electron fraction of ejecta ($Y_e$, number of protons per nucleon which controls the final element abundances). Thus, it is crucial to understand the properties of the ejecta, such as the mass, density distribution, and composition, to predict the kilonova light curves.

These properties of the ejecta depend on the  masses (and spins) of the binary components, and NS equation of state, and mass ejection mechanism. In particular, the ejecta properties reflect the fate of the system after the merger. Thus, the kilonova light curves will carry the physical information of the merged binary and the post-merger evolution of the system as complementary information to that inferred by the GW data analysis. For example, a number of studies have shown that the optical and infrared EM counterparts to GW170817 are consistent with kilonova models composed of multiple components~\citep[e.g.,][]{Kasliwal:2017ngb,Cowperthwaite:2017dyu,Kasen:2017sxr,Villar:2017wcc,Kawaguchi:2018ptg} suggesting that a substantial fraction is likely to be lanthanide-free. The presence of lanthanide-free ejecta indicates that the remnant NS is temporarily survived before it collapsed eventually to a black hole (BH).

While detailed observational data obtained for the kilonova associated with GW170817 would be a standard guideline, next events would not be always similar to the previous event. As we discuss in the next section, kilonova light curves could show a large diversity reflecting the variety in the ejecta properties that depend on the binary parameters or the binary composition. Indeed, diversity of kilonova light curves has been suggested in the previous studies on the kilonova candidates found in the archive observational data of short gamma-ray-burst afterglows~\citep{Gompertz:2017mbv,Ascenzi:2018mbh,Rossi:2019fnm}. Thus, in this work, we perform radiative transfer simulations for kilonova light curves in various situations predicted by numerical-relativity simulations to investigate their possible diversity. In our previous study~\citep{Kawaguchi:2018ptg}, we performed an axisymmetric radiative transfer simulation for kilonovae taking the radiative transfer effect of photons in the multiple ejecta components of non-spherical morphology into account, and demonstrated that such photon transfer effect could have a large impact on predicting the light curves. In this paper, we also show how this effect play roles by performing the calculations for a variety of ejecta models. 

This paper is organized as follows: In Section~\ref{sec:ejecta}, we briefly summarize the variety in the ejecta properties predicted by recent numerical simulations. In Section~\ref{sec:method}, we describe our method for radiative transfer simulations and the setups for ejecta profile, and summarize the models which we study in this paper. In Section~\ref{sec:multi}, we study how large the impact of the radiative transfer effect of photons in the multiple ejecta components to the light curve predictions could be, and show the dependence of the light curves on the ejecta parameters. An interpretation of the optical and infrared EM counterparts to GW170817 based on our new calculation is also discussed in Section~\ref{sec:gw170817}. In Section~\ref{sec:variety}, we show a variety of NS merger models, such as the cases of prompt collapse to a BH from NS mergers, high-velocity ejecta possibly accelerated by magnetars, and a black hole-neutron star (BH-NS) merger. We summarize our findings of this work in Section~\ref{sec:summary}. Throughout the paper, magnitudes are given in the AB magnitude system.

\section{Variety in the ejecta properties}\label{sec:ejecta}

\begin{figure*}
 	 \includegraphics[width=1.\linewidth]{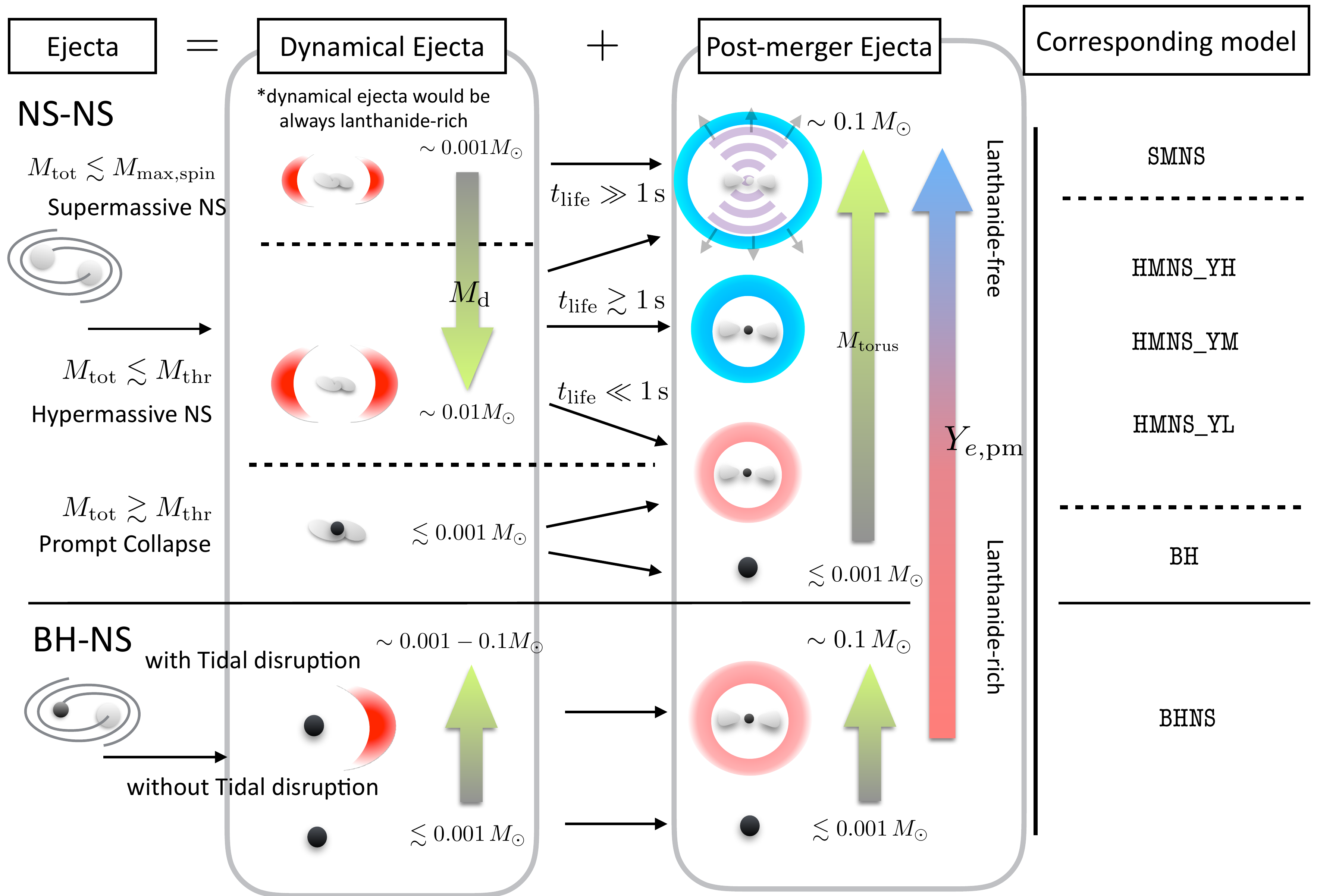}
 	 \caption{Schematic picture for the post-merger evolution and the typical properties of ejecta for NS-NS and BH-NS binaries in various situations. We note that our ejecta model is consist of two parts; the dynamical ejecta with non-spherical geometry and post-merger ejecta with spherical geometry. $M_{\rm tot}$, $M_{\rm max,spin}$, $M_{\rm thr}$, $M_{\rm d}$, $M_{\rm torus}$, and $t_{\rm life}$ are the total mass of the binary, maximum mass of a rigidly rotating NS, threshold total mass for the prompt collapse, dynamical ejecta mass, remnant torus mass, and the timescale for the remnant to collapse to a BH, respectively. The correspondence between each situation and kilonova models listed in Table~\ref{tb:model} is summarized in the right side of the figure. We note that this figure shows only a simplified overview for the typical scenarios, and quantitative properties of the post-merge evolution and mass ejection depend on the detail of the binary parameters, such as the mass ratio, spins, and equation of state of NS (see references mentioned in the main text).}\label{fig:pic}
\end{figure*}

In this section, we briefly summarize the variety in the ejecta properties predicted by numerical-relativity simulations for NS mergers. For the last decade, many numerical simulations for NS-NS mergers have been performed to investigate the properties of the ejected material. Those work reveal that there are broadly two types of mass ejection mechanisms. One is called dynamical mass ejection, which occurs within $\sim10\,{\rm ms}$ after the onset of mergers. In this timescale, the NS material is ejected by the angular momentum transport via tidal force of the merger remnant or by the shock heating which takes place in the collisional surface of the NSs~\citep{Hotokezaka:2012ze,Bauswein:2013yna,Sekiguchi:2015dma,Sekiguchi:2016bjd,Radice:2016dwd,Dietrich:2016hky,Bovard:2017mvn}. Second is called post-merger ejection, which occurs in the timescale up to $\sim 0.1$--$10\,{\rm s}$ after the mergers. Post-merger ejecta is launched from the remnant of the merger, such as a system composed of a massive NS or BH with a massive torus, driven by amplified magnetic fields or effective viscous heating due to magnetic turbulence~\citep{Dessart:2008zd,Metzger:2014ila,Perego:2014fma,Just:2014fka,Siegel:2017nub,Shibata:2017xdx,Lippuner:2017bfm,Fujibayashi:2017puw}.

Figure~\ref{fig:pic} summarizes the possible scenarios for the post-merger evolution and the typical properties of ejecta for a variety of NS mergers. For a NS-NS merger, the fate after the merger can be broadly categorized into three cases: 

\paragraph{Prompt collapse to a BH: }One is the case in which the NSs collapse promptly to a BH. We refer to this case as a prompt collapse case in the following. For most of the prompt collapse cases, both dynamical ejecta and remnant torus mass are suppressed, and typically they are $\lesssim 0.001\,M_\odot$ and $\lesssim 0.01\,M_\odot$, respectively, unless the mass ratio is $\alt 0.8$~\citep{Kiuchi:2009jt,Hotokezaka:2012ze}. Due to the absence of neutrino radiation from the remnant, the electron fraction in the post-merger ejecta would remain to be low, and as a consequence, the ejecta would be lanthanide-rich for the prompt-collapse cases~\citep{Metzger:2014ila,Wu:2016pnw,Lippuner:2017bfm,Siegel:2017nub,Fernandez:2018kax}. 

\paragraph{Hypermassive massive neutron star (HMNS): }
Second is the case in which a hypermassive NS (HMNS) is formed temporarily and the remnant collapses to a BH after surviving for more than $10\,{\rm ms}$. For this case, mass of the dynamical ejecta and the remnant torus can be massive up to $\sim 0.001$--$0.01\,M_\odot$ and $\sim 0.1\,M_\odot$, respectively~\citep{Hotokezaka:2012ze,Bauswein:2013yna,Sekiguchi:2015dma,Sekiguchi:2016bjd,Radice:2016dwd,Dietrich:2016hky,Bovard:2017mvn}. $Y_e$ of the dynamical ejecta can be raised due to electron/positron capture and the neutrino irradiation from the remnant~\citep{Sekiguchi:2015dma}. The previous studies show that the dynamical ejecta in the polar region can be lanthanide-free while that in the equatorial plane would remain to be lanthanide-rich. $Y_e$ of the post-merger ejecta depends strongly on the lifetime of the remnant NS. The previous studies also show that post-merger ejecta would be lanthanide-free if the remnant NS is sufficiently long-surviving ($t_{\rm life}\gtrsim1\,{\rm s}$, where $t_{\rm life}$ is the timescale for the remnant to collapse to a BH), while a substantial amount of lanthanide is synthesized if the remnant collapses to a BH in a short timescale~\citep{Metzger:2014ila,Perego:2014fma,Wu:2016pnw,Siegel:2017nub,Fernandez:2018kax,Lippuner:2017bfm,Fujibayashi:2017puw}.
 
\paragraph{Long-lived super massive neutron star (SMNS): }
Third is the case in which the remnant NS survives for a long period ($t_{\rm life}\gg1{\rm s}$) or does not collapse to a BH. Such a situation can be realized if the total mass of the binary is close to or smaller than the maximum mass of a rigidly rotating NS (a supermassive NS; SMNS). For such a case, the mass of the dynamical ejecta would be relatively small (order of $10^{-3}\,M_\odot$) unless the mass ratio of the binary is far from unity~\citep{Hotokezaka:2012ze,Bauswein:2013yna,Foucart:2015gaa,Radice:2016dwd,Dietrich:2016hky,Bovard:2017mvn}. On the other hand, the post-merger ejecta could be massive ($\sim 0.01-0.1\,M_\odot$) due to large remnant torus mass, and it would be lanthanide-free due to neutrino irradiation~\citep{Fujibayashi:2017puw}.

In addition, the rotational kinetic energy  of the remnant NS could be an additional energy source to the ejecta by releasing it through the EM radiation, and could modify the light curves for the early phase $\lesssim1\,{\rm days}$~\citep{Metzger:2013cha,Shibata:2017xdx,Margalit:2017dij}. We note that the efficiency and timescale for releasing the rotational kinetic energy of the remnant NS to the ejecta are currently quite uncertain. In particular, if the timescale of the energy injection is much shorter than the timescale of the kilonova emission, $\sim 1$--$10$ days, energy injected into the ejecta would be lost by adiabatic expansion, and would not be directly reflected to the light curves. However, even for such a case, the light curves could show different feature from the case in the absence of the energy injection from the remnant, because the ejecta profile would be modified by the increase in its kinetic energy. Indeed, the rotational kinetic energy of the NS could be as large as $\approx1$-$2\times10^{53}\,{\rm erg}$ for the case of mass shedding limit~\citep[e.g.,][]{Shibata:2017xdx,Margalit:2017dij,Shibata:2019ctb}, and it is sufficiently large to modify the velocity distribution of the ejecta even if only a small fraction of the rotational kinetic energy is converted to the kinetic energy of ejecta, which is typically $\sim10^{49}$--$10^{51}\,{\rm erg}$. In this work, for simplicity, we focus on the case that the activity of the remnant NS contributes only to changing the ejecta profile. There may also be a case that the remnant activity directly affects the light curves~\citep{Wollaeger:2019igm}.

\paragraph{Black hole-neutron star merger (BH-NS): }
BH-NS mergers can be also associated with the EM counterparts if the NS is tidally disrupted before the merger~\citep{Rosswog:2005su,Shibata:2007zm,Etienne:2008re,Lovelace:2013vma,Kyutoku:2015gda,Foucart:2019bxj,Foucart:2018rjc,Kawaguchi:2016ana}. Masses of dynamical ejecta and the remnant torus can be much more and/or less massive than those formed in a NS-NS merger, and their values depend strongly on the parameters of the binary, such as the NS radius, mass ratio of the BH to the NS, and, in particular, the BH spin. Because of the absence of the strong shock heating effect in the merger and post-merger processes, $Y_e$ of the dynamical ejecta keeps the original value of the NS, $Y_e\lesssim0.1$~\citep{Rosswog:2012wb,Just:2014fka,Foucart:2016vxd,Kyutoku:2017voj}. $Y_e$ of the remnant torus and that of the post-merger ejecta could raise due to viscous heating, but yet, a substantial amount of the post-merger ejecta would be lanthanide-rich as in the prompt collapse of NS-NS cases.

\section{Method}\label{sec:method}

\subsection{Radiative Transfer simulation}

We calculate the light curves of kilonovae/macronovae by a wavelength-dependent radiative transfer simulation code~\citep{Tanaka:2013ana,Tanaka:2017qxj,Tanaka:2017lxb}. The photon transfer is calculated by a Monte Carlo method for given ejecta profiles of density, velocity, and element abundance. The nuclear heating rates are determined employing the results of r-process nucleosynthesis calculations by~\cite{Wanajo:2014wha}. We also consider the time-dependent thermalization efficiency following an analytic formula derived by~\cite{Barnes:2016umi}. Axisymmetry is imposed for the matter profile, such as the density, temperature, abundance distribution. Special-relativistic effects on photon transfer and light travel time effects are fully taken into account.

For photon-matter interaction, we consider the same physical processes as in~\cite{Tanaka:2013ana,Tanaka:2017qxj,Tanaka:2017lxb}; bound-bound, bound-free, and free-free transitions and electron scattering for a transfer of optical and infrared photons. For the bound-bound transitions, which have a dominant contribution in the optical and infrared wavelengths, we use the formalism of the expansion opacity~\citep{1993ApJ...412..731E,Kasen:2006ce}. For atomic data, the updated line list calculated in~\cite{Tanaka:2019iqp} is employed. This line list is constructed by an atomic structure calculation for all the elements from $Z=26$ to $Z=92$, and supplemented by Kurucz's line list for $Z < 26$~\citep{1995all..book.....K}, where $Z$ is the atomic number. The updated atomic data include up to triply ionization for all the ions. The ionization and excitation states are calculated under the assumption of local thermodynamic equilibrium (LTE) by using the Saha ionization and Boltzmann excitation equations.

\subsection{Ejecta profile}
\begin{figure}
 	 \includegraphics[width=1.1\linewidth]{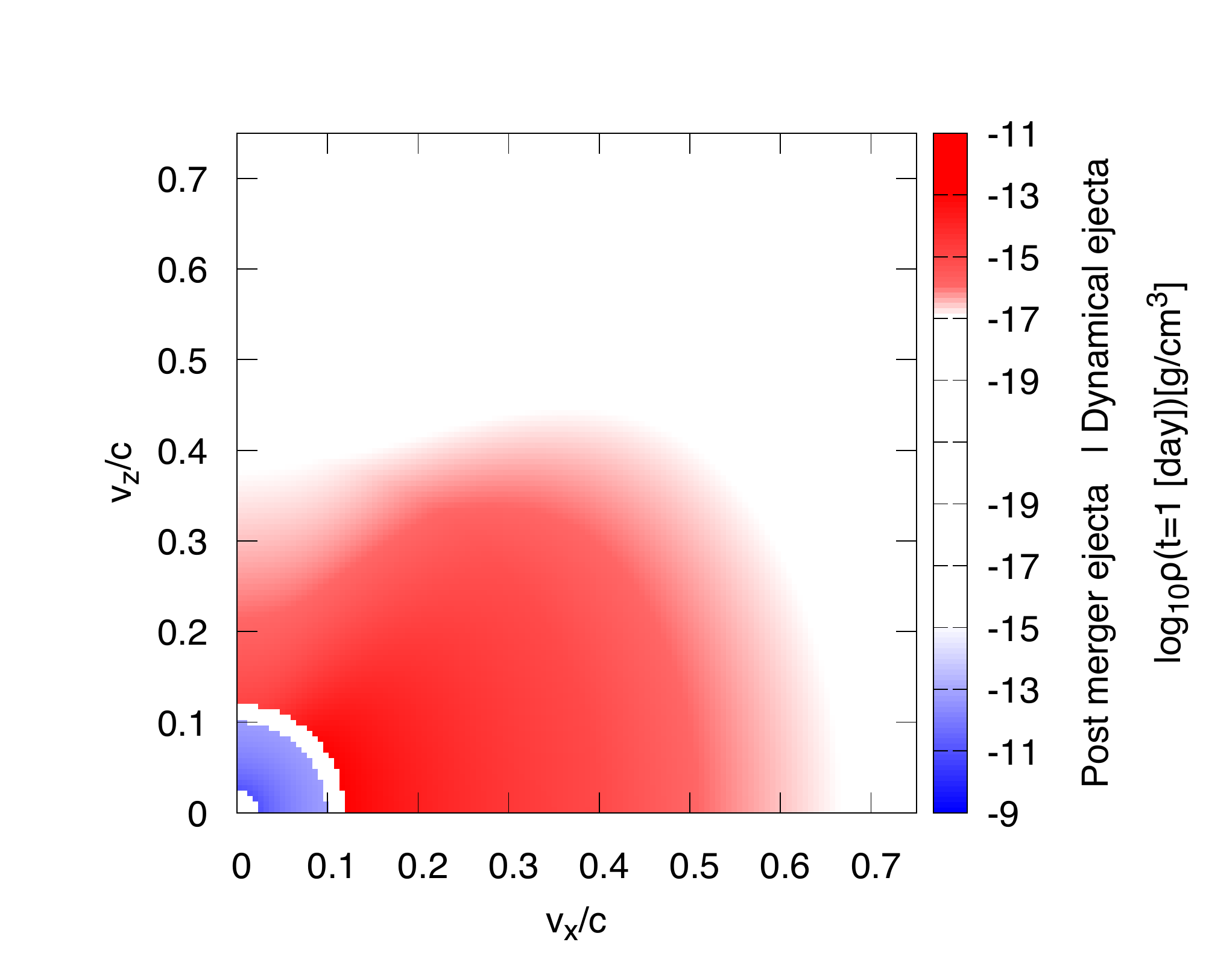}\\
 	 \includegraphics[width=1.1\linewidth]{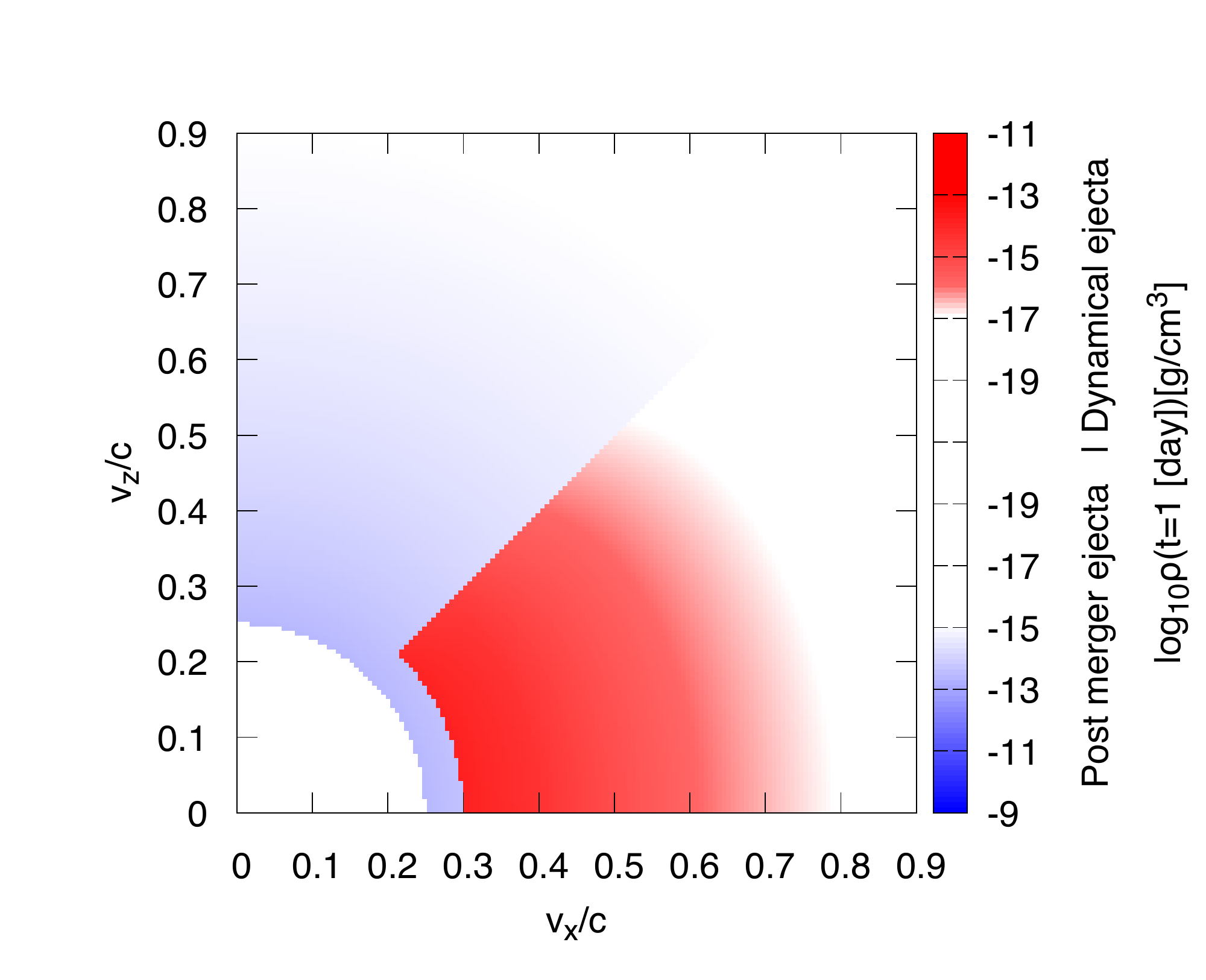}
 	 \caption{Density profile of the ejecta employed in the radiative transfer simulation for NS-NS merger models. The red and blue regions denote the dynamical and post-merger ejecta, respectively. Homologous expansion of the ejecta and axisymmetry around the rotational axis ($z$-axis) are assumed in the simulation. The top panel denotes the case of the fiducial model ({\tt HMNS\_YH}) in Table~\ref{tb:model} for which we assume that the outer edge of the post-merger ejecta is slower than the inner edge of the dynamical ejecta. The bottom panel denotes the case of {\tt SMNS\_DYN0.01} in Table~\ref{tb:model}. In this model, we suppose that the post-merger ejecta is accelerated by the energy injection from the merger remnant.}
	 \label{fig:dens}
\end{figure}
\begin{figure}
 	 \includegraphics[width=0.9\linewidth]{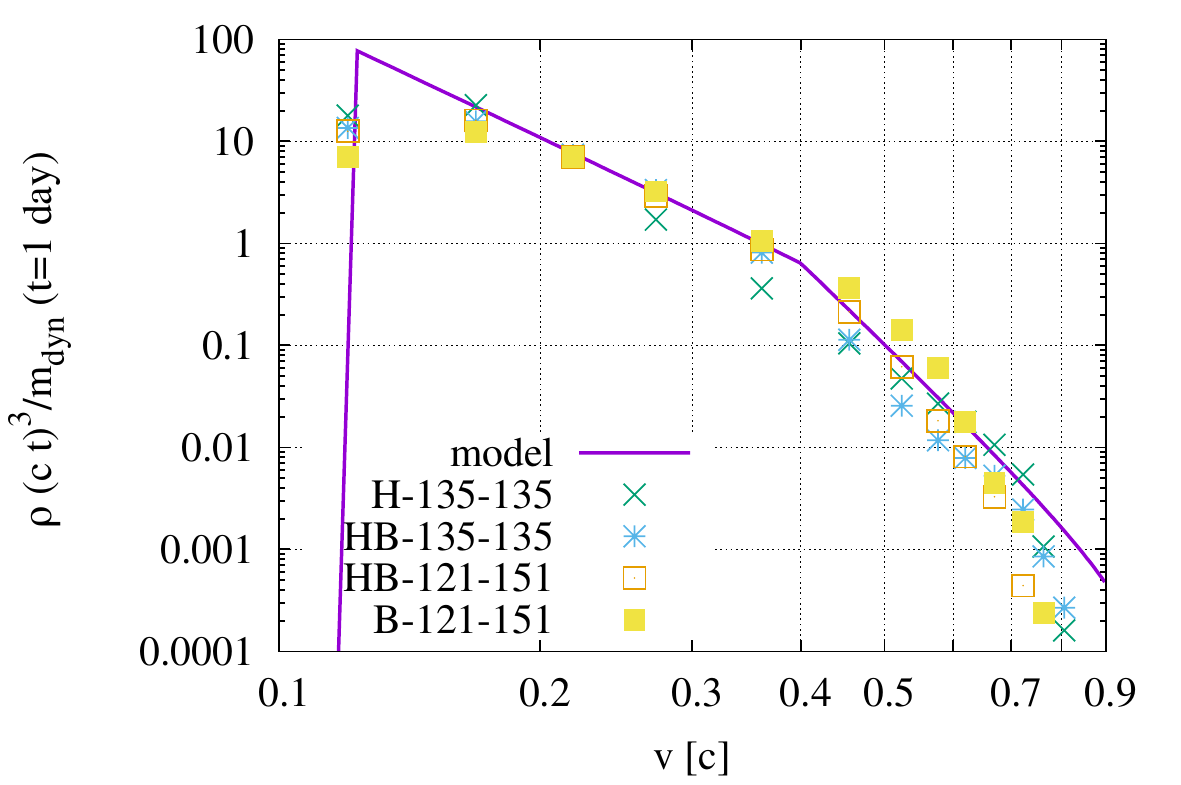}
 	 \caption{Comparison of the radial density profile of the dynamical ejecta employed in this work (the blue lines) and that obtained by numerical relativity simulations (data points)~\citep{Kiuchi:2017pte,Hotokezaka:2018gmo}.}\label{fig:dyn_dens}
\end{figure}
\begin{figure}
 	 \includegraphics[width=1.1\linewidth]{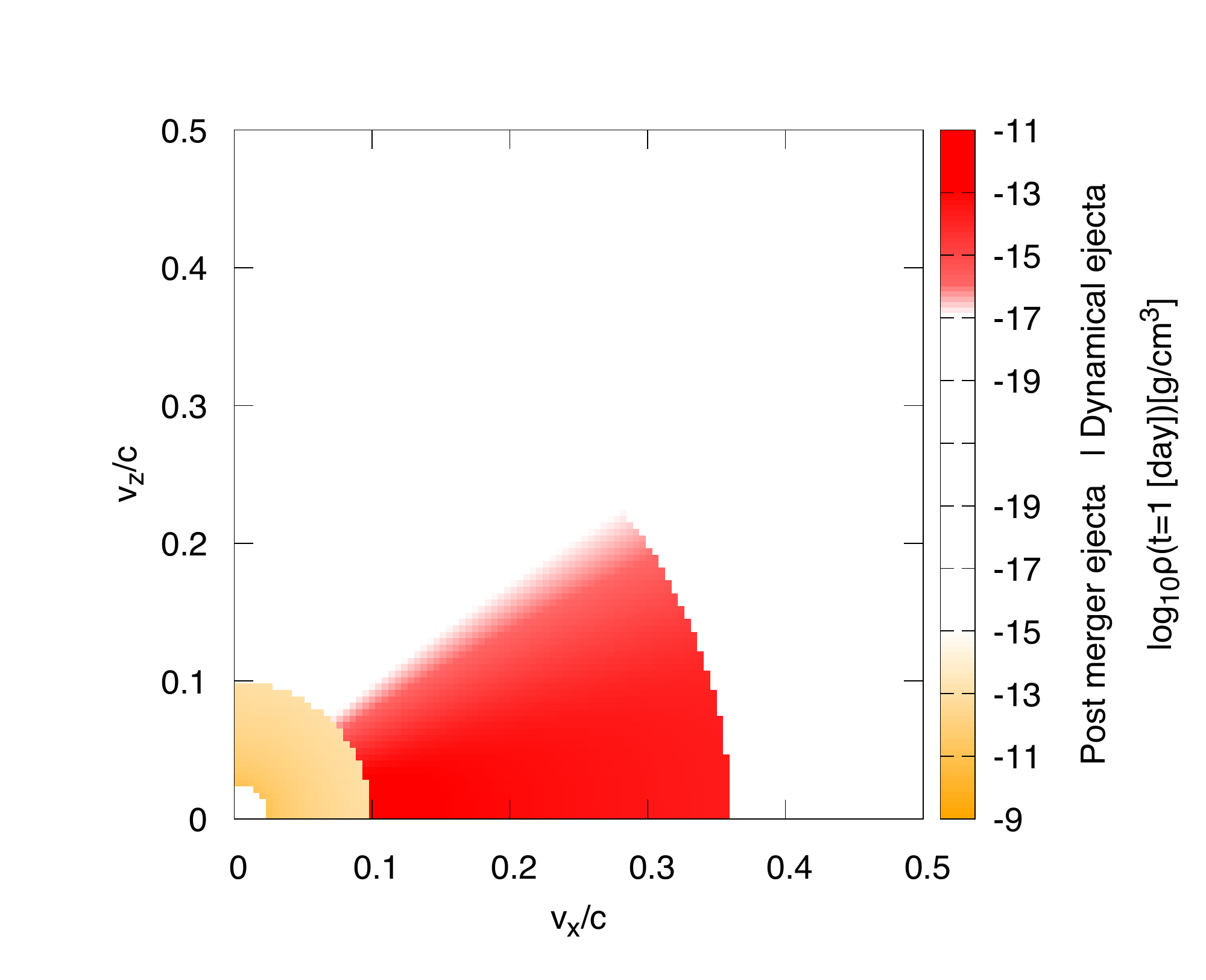}\
 	 \caption{Density profile of the ejecta employed in the radiative transfer simulation for the BH-NS merger ejecta. The density profile of {\tt BHNS\_A} in Table~\ref{tb:model} is shown as an example. The red and orange regions denote the dynamical and post-merger ejecta, respectively.}\label{fig:dens_bhns}
\end{figure}

In this work, we employ ejecta models that consist of two parts; the dynamical ejecta with non-spherical geometry and post-merger ejecta with spherical geometry. In particular, we employ three different types of ejecta profiles. First, we consider the fiducial case in which the post-merger ejecta is slower than the dynamical ejecta. Second, we consider the case in which both ejecta are supposed to be accelerated by an additional energy injection. Third, we consider an ejecta profile model for a BH-NS merger.

Our first ejecta model is composed of a homologously expanding ejecta with its velocity distributing from $v=v_{\rm pm,min}$ to $0.9\,c$. Here, the post-merger ejecta has the velocity from $v=v_{\rm pm,min}$ to $v_{\rm pm,max}$, and dynamical ejecta from $v=v_{\rm d,min}$ to $0.9\,c$, $v=r/t$ the velocity of the fluid elements, $r$ the radial coordinate, and $t$ time measured from the onset of a merger. The density profile of each ejecta component is given by
\begin{equation}
	\rho\left(r,t\right)\propto
	\left\{\begin{array}{cc}
		r^{-3} t^{-3}&, v_{\rm pm,min}\le r/t \le v_{\rm pm,max}\\
		\eta\left(\theta\right)r^{-4} t^{-3}&, v_{\rm d,min}\le r/t\le 0.4\,c,\\
		\eta\left(\theta\right)r^{-8}t^{-3}&, 0.4\,c\le r/t\le 0.9\,c,
	\end{array}
	\right.\label{eq:dens}
\end{equation}
and a floor value of $10^{-20} (t/{\rm day})^{-3} {\rm g/cm^3}$ is employed for the density in $v<v_{\rm pm,min}$ and $v_{\rm pm,max}\le v \le v_{\rm d,min}$ (see Figure~\ref{fig:dens}). We employ a broken-power law profile for the radial profile of the dynamical ejecta to model the results obtained by numerical-relativity simulations~\citep{Kiuchi:2017pte,Hotokezaka:2018gmo}. $\eta\left(\theta\right)$ is a function defined by
\begin{equation}
	\eta\left(\theta\right)=(1-\Theta(\theta))f_{\rm d,pol}+\Theta(\theta),
\end{equation}
\begin{equation}
	\Theta(\theta)=\frac{1}{1+{\rm exp}\left[-10\left(\theta-\pi/4\right)\right]},
\end{equation}
which is introduced to describe the angular distribution of the dynamical ejecta with $\theta$ an angle measured from the polar axis. By employing this function, the density in the polar region ($0\le\theta\le\pi/4$) is smaller than that in the equatorial region ($\pi/4\le\theta\le\pi/2$) approximately by a factor of $f_{\rm d,pol}$ (we assume the reflection symmetry with respect to the equatorial plane). We employ $0.025\,c$ as a fiducial value for $v_{\rm pm,min}$, while the results depend only weakly on $v_{\rm pm,min}$ as far as $v_{\rm pm,min}=0.025$--$0.05\,c$. The angular dependence of the density profile and element abundances is not incorporated for the post-merger ejecta for simplicity because it is not significant as is the case for the dynamical ejecta~\citep{Fujibayashi:2017puw,Fernandez:2018kax}.

Following the previous study~\citep{Kawaguchi:2018ptg}, element abundances are determined employing the results of r-process nucleosynthesis calculations by~\cite{Wanajo:2014wha}. For the dynamical ejecta in the equatorial region, we employ elements abundances resulting from a flat $Y_e$ distribution in $0.1$--$0.3$. Taking into account the results of ~\cite{Radice:2016dwd,Sekiguchi:2015dma,Sekiguchi:2016bjd,Perego:2017wtu}, we gradually change the element abundances to that of a flat $Y_e$ distribution in $0.35$--$0.44$ as the location approaches the polar direction to describe the $Y_e$ angular dependence of the dynamical ejecta. Specifically, the mass fraction of element $Z$, $X_Z$, is given by 
\begin{equation}
		(1-\Theta(\theta))X_Z(Y_e:0.35{\rm -}0.44)+\Theta(\theta)X_Z(Y_e:0.1{\rm -}0.3),
\end{equation} 
where $X_Z(Y_e:0.35{\rm -}0.44)$ and $X_Z(Y_e:0.1{\rm -}0.3)$ are the mass fractions calculated assuming flat $Y_e$ distributions in $0.35$--$0.44$ and in $0.1$--$0.3$, respectively. For post-merger ejecta, we employ either flat $Y_e$ distributions in $0.3$--$0.4$, $0.2$--$0.4$, or $0.1$--$0.3$ to study the dependence of light curves on the $Y_e$ distribution. We note that the lanthanide mass fractions of flat $Y_e$ distributions in $0.3$--$0.4$, $0.2$--$0.4$, and $0.1$--$0.3$ are $\ll10^{-3}$, $\approx0.025$, and $\approx0.14$, respectively.

As mentioned above, the post-merger ejecta could be accelerated due to an energy injection from the remnant NS after the merger. For such a situation, the outer edge of the post-merger ejecta can interact with the inner edge of the dynamical ejecta, and particularly, a fraction of the post-merger ejecta can spread into the polar region of the dynamical ejecta. Hydrodynamical simulations would be needed to obtain a realistic distribution of the ejecta. However, such a task is beyond the scope of this work. Instead, for simplicity, we employ the density profile described in Eq.~(\ref{eq:dens}) allowing the outer edge of the post-merger ejecta, $v_{\rm pm,max}$, to be larger than the inner edge of the dynamical ejecta, $v_{\rm d,min}$, but truncating and replacing the post-merger ejecta by the dynamical ejecta for $\theta>\pi/4$ and $v>v_{\rm d,min}$ to keep the dense part of dynamical ejecta (see Figure~\ref{fig:dens}). This is our second choice of the ejecta profile.

For a BH-NS merger, the ejecta profile could be different from those for NS-NS mergers, and thus, we consider the third profile. For this case, the dynamical ejecta is typically expected to be more confined in the equatorial plane, and the radial density profile is more shallower than that formed in a NS-NS merger. Based on the results of numerical-relativity simulations~\citep{Foucart:2016vxd,Kyutoku:2017voj}, we employ the following density profile for the BH-NS ejecta models:
\begin{equation}
	\rho\left(r,t\right)\propto
	\left\{\begin{array}{cc}
		r^{-3} t^{-3}&, 0.025\,c\le r/t \le 0.1\,c\\
		{\tilde \Theta}\left(\theta\right)r^{-2} t^{-3}&, 0.1\,c\le r/t\le 0.36\,c
	\end{array}
	\right.,\label{eq:dens_bhns}
\end{equation}
where ${\tilde \Theta}\left(\theta\right)$ is given by
\begin{equation}
	{\tilde \Theta}(\theta)=\frac{1}{1+{\rm exp}\left[-20\left(\theta-1.2\,{\rm [rad]}\right)\right]}.
\end{equation}
Here, the outer edge of the dynamical ejecta ($v_{\rm d,max}=0.36\,c$) is determined from the condition that the average velocity of the dynamical ejecta, $\sqrt{2E_{\rm K,d}/M_{\rm d}}$, is $0.25\,c$ with $E_{\rm K,d}$ the kinetic energy of the dynamical ejecta~\citep{Foucart:2016vxd,Kyutoku:2017voj}. For BN-NS mergers,  collisional shock heating of the NS or neutrino irradiation from the merger remnant is weak, and hence, substantial amount of dynamical ejecta and the post-merger ejecta would have low $Y_e$ values. Taking the prediction obtained by numerical simulations into account~\citep{Rosswog:2012wb,Just:2014fka,Foucart:2016vxd,Kyutoku:2017voj}, flat $Y_e$ distributions in $0.09$--$0.11$\footnote{We note that the $Y_e$ range of $0.09$--$0.11$ is employed due to the limitation in available tables for element abundances and heating rate, while the numerical simulations~\citep{Rosswog:2012wb,Just:2014fka,Foucart:2016vxd,Kyutoku:2017voj} predict lower values for $Y_e$ of the dynamical ejecta (typically the lowest value could be $\approx0.05$). Nevertheless, we expect that the resulting lanthanide fraction and heating rate would be different at most $20$--$30\%$~\citep{Lippuner:2015gwa}.} and in $0.1$--$0.3$ are employed for the elements abundances of the dynamical ejecta and the post-merger ejecta, respectively. 

\subsection{Limitation}

Before showing the numerical results, we clarify the limitation of our light curve prediction. Employing new line list, our calculations are applicable for the ejecta of the temperature $\alt20,000\,{\rm K}$~\citep{Tanaka:2019iqp}. However, ejecta temperature can be higher than $20,000\,{\rm K}$ for $\lesssim1$ day after the merger particularly for the case that post-merger ejecta is lanthanide-rich or slow ($\lesssim0.05\,c$). Moreover, in the late time of the calculation, assuming the LTE, which we impose to determine the temperature, emissivity, etc., could be no longer valid. For typical cases which we study in this work, the post-merger ejecta becomes completely optically thin for $\gtrsim10\,{\rm days}$ for lanthanide-free cases. Thus, we focus on the results mainly for $1\,{\rm days}\le t\le 10\,{\rm days}$.

\subsection{Models}

\begin{table*}
\center
	\caption{Summary of Ejecta Models employed for radiative transfer simulations. The columns describe the model name, mass of post-merger ejecta $M_{\rm pm}$, mass of dynamical ejecta $M_{\rm d}$, thickness parameter for dynamical ejecta in the polar region $f_{\rm d,pol}$, velocity range of post-merger ejecta $v_{\rm pm,min}$--$v_{\rm pm,max}$, velocity range of dynamical ejecta $v_{\rm d,min}$--$v_{\rm d,max}$, $Y_e$ distribution of post-merger ejecta, and $Y_e$ distribution of dynamical ejecta in equatorial plane/polar axis, respectively. The average velocity calculated by $v_{\rm ave}=\sqrt{2E_{\rm K}/M}$ is shown in the parenthesis of the columns for ejecta velocity, where $E_{\rm K}$ and $M$ are the kinetic energy and mass of ejecta, respectively. * denotes that the same value as for the fiducial model ({\tt HMNS\_YH}) is employed.}\label{tb:model}
	\begin{tabular}{c|ccccccc}
		Model	&	$M_{\rm pm}\,[M_\odot]$&	$M_{\rm d}\,[M_\odot]$&	$f_{\rm d,pol}$&	$v_{\rm pm}\,(v_{\rm pm,ave})\,[c]$&	$v_{\rm d}\,(v_{\rm d,ave})\, [c]$		&	$Y_{e,{\rm pm}}$	&	$Y_{e,{\rm d}}$ (equatorial/polar)	\\\hline
{\tt HMNS\_YH} (Fiducial)	&	0.03	&	0.01	&	0.01		& 0.025--0.1	 (0.06)	& 0.12--0.9 (0.25)	&$0.3$--$0.4$	&$0.1$--$0.3$ / $0.35$--$0.44$\\
{\tt {\tt HMNS\_YH\_DYN0.003}}	&	*	&	0.003	&	*		& *		& *	&*	&*\\
{\tt {\tt HMNS\_YH\_PM0.01}}	&	0.01	&	*		&	*		& *		& *	&*	&*\\
{\tt {\tt HMNS\_YH\_VL}}	&	*	&	*	&	*		& 0.025--0.05 (0.04)	 & 0.12--0.9 (0.25)	&*	&*\\
{\tt {\tt HMNS\_YH\_VH}}	&	*	&	*	&	*		& 0.025--0.2 (0.10) 	& 0.2--0.9 (0.33) 	&*	&*\\
{\tt {\tt HMNS\_YM}}	&	*	&	*	&	*		& *		& *	&$0.2$--$0.4$	&*\\
{\tt {\tt HMNS\_YL}}	&	*	&	*	&	*		& *		& *	&$0.1$--$0.3$	&*\\\hline
{\tt {\tt GW170817\_YM}}	&	0.02	&	0.003	&	0.0		& 0.025--0.1 (0.06)		& 0.12--0.9 (0.25)	&$0.2$--$0.4$	&$0.1$--$0.3$ / $0.35$--$0.44$\\
{\tt {\tt GW170817\_YH}}	&	0.02	&	0.003	&	0.0		& 0.025--0.1 (0.06)		& 0.12--0.9 (0.25)	&$0.3$--$0.4$	&$0.1$--$0.3$ / $0.35$--$0.44$\\\hline
{\tt {\tt BH\_PM0.001}}	&	0.001	&	0.001	&	0.0	& 0.025--0.1 (0.06)		& 0.12--0.9 (0.25)	&$0.1$--$0.3$	&$0.1$--$0.3$ / $0.35$--$0.44$\\
{\tt {\tt BH\_PM0.01}}	&	0.01		&	0.001	&	0.0	& 0.025--0.1 (0.06)		& 0.12--0.9 (0.25)	&$0.1$--$0.3$	&$0.1$--$0.3$ / $0.35$--$0.44$\\\hline
{\tt {\tt SMNS\_DYN0.01}}	&	0.05	&	0.01&	0.01		& 0.25--0.9 (0.53)		& 0.3--0.9 (0.41)	&$0.3$--$0.4$	&$0.1$--$0.3$ / $0.35$--$0.44$\\
{\tt {\tt SMNS\_DYN0.003}}	&	0.05	&	0.003 &	0.01		& 0.25--0.9 (0.53)		& 0.3--0.9 (0.41)	&$0.3$--$0.4$	&$0.1$--$0.3$ / $0.35$--$0.44$\\\hline
{\tt {\tt BHNS\_A}}	&	0.02		&	0.02		&	--	& 0.025--0.1 (0.06)		& 0.1--0.36 (0.25)		&$0.1$--$0.3$	&$0.09$--$0.11$\\
{\tt {\tt BHNS\_B}}	&	0.04		&	0.01		&	--	& 0.025--0.1	 (0.06)	& 0.1--0.36 (0.25)		&$0.1$--$0.3$	&$0.09$--$0.11$\\
{\tt {\tt BHNS\_DYN}}	&	--		&	0.02		&	--	& --	 	& 0.1--0.36 (0.25)		& 	--	&$0.09$--$0.11$\\\hline
{\tt {\tt PM\_YH}}	&	0.03	&	--	&	--		& 0.025--0.1	 (0.06)		& --	&$0.3$--$0.4$	&--\\
{\tt {\tt PM\_YM}}	&	0.03	&	--	&	--		& 0.025--0.1 (0.06)		& --	&$0.2$--$0.4$	&--\\
{\tt {\tt PM\_YL}}	&	0.03	&	--	&	--		& 0.025--0.1 (0.06)		& --	&$0.1$--$0.3$	&--\\\hline
{\tt {\tt DYN0.01}}	&	--	&	0.01		&	0.01		& --		& 0.12--0.9 (0.25)	&--	&$0.1$--$0.3$ / $0.35$--$0.44$\\
{\tt {\tt DYN0.003}}	&	--	&	0.003	&	0.01		& --		& 0.12--0.9 (0.25)	&--	&$0.1$--$0.3$ / $0.35$--$0.44$\\\hline
	\end{tabular}
\end{table*}

In Table~\ref{tb:model}, we summarize the models and their parameters studied in this paper (see also Figure~\ref{fig:pic}). {\tt HMNS\_YH} corresponds to a model for which high-$Y_e$ post-merger ejecta is supposed to be formed due to neutrino irradiation from a long-surviving remnant NS. We set {\tt HMNS\_YH} as the fiducial setup. {\tt HMNS\_YH\_DYN0.003} and {\tt HMNS\_YH\_PM0.01} are the models with smaller dynamical ejecta mass and smaller post-merger ejecta mass than the fiducial model, respectively. These models are employed to study the mass dependence of the light curves. {\tt HMNS\_YH\_VL} and {\tt HMNS\_YH\_VH} are the models with lower velocity for post-merger ejecta and higher velocity for both ejecta components than the fiducial model, respectively. {\tt HMNS\_YM} and {\tt HMNS\_YL} are the models with same mass and velocity parameters as in the fiducial model but with low-$Y_e$ distributions for post-merger ejecta ($Y_e=0.2$--$0.4$ and $0.1$--$0.3$, respectively) employed to study the light curves in the presence of lanthanides in the post-merger ejecta. They model the cases in which the remnant collapses to a BH in a short timescale (a few ms) after the merger. {\tt GW170817\_YM} and {\tt GW170817\_YH} are the models of which ejecta parameters are chosen to reproduce the peak brightness of the optical and infrared counterpart to GW170817.

{\tt BH\_PM0.001} and {\tt BH\_PM0.01} are the models with lanthanide-rich post-merger ejecta of small masses. They model the cases in which the NSs collapse to a BH promptly after the merger and the dynamical ejecta mass and the torus mass of the remnant are suppressed as a consequence~\citep{Kiuchi:2009jt,Hotokezaka:2012ze}. In reference to the results of the prompt collapse case in previous studies~\citep[e.g.,][]{Kiuchi:2009jt,Hotokezaka:2012ze,Kiuchi:2019lls}, $0.001\,M_\odot$  and $0.01\,M_\odot$ are employed as the post-merger ejecta mass for {\tt BH\_PM0.001} and {\tt BH\_PM0.01}, respectively, while the dynamical ejecta mass is set to be $0.001\,M_\odot$ for both {\tt BH\_PM0.001} and {\tt BH\_PM0.01}. According to previous studies~\citep{Metzger:2014ila,Wu:2016pnw,Siegel:2017nub,Fernandez:2018kax,Lippuner:2017bfm}, post-merger ejecta is expected to be lanthanide-rich due to the absence of neutrino irradiation from the remnant NS. Thus, we employ a flat distribution of $Y_e=0.1$--$0.3$ for the post-merger ejecta.

{\tt SMNS\_DYN0.01} and {\tt SMNS\_DYN0.003} are the models composed of high-velocity post-merger ejecta. For these models, we suppose that the ejecta is accelerated by the energy injection from a long-surviving SMNS. In this study, we consider that the average velocity of the post-merger ejecta and dynamical ejecta are $\approx0.5\,c$ and $\approx0.4\,c$, respectively (see Table~\ref{tb:model}). These values correspond to the cases for which $\approx1\times10^{52}\,{\rm erg}$ of the rotational kinetic energy of remnant NS is converted to the ejecta kinetic energy. We set the post-merger ejecta mass to be $0.05\,M_\odot$ because the remnant torus mass would be large for a NS-NS merger which results in the long-surviving SMNS. $0.01\,M_\odot$ and $0.003\,M_\odot$ are employed for the dynamical ejecta mass for {\tt SMNS\_DYN0.01} and {\tt SMNS\_DYN0.003}, respectively. We note that the latter would be consistent with the prediction of numerical simulations because relatively small dynamical ejecta mass is typically predicted for a NS-NS binary with small total mass~\citep{Foucart:2015gaa}.

{\tt BHNS\_A}, {\tt BHNS\_B} and {\tt BHNS\_DYN} are the models for the BH-NS ejecta of density profile described in Eq.~(\ref{eq:dens_bhns}). As mentioned above, mass of the dynamical ejecta and the remnant torus depend strongly on the binary parameters. In this paper, we specifically study the models with $(M_{\rm pm},M_{\rm d})=(0.02\,M_\odot,0.02\,M_\odot)$ ({\tt BHNS\_A}) and $(0.04\,M_\odot,0.01\,M_\odot$) ({\tt BHNS\_B}) in reference to the results of ALF2-Q7a75 and H4-Q3a0 in~\cite{Kyutoku:2015gda}, respectively, for which massive ejecta and remnant torus are formed as a result of NS tidal disruption. Here, we assume that $\approx40\%$ of the remnant torus mass is ejected as the post-merger ejecta. In addition to these models, we also show the results of the BH-NS model only with the dynamical ejecta of mass $0.02\,M_\odot$ employing the ejecta density profile described in Eq.~(\ref{eq:dens_bhns}) ({\tt BHNS\_DYN}) to understand the basic properties of the light curves by the BH-NS dynamical ejecta.

{\tt PM\_YH}, {\tt PM\_YM}, {\tt PM\_YL}, {\tt DYN0.01}, and {\tt DYN0.003} are the models only with post-merger ejecta or dynamical ejecta for NS-NS mergers. These models are calculated to understand the basic properties of the light curves determined by each ejecta component. The results for these models are summarized in the appendix. 

\section{Effects of multiple ejecta components}\label{sec:multi}
In this section, we study how large the effect of radiative transfer of photons in the multiple ejecta components could be and show the dependence of light curves on the ejecta parameter.

\subsection{Fiducial model}
\begin{figure*}
 	 \includegraphics[width=1.\linewidth]{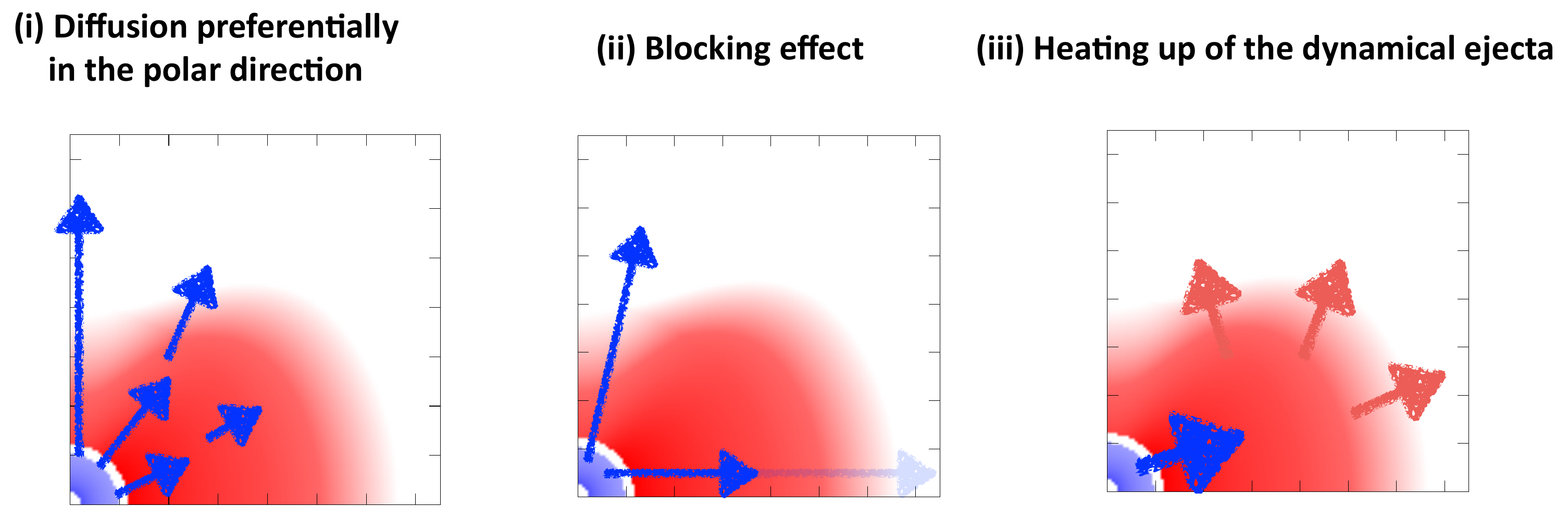}
 	 \caption{Schematic picture for the effects which are incorporated by taking the radiative transfer of photons in the multiple ejecta components into account (see the main text for the detail). (Left panel)  Preferential diffusion of photons in the polar direction. (Middle panel) Blocking effect of the emission from post-merger ejecta by the dynamical ejecta. (Right panel) Heating up of the dynamical ejecta by the post-merger ejecta. }
	 \label{fig:mc_eff}
\end{figure*}

The effect of radiative transfer of photons in the  multiple ejecta components has a large impact on the resulting light curves. There are three effects which take place by solving the photon radiation transfer in the multiple ejecta components consistently. The first one is the enhancement of the flux in the polar direction (see the left panel of Figure~\ref{fig:mc_eff}). In the presence of optically thick dynamical ejecta concentrated near the equatorial region, photons emitted from the post-merger ejecta in an early time $(\lesssim\,3$--$5\,{\rm days})$ preferentially diffuse in the polar direction, and this effectively enhances the energy flux observed from the polar region. The second is the blocking effect of the emission from post-merger ejecta by the dynamical ejecta~\citep{Kasen:2014toa}(see the middle panel of Figure~\ref{fig:mc_eff}). Because the post-merger ejecta is surrounded by an optically-thick dynamical ejecta, photons emitted from the post-merger ejecta cannot directly diffuse to the equatorial direction. This results in the strong suppression of the energy flux in the optical wavelengths observed from the equatorial direction. The third is the heating up of the dynamical ejecta by the post-merger ejecta (see the right panel of Figure~\ref{fig:mc_eff}). A fraction of photons emitted from the post-merger ejecta is absorbed and becomes additional heating source for the dynamical ejecta. Absorbed photons are re-emitted from the dynamical ejecta and enhance the flux particularly in the infrared wavelengths. 

\begin{figure*}
 	 \includegraphics[width=.5\linewidth]{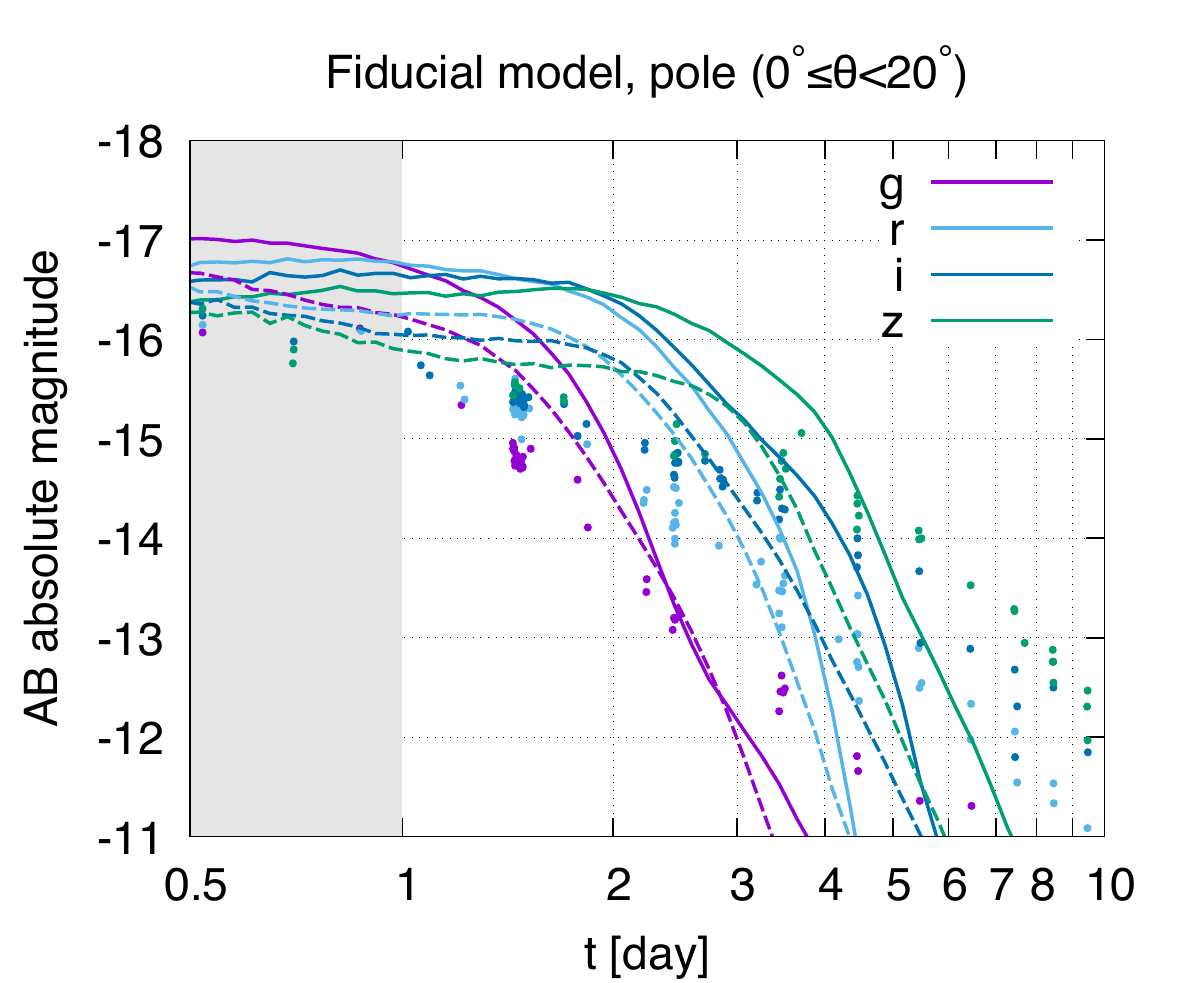}
 	 \includegraphics[width=.5\linewidth]{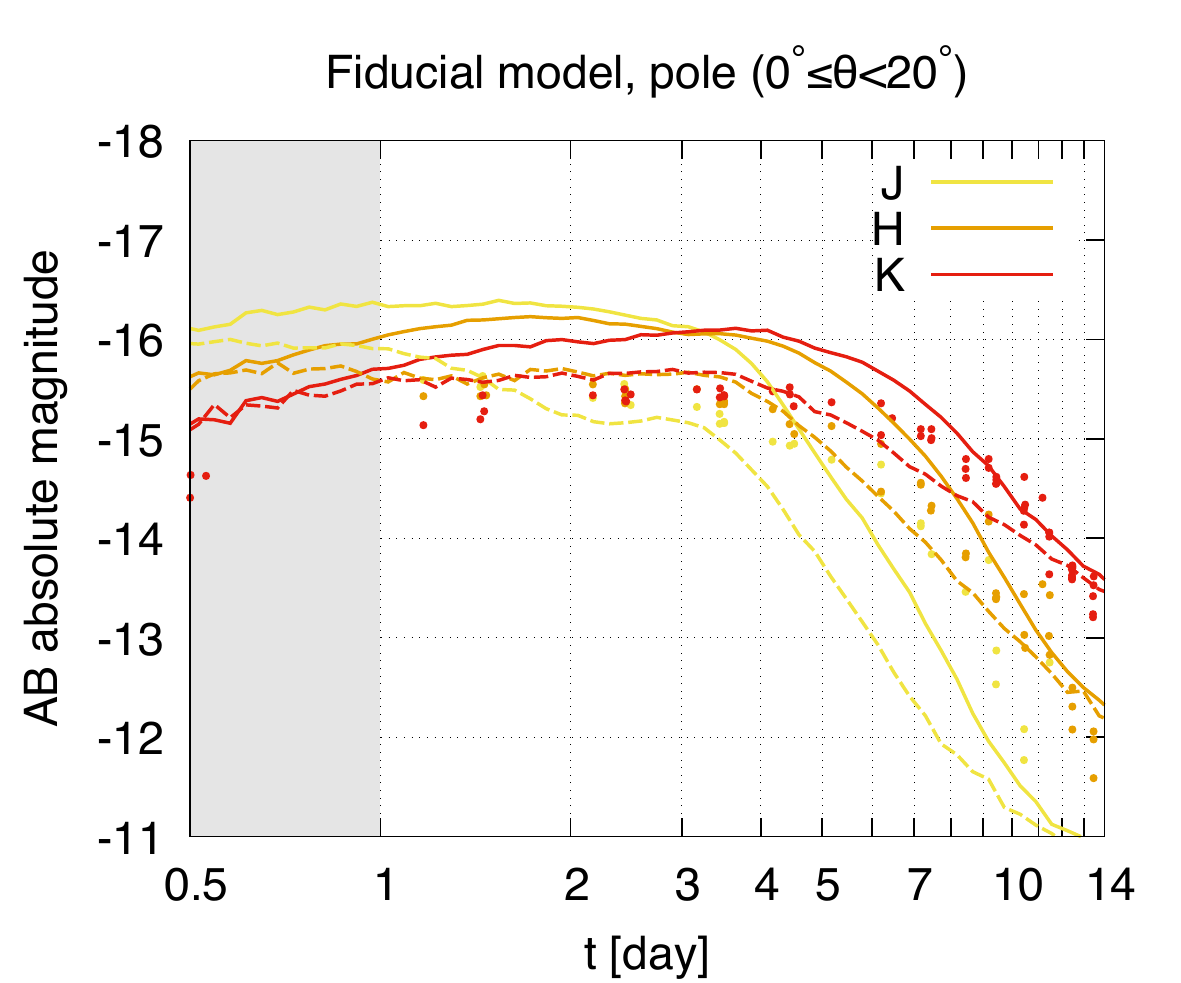}\\
 	 \includegraphics[width=.5\linewidth]{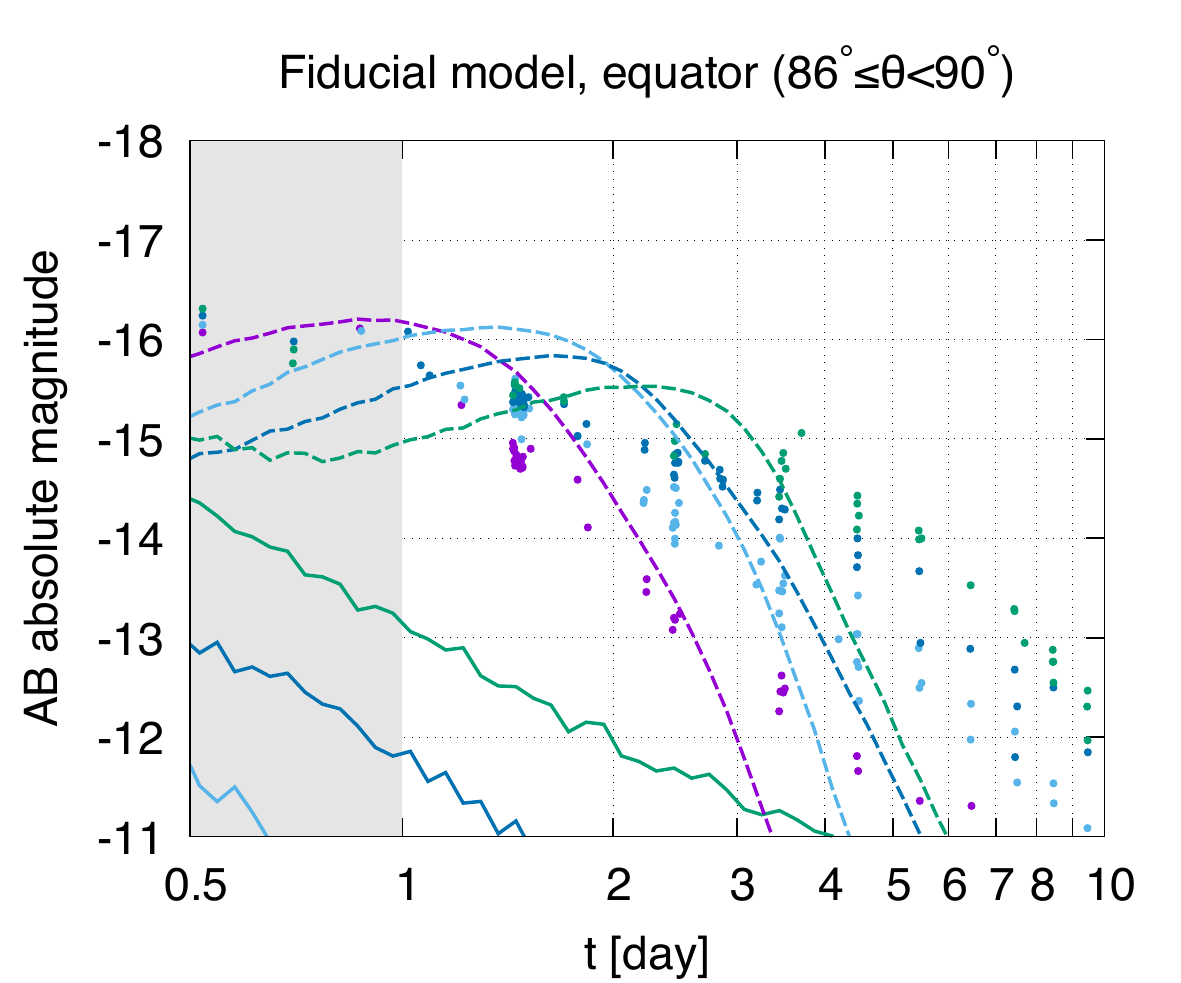}
 	 \includegraphics[width=.5\linewidth]{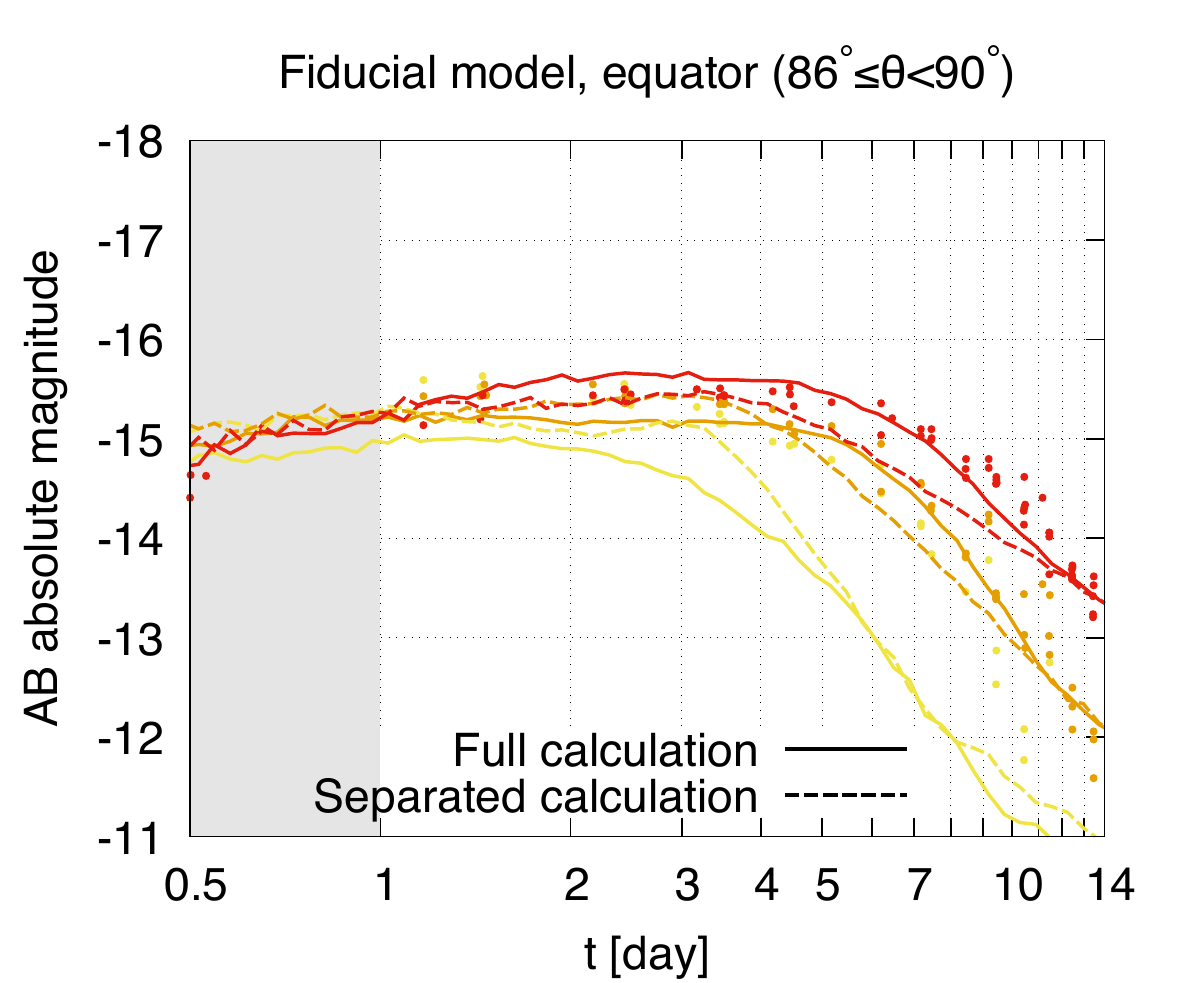}
 	 \caption{The {\it grizJHK}-band light curves for the fiducial model ({\tt HMNS\_YH}). The solid and dashed curves denote the light curves obtained by the calculation in which both post-merger and dynamical ejecta are solved together and in which each ejecta component are calculated separately and superimposed, respectively. Top and bottom panels denote the light curves observed from the polar ($0^\circ\le\theta\le20^\circ$) and equatorial direction ($86^\circ\le\theta\le90^\circ$), respectively. Left and right panels denote the {\it griz} and {\it JHK}-band light curves, respectively. The optical and infrared data points in GW170817 taken from~\cite{Villar:2017wcc} are shown as a reference assuming $40\,$Mpc for the distance to the event. The early parts of the light curves shedded in grey color ($t\le 1\,{\rm days}$) are the phase in which our calculations are not reliable due to too high ejecta temperature.}
	 \label{fig:mag_comp}
\end{figure*}
As an illustration, we show the results of the radiative transfer simulation for the {\tt HMNS\_YH} model in Figure~\ref{fig:mag_comp}. The {\it grizJHK}-band light curves observed from the polar direction ($0^\circ\le\theta\le20^\circ$, the top panels) and the equatorial direction  ($86^\circ\le\theta\le90^\circ$, the bottom panels) are shown in Figure~\ref{fig:mag_comp}. Here, the solid and dashed curves denote the light curves obtained by the simulation in which the radiative transfer effect of photons in the multiple ejecta components is taken into account (which we refer to as the full calculation in the following) and the calculation in which each ejecta component is treated separately and superimposed afterward, respectively. For the light curves observed from the polar direction, both optical and infrared emission is brighter for the full calculation than for the superimposed model. This clearly shows that photons preferentially diffuse out toward the polar direction. The photon fluxes are enhanced typically by $\approx0.5$--$1\,{\rm mag}$ particularly for the early phase ($t\lesssim7\,{\rm days}$). This indicates that ejecta mass could be overestimated by $\approx50$--$100\%$ if the radiative transfer effect of photons in the multiple ejecta components is not taken into account for the mass estimation from observed light curves. 

The enhancement of the {\it JHK}-band flux is most significant after $\approx 2\,{\rm days}$, while it is less significant in the earlier time. This implies that the {\it JHK}-band light curves after $\approx 2\,{\rm days}$ of the full calculation are dominated by the emission from the post-merger ejecta reprocessed in the dynamical ejecta. On the other hand, the less significant enhancement of the {\it JHK}-band flux for $\lesssim 2\,{\rm days}$ indicates that the emission powered by the radioactive heating in the dynamical ejecta also has a significant contribution to the {\it JHK}-band light curves in such an early phase. This reflects the fact that, for the fiducial model, the photon reprocessing is not efficient for the early phase due to a long diffusion timescale of the dynamical ejecta of mass $\gtrsim 10^{-3}\,M_\odot$ (see Equation~(\ref{eq:tpeak})).

\begin{figure}
 	 \includegraphics[width=1.\linewidth]{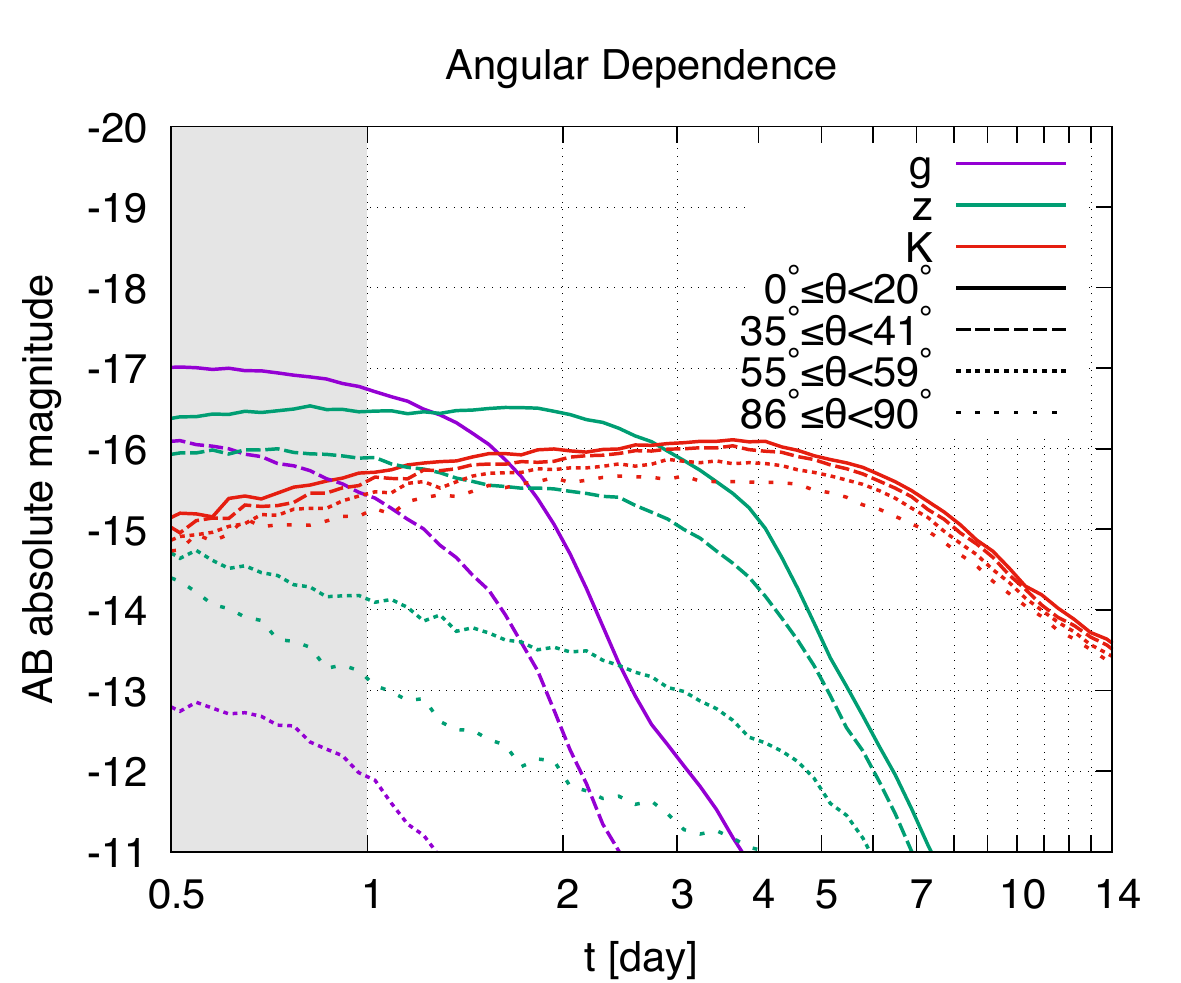}
 	 \caption{Angular dependence of the {\it gzK}-band light curves for the fiducial model ({\tt HMNS\_YH}). The solid, dashed, densely dotted, and sparsely dotted curves denote the light curves observed from $0^\circ\le\theta<20^\circ$, $35^\circ\le\theta\le41^\circ$, $55^\circ\le\theta<59^\circ$, and $86^\circ\le\theta<90^\circ$, respectively.}
	 \label{fig:mag_ang}
\end{figure}
For the light curves observed from the equatorial direction, the {\it griz}-band emission is strongly suppressed in the full calculation, while that in the separately calculated model is as bright as that observed from the polar direction. This is due to the blocking effect of optical photons from post-merger ejecta by dynamical ejecta~\citep{Kasen:2014toa,Bulla:2019muo}. Indeed, the suppression of the optical bands becomes significant for $\theta\ge45^\circ$, which reflects the employed opening angle of the dynamical ejecta (see Figure~\ref{fig:mag_ang}). On the other hand, the {\it HK}-band emission is brighter for the full calculation than the other, particularly after $\approx3\,{\rm days}$. This is due to the heating of the dynamical ejecta by the post-merger ejecta. The angular dependence of the infrared light curves is much weaker than that of the optical light curves. This implies that infrared emission is dominated by photons emitted from or reprocessed in the dynamical ejecta.

\subsection{Mass dependence}

\begin{figure*}
 	 \includegraphics[width=.5\linewidth]{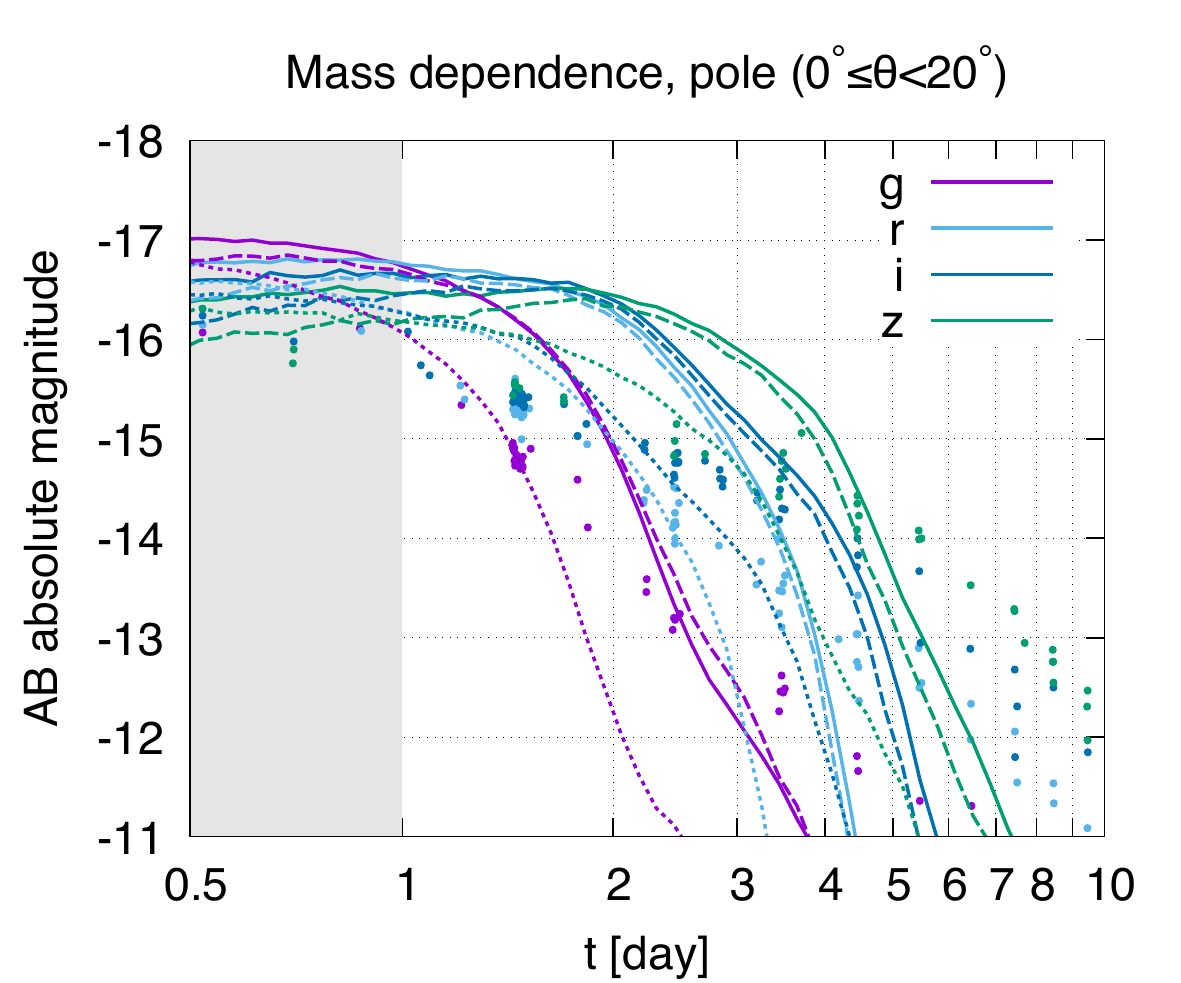}
 	 \includegraphics[width=.5\linewidth]{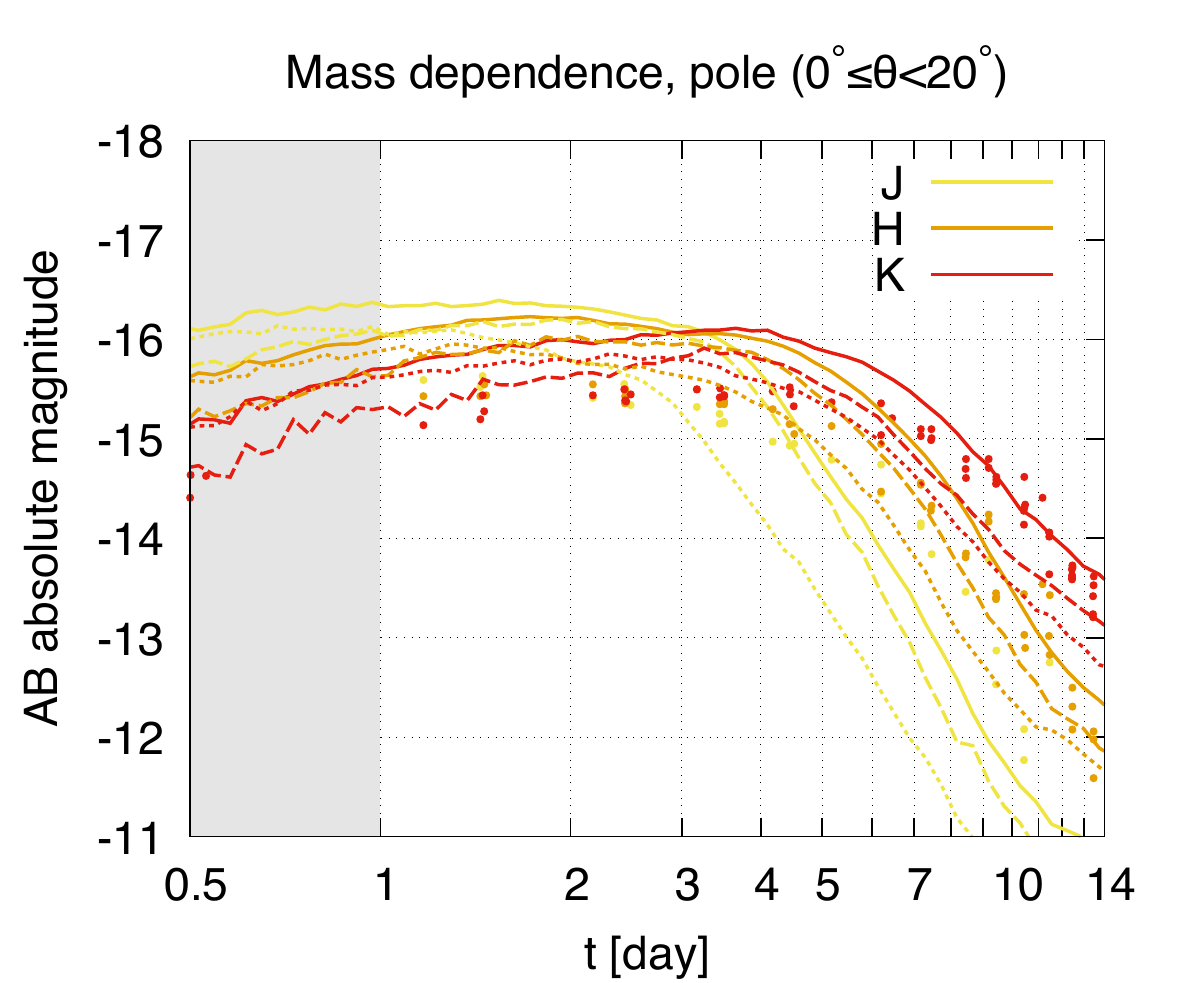}\\
 	 \includegraphics[width=.5\linewidth]{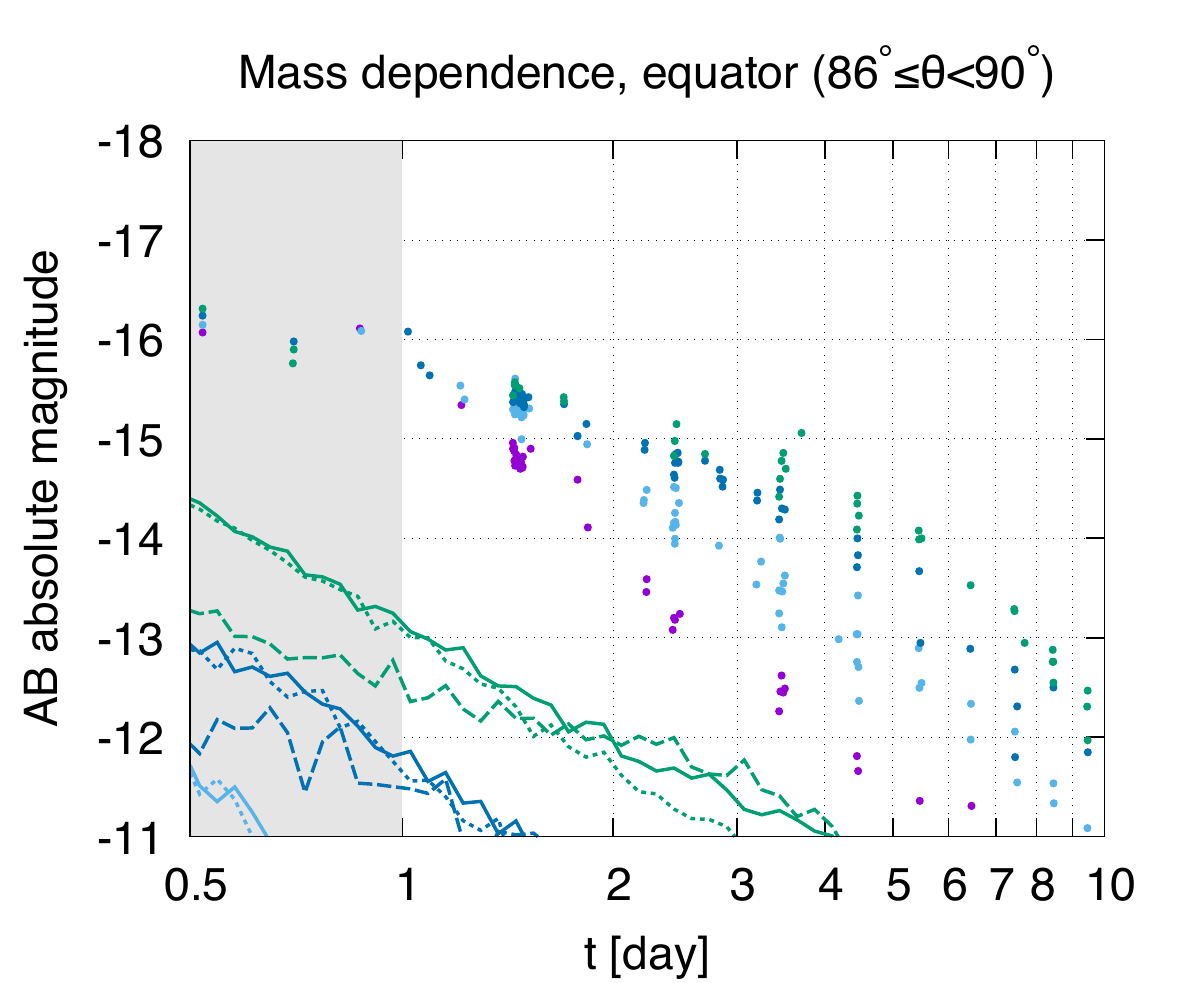}
 	 \includegraphics[width=.5\linewidth]{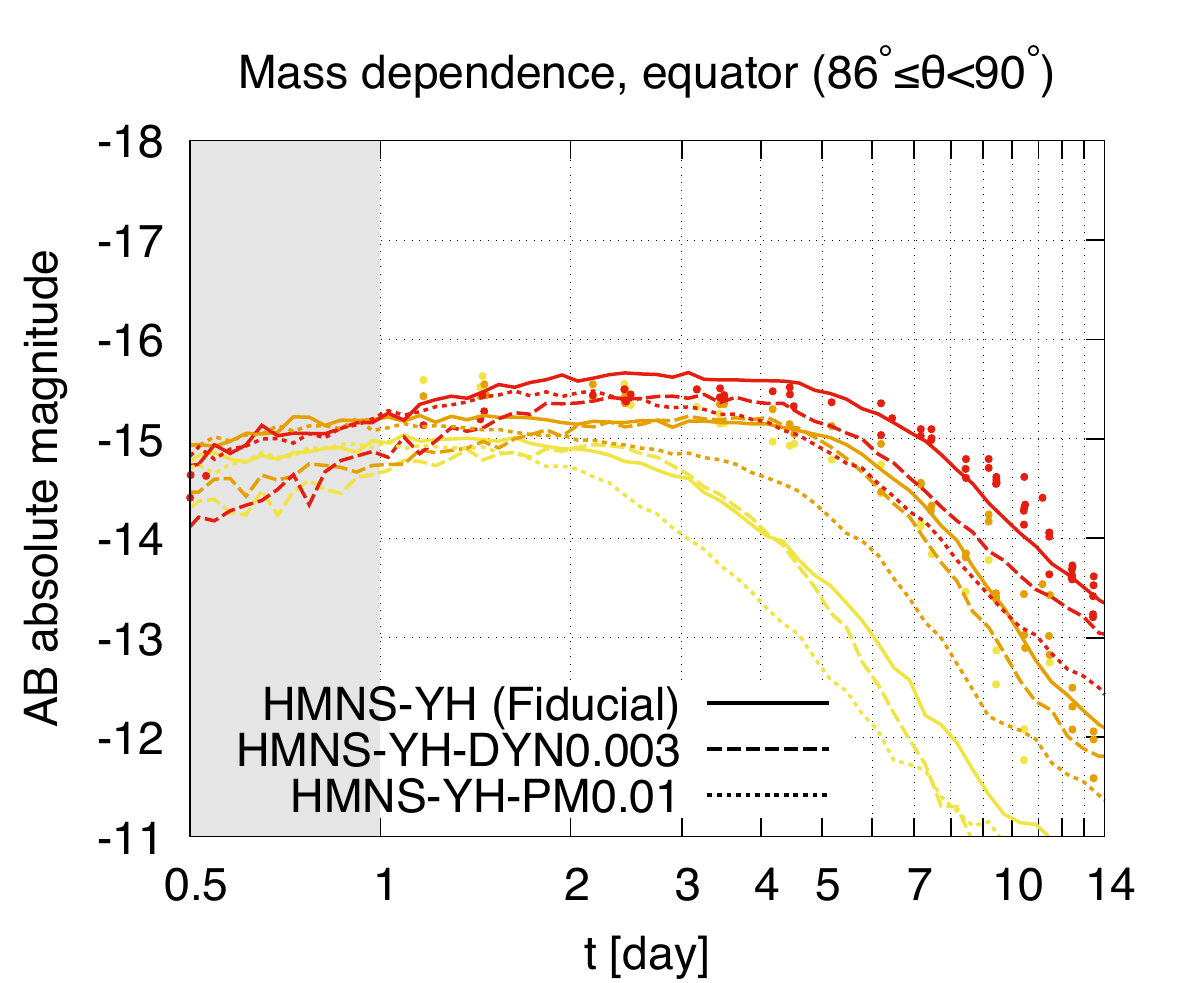}
 	 \caption{The {\it grizJHK}-band light curves for the fiducial model and the models with different ejecta mass. The solid, dashed and dotted curves denote the fiducial model ({\tt HMNS\_YH}), the model with small dynamical ejecta mass ({\tt HMNS\_YH\_DYN0.003}), and the model with small post-merger ejecta mass ({\tt HMNS\_YH\_PM0.01}), respectively. For a reference, we also plot the data points of GW170817~\citep{Villar:2017wcc}.}\label{fig:mag_dc2}
\end{figure*}

Figure~\ref{fig:mag_dc2} shows the {\it grizJHK}-band light curves for a small dynamical ejecta mass model ({\tt HMNS\_YH\_DYN0.003}) and a small post-merger ejecta model ({\tt HMNS\_YH\_PM0.01}) as well as the fiducial model ({\tt HMNS\_YH}) as a reference. We found that the {\it griz}-band light curves show approximately the same brightness as in the fiducial case for $t\lesssim2\,{\rm days}$ due to the enhancement of the optical brightness by photon diffusion preferentially in the polar direction as long as $M_{\rm d}\ge0.001\,M_\odot$. On the other hand, such an effect fades earlier than the fiducial case for the small dynamical ejecta mass case due to its shorter diffusion timescale. Indeed, for the light curves observed from the polar direction, the {\it griz}-band brightness for the small dynamical ejecta mass ({\tt HMNS\_YH\_DYN0.003}) and fiducial model agrees with each other within $\approx0.2\,{\rm mag}$ for $1\,{\rm day}\le t \le2.5\,{\rm days}$, while the brightness becomes fainter after $\approx2.5\,{\rm days}$. Note that, exceptionally, the {\it g}-band emission is slightly brighter than the fiducial model particularly for $t\gtrsim2\,{\rm days}$ due to the decrease in the density of dynamical ejecta in the polar region. 

The brightness of the {\it JHK}-band light curves does not necessarily reflect the mass of the dynamical ejecta but could depend strongly on the mass of post-merger ejecta. The {\it JHK}-band emission of the small dynamical ejecta mass model ({\tt HMNS\_YH\_DYN0.003}) is only by $\approx0.5\,{\rm mag}$ fainter than that of the fiducial model, though the dynamical ejecta mass is different by a factor of $\approx3$. This is because photons emitted from the post-merger ejecta are reprocessed more efficiently from the earlier phase for the smaller dynamical ejecta mass model due to shorter diffusion timescale. Nevertheless, the {\it JHK}-band brightness still could provide the upper limit to the dynamical ejecta mass, and the results above indicate that using data observed for the early phase $\lesssim2\,{\rm days}$ might be the better choice for constraining the dynamical ejecta mass because the contribution of reprocessed photons from the post-merger ejecta is relatively less significant than that for the later phase.

%Indeed, the {\it JHK}-band emission of {\tt HMNS\_YH\_DYN0.003} is brighter by $\approx1\,{\rm mag}$ than those of the dynamical ejecta only model with the same mass ({\tt DYN0.003}) even at $t=1\,{\rm day}$ (see Figure~\ref{fig:mag_dyn}). 

The brightness of the light curves for a small post-merger ejecta mass case is typically faint due to the decrease in the heating source ({\tt HMNS\_YH\_PM0.01}). Indeed, the {\it griz}-band emission for {\tt HMNS\_YH\_PM0.01} is fainter than that of the fiducial model and the small dynamical ejecta mass model ({\tt HMNS\_YH\_DYN0.003}). The {\it JHK}-band emission for the later phase ($\gtrsim2\,{\rm days}$) is also fainter for {\tt HMNS\_YH\_PM0.01} than the other models because it is dominated by the emission from the post-merger ejecta reprocessed in the dynamical ejecta. Exceptionally, the {\it JHK}-band emission for {\tt HMNS\_YH\_PM0.01} is slightly brighter than {\tt HMNS\_YH\_DYN0.003} in the early time $\lesssim2\,{\rm days}$. This is due to the fact that {\tt HMNS\_YH\_PM0.01} has larger dynamical ejecta mass than {\tt HMNS\_YH\_DYN0.003} and that the radioactive heating in the dynamical ejecta also contributes to the {\it JHK}-band light curves for $\lesssim2\,{\rm days}$. 

For the light curves observed from the equatorial direction, the {\it griz}-band energy flux is strongly suppressed due to the blocking effect of the dynamical ejecta. All the cases exhibit approximately the same light curves and are not distinguishable by the optical light curves. Also, {\it JHK}-band emission observed in the equatorial direction is fainter by $\approx1\,{\rm mag}$ than in the polar direction but the light curves show the same feature as those observed from the polar direction, and the dependence on the ejecta mass is clearly reflected.

\subsection{Velocity dependence}

\begin{figure*}
 	 \includegraphics[width=.5\linewidth]{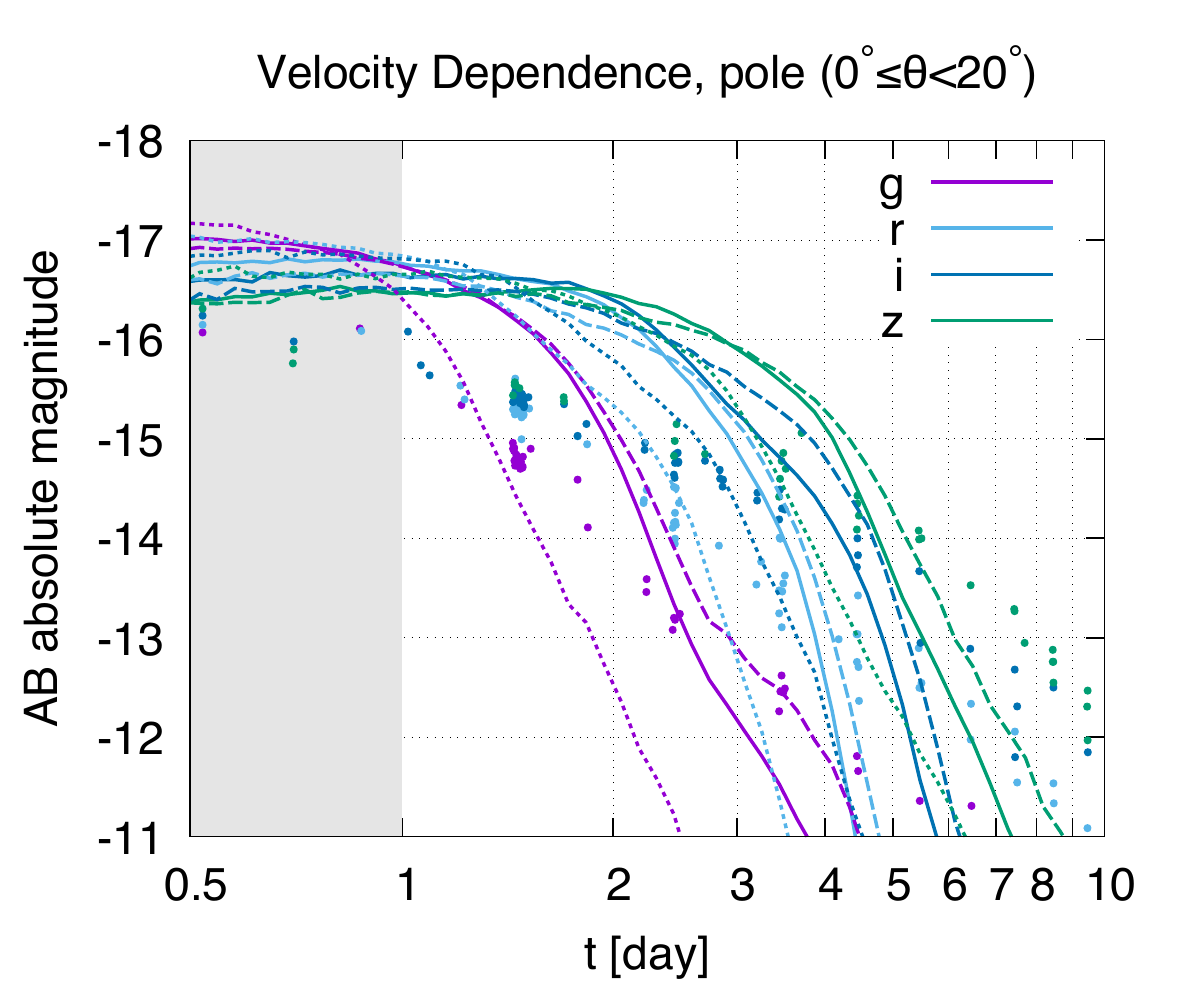}
 	 \includegraphics[width=.5\linewidth]{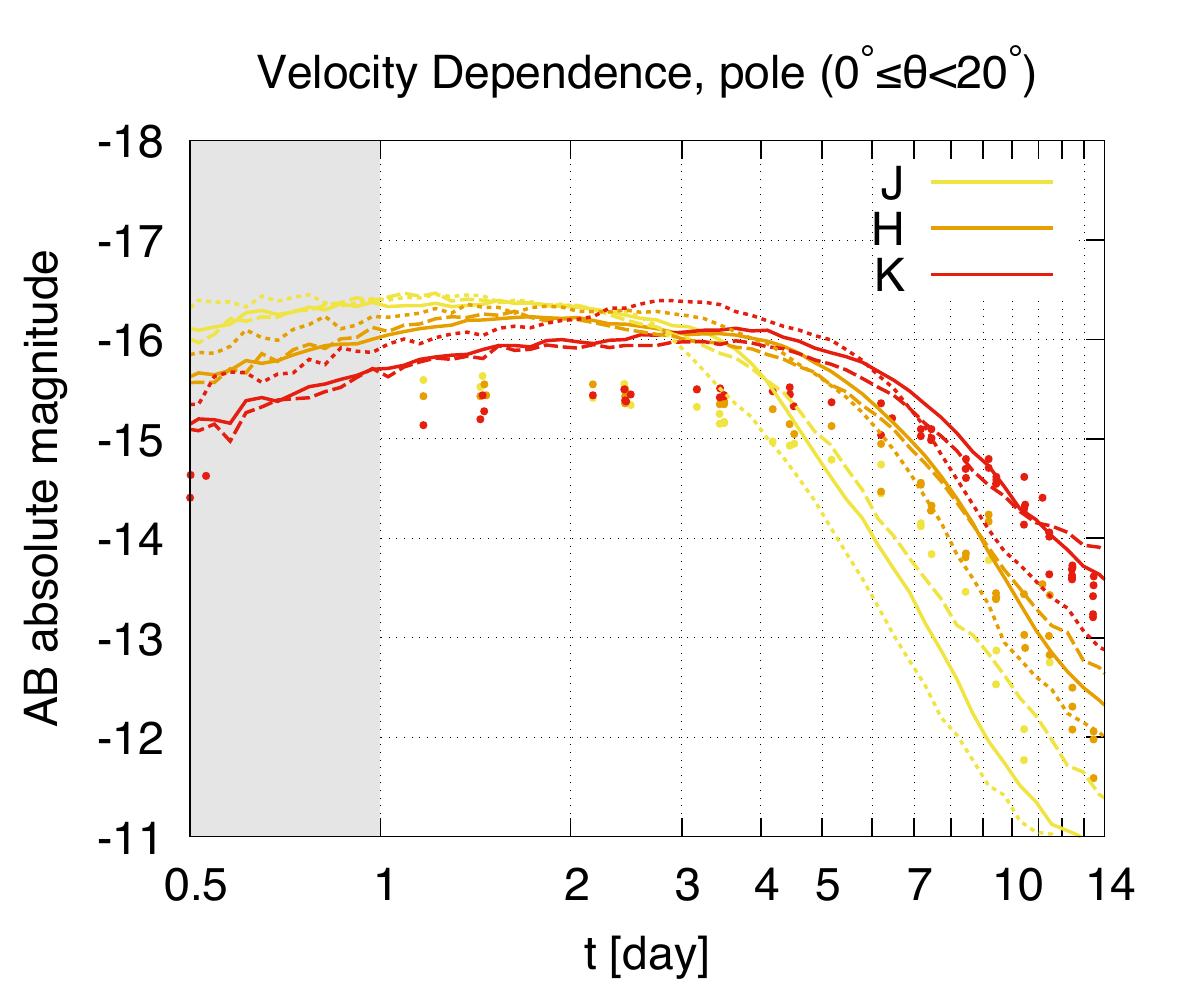}\\
 	 \includegraphics[width=.5\linewidth]{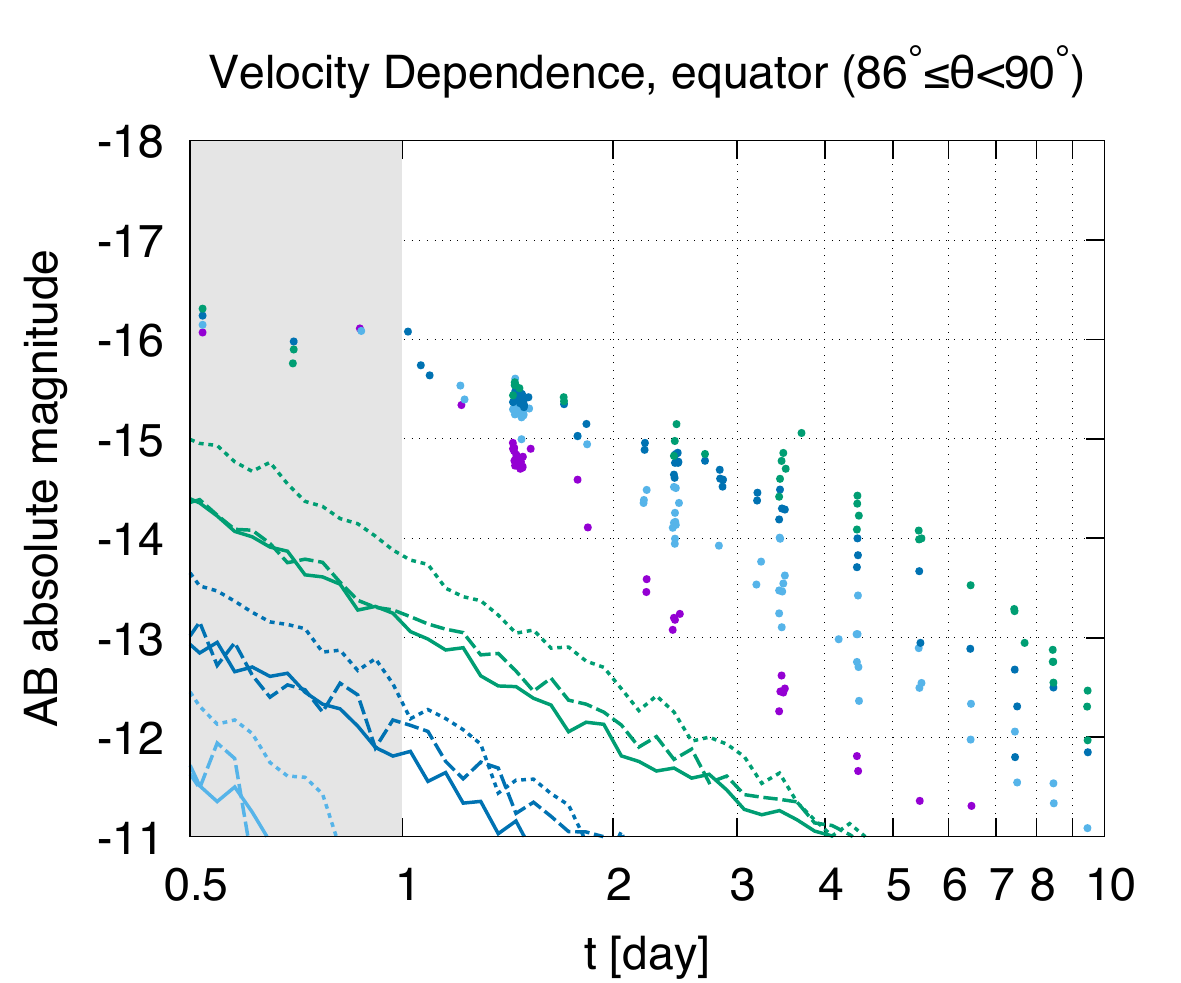}
 	 \includegraphics[width=.5\linewidth]{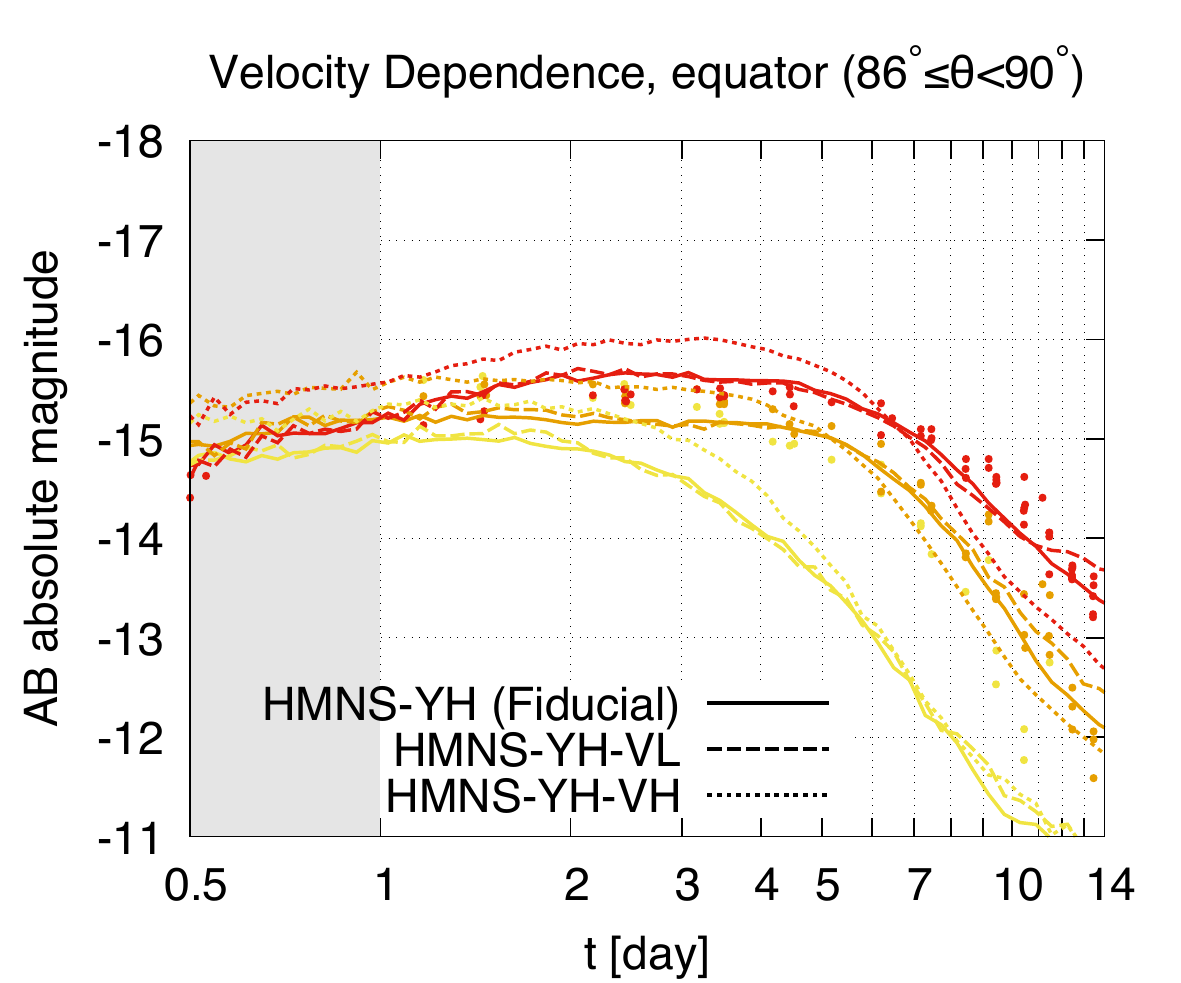}
 	 \caption{The {\it grizJHK}-band light curves for the fiducial model and the models with the different ejecta velocity. The solid, dashed, and dotted curves denote the fiducial model ({\tt HMNS\_YH}), the model with slow-velocity post-merger ejecta ({\tt HMNS\_YH\_VL}), and the model with high-velocity post-merger and dynamical ejecta ({\tt HMNS\_YH\_VH}), respectively. For a reference, we also plot the data points of GW170817~\citep{Villar:2017wcc}.}
	 \label{fig:mag_dc4}
\end{figure*}

Figure~\ref{fig:mag_dc4} compares the {\it grizJHK}-band light curves among the models with the same ejecta mass as in the fiducial model ({\tt HMNS\_YH}) but with different velocity parameters. The model with slow-velocity post-merger ejecta ({\tt HMNS\_YH\_VL}) shows slightly longer-lasting light curves than the fiducial model due to the longer diffusion timescale (e.g., Equation~\ref{eq:tpeak}). On the other hand, the light curves for the model of which both post-merger and dynamical ejecta have higher velocity than the fiducial model ({\tt HMNS\_YH\_VH}) decline much earlier than the fiducial model due to the short diffusion timescale of ejecta, and the rapid decline is particularly significant for the {\it griz}-band light curves, although the peak magnitudes in all the {\it grizJHK}-bands are approximately the same as in the fiducial model for {\tt HMNS\_YH\_VL} and the difference from the fiducial model is smaller than $\approx0.2\,{\rm mag}$. The peak brightness of {\tt HMNS\_YH\_VH} also agrees with that of the fiducial model within $\approx 0.3\,{\rm mag}$ in all the {\it grizJHK}-bands. These results indicate that the difference in the velocity profile is reflected most significantly in the decline rates of the light curves particularly in the optical (and perhaps, also, ultraviolet) bands. We note that the rapid decline of the optical light curves for {\tt HMNS\_YH\_VH} may reflect the fact that the opacity depends strongly on the temperature near $T\approx 3000\,{\rm K}$ and the temperature of the post-merger ejecta for {\tt HMNS\_YH\_VH} decreases more steeply than for other models.

For the light curves observed from the equatorial direction, the difference from the fiducial model is only remarkable for {\tt HMNS\_YH\_VH}. This reflects the fact that the light curves observed from the equatorial direction are dominated by the emission from the dynamical ejecta because the density profiles of the dynamical ejecta for {\tt HMNS\_YH\_VL} and the fiducial model are the same.

\subsection{$Y_e$ dependence}

\begin{figure*}
 	 \includegraphics[width=.5\linewidth]{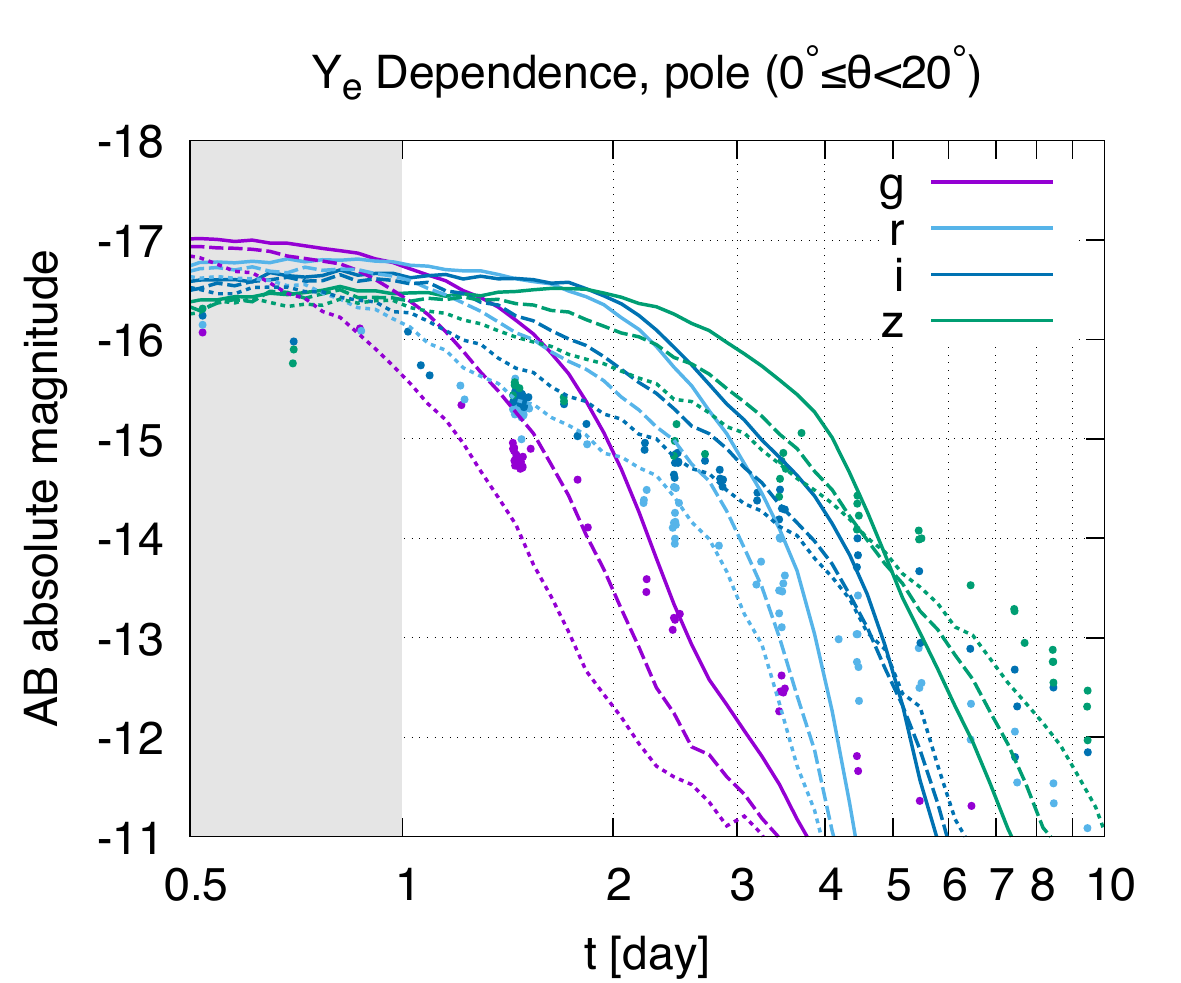}
 	 \includegraphics[width=.5\linewidth]{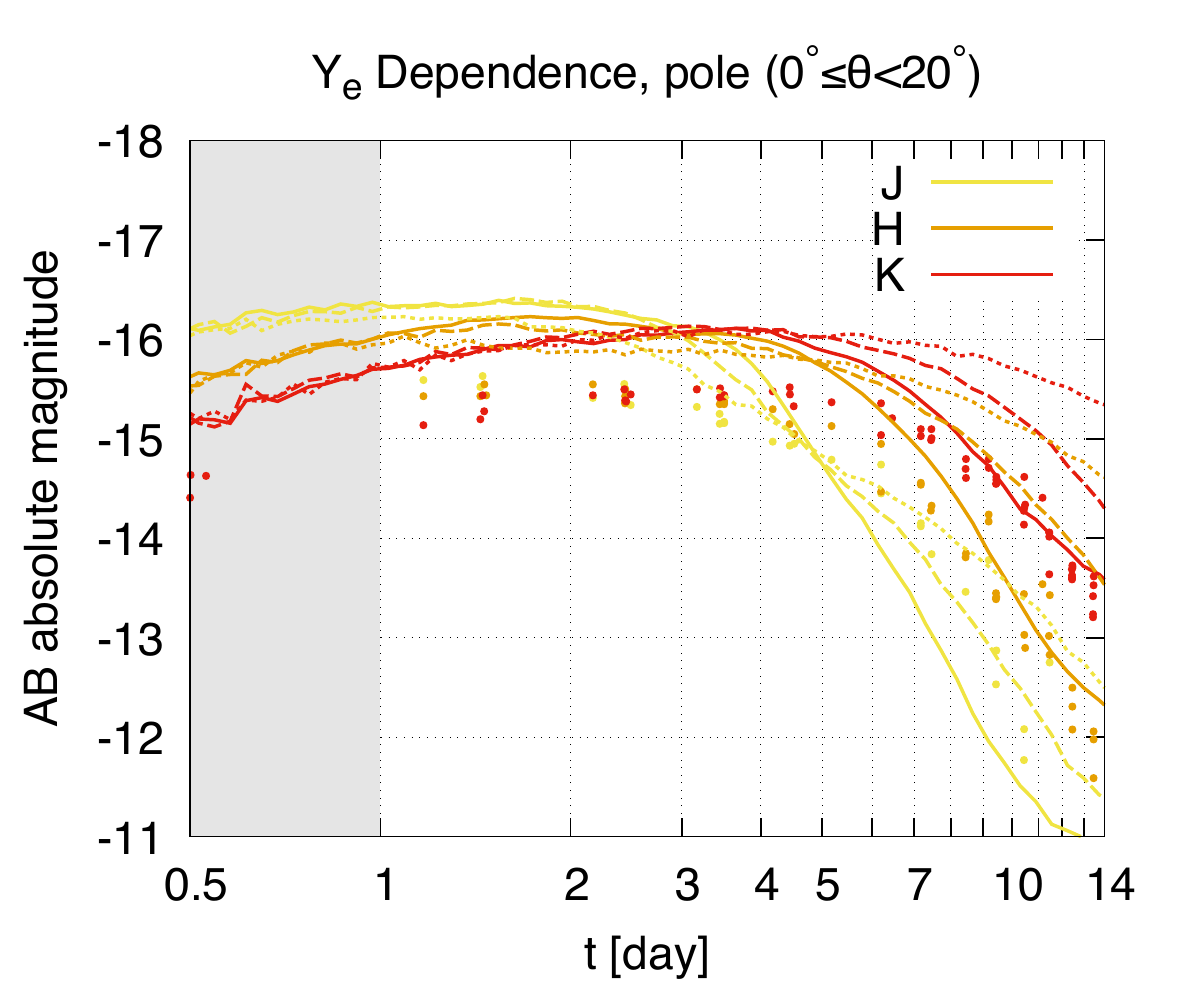}\\
 	 \includegraphics[width=.5\linewidth]{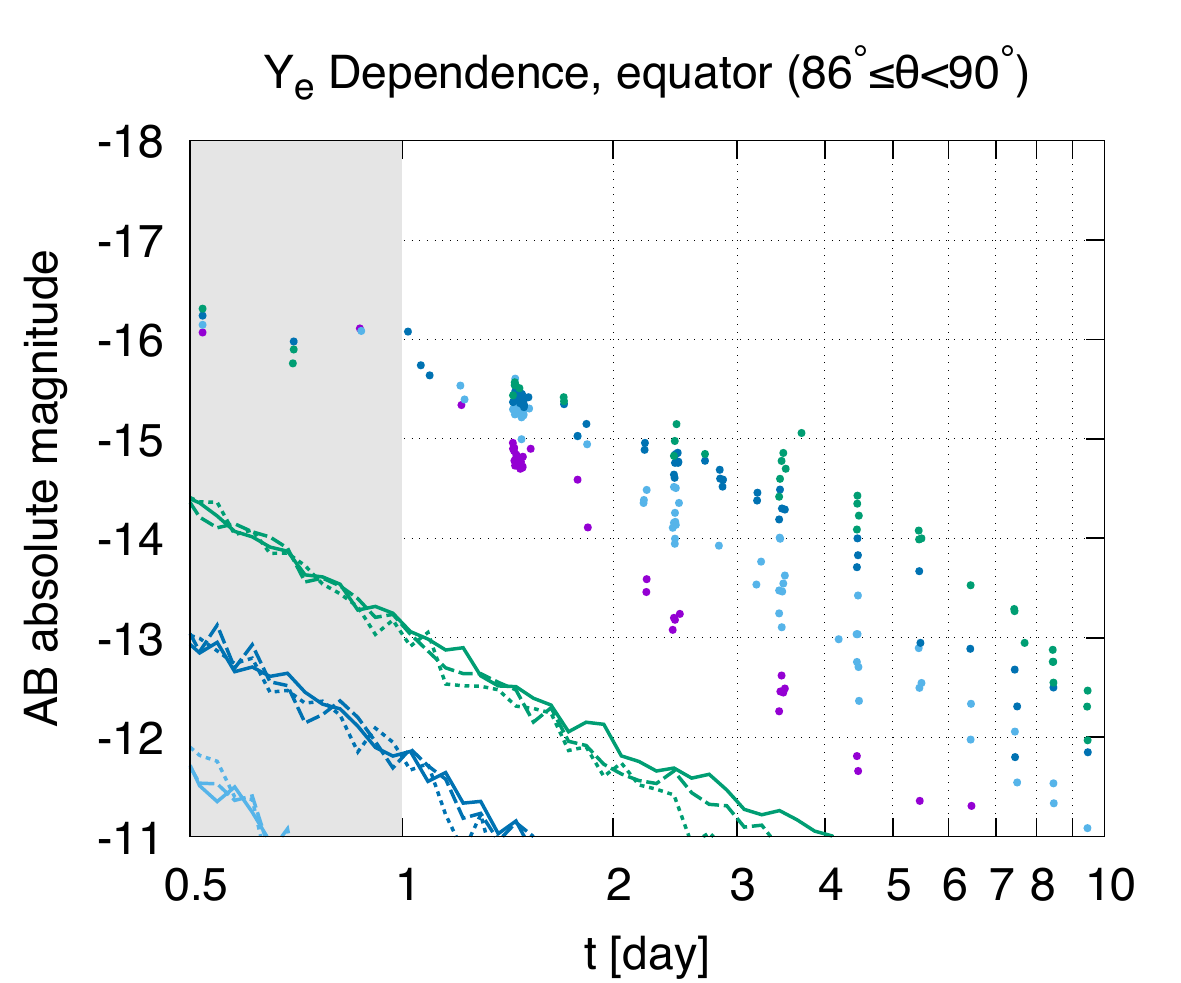}
 	 \includegraphics[width=.5\linewidth]{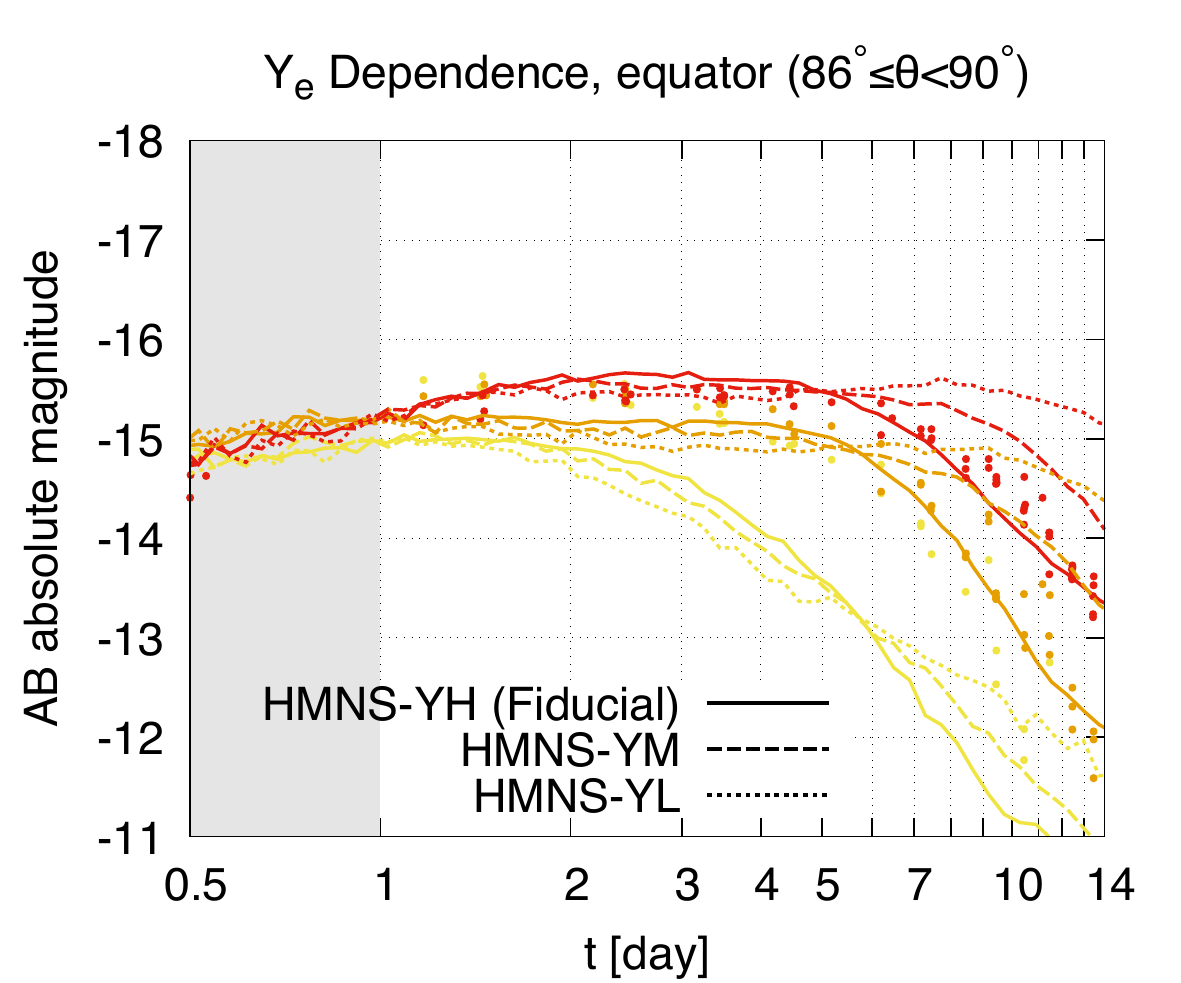}
 	 \caption{The {\it grizJHK}-band light curves for lanthanide-free ({\tt HMNS\_YH}; Solid curves, $X_{\rm pm,lan}\ll10^{-3}$), mildly lanthanide-rich ({\tt HMNS\_YM}; Dashed curves, $X_{\rm pm,lan}\approx0.025$), and highly lanthanide-rich ({\tt HMNS\_YL}; Dotted curves, $X_{\rm pm,lan}\approx0.14$) post-merger ejecta. Here, $X_{\rm pm, lan}$ denotes the lanthanide mass fraction of the post-merger ejecta. For a reference, we also plot the data points of GW170817~\citep{Villar:2017wcc}.}
	 \label{fig:mag_dc1}
\end{figure*}

\begin{figure}
 	 \includegraphics[width=1\linewidth]{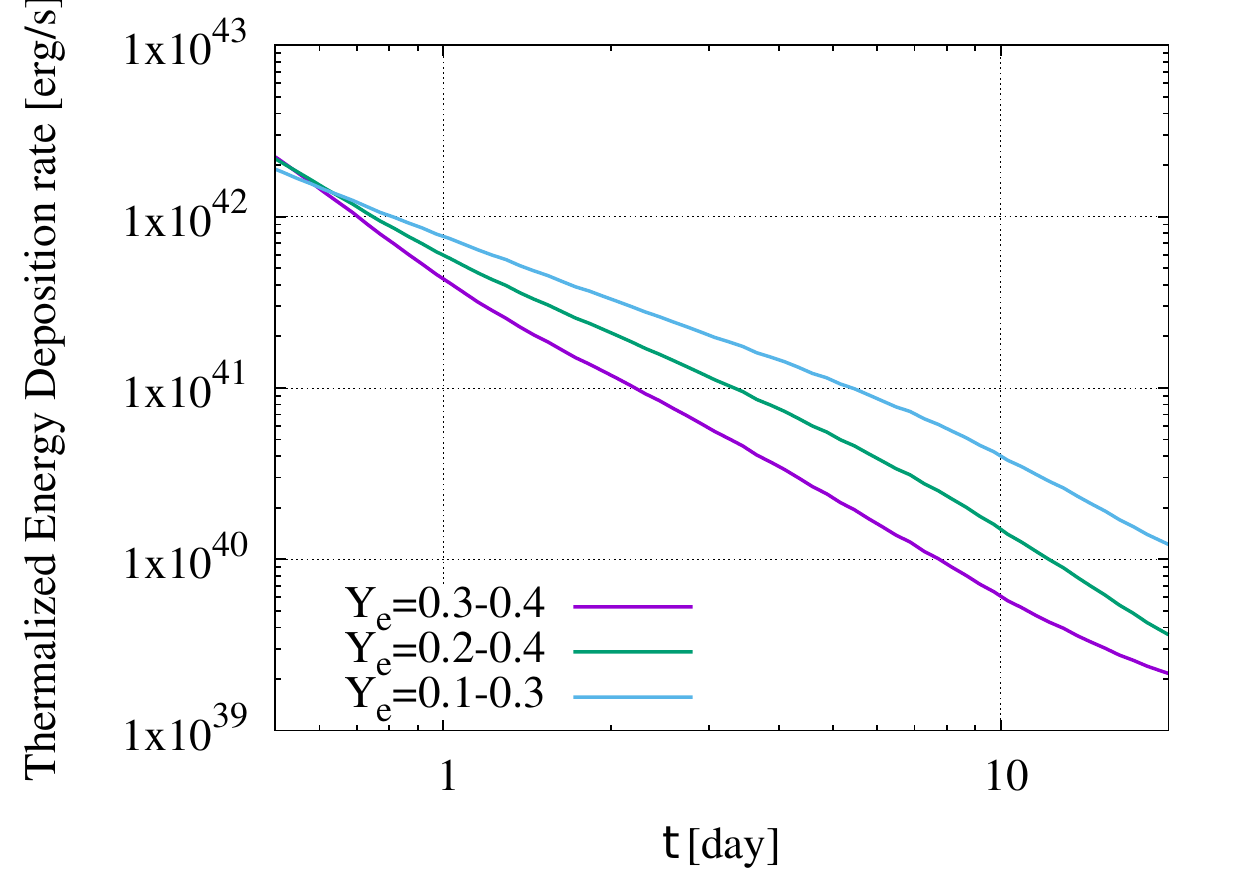}\
	 \caption{Comparison of the energy deposition rates among the post-merger ejecta models with different $Y_e$ distributions ({\tt PM\_YH}, {\tt PM\_YM}, and {\tt PM\_YL}). The energy deposition rates are shown after thermalization efficiency is taken into account.}\label{fig:dep_pm}
\end{figure}

Figure~\ref{fig:mag_dc1} shows the {\it grizJHK} light curves of the models with lanthanide-free ({\tt HMNS\_YH}; Solid curves, $X_{\rm pm,lan}\ll10^{-3}$), mildly lanthanide-rich ({\tt HMNS\_YM}; Dashed curves, $X_{\rm pm,lan}\approx0.025$), and highly lanthanide-rich ({\tt HMNS\_YL}; Dotted curves, $X_{\rm pm,lan}\approx0.14$) post-merger ejecta (Here, $X_{\rm pm, lan}$ denotes the lanthanide mass fraction of the post-merger ejecta). For the light curves observed from the polar direction, the {\it griz}-band emission become faint and show shallow decline as the lanthanide fraction of the post-merger ejecta increases. These can be understood by the fact that diffusion timescale of the post-merge ejecta is longer for the lower-$Y_e$ models due to the large opacities. We note that the $Y_e$ dependence of the heating rate is also an important reason for the difference in the light curves particularly for the late phase ($t\gtrsim5\,{\rm days}$) (see Figure~\ref{fig:dep_pm}). 

We find that the optical light curves observed from the polar direction are dominated by the emission from the post-merger ejecta as long as its mass is larger than that of dynamical ejecta. This indicates that the observed brightness in the {\it griz}-band would be used to constrain the post-merger ejecta mass. We note that, as is found in the fiducial model, the optical emission for the lanthanide-rich post-merger ejecta ({\tt HMNS\_YM} and {\tt HMNS\_YL}) is also brighter than that for the models purely composed of the post-merger ejecta ({\tt PM\_YM} and {\tt PM\_YL}) due to the preferentially diffusion of the photon in the polar direction (see Figure~\ref{fig:mag_pm}). Thus, the enhancement of the brightness due to diffusion in the preferential direction should be taken into account for the ejecta mass estimation also for the case with lanthanide-rich post-merger ejecta. 

For all the models, the brightness in the {\it JHK}-bands observed from the polar direction agrees with each other within $\approx0.1\,{\rm mag}$ for $\lesssim3\,{\rm days}$. On the other hand, the low-$Y_e$ models show brighter {\it JHK}-band emission than the fiducial model after $\approx5\,{\rm days}$, and much brighter emission is seen in the lanthanide-rich model ({\tt HMNS\_YL}). This indicates that the lanthanide fraction of the post-merger ejecta would be reflected in the {\it JHK}-band light curves for the late phase. However, we note that this enhancement in the {\it JHK}-band brightness for the late phase for the lanthanide-rich post-merger ejecta models is not only due to the bright infrared emission directly emitted from the post-merger ejecta but also due to the strong heating to the dynamical ejecta by the post-merger ejecta because the heating rate is larger for the lower-$Y_e$ model particularly for the late phase (see Figure~\ref{fig:dep_pm}). Indeed, we found that the {\it JHK}-band emission at $\gtrsim5\,{\rm days}$ for the fiducial model is as bright as that for the mildly lanthanide-rich case ({\tt HMNS\_YM}) shown in Figure~\ref{fig:mag_dc1} if the same heating rate for the post-merger ejecta is employed. This indicates that employing reliable heating rate model particularly for the late phase would be crucial for measuring the lanthanide fractions of ejecta from the infrared light curves.

As is the case for the ejecta mass dependence, $Y_e$ dependence of the light curves observed from the equatorial direction is only remarkable in the infrared light curves. The {\it griz}-band emission is strongly suppressed and that among different $Y_e$ models are not distinguishable. On the other hand, the difference in the {\it JHK}-band light curves among different $Y_e$ models is clearly seen as in those observed from the polar direction for $t\gtrsim 6\,{\rm days}$.

\section{Model parameters for interpreting GW170817}\label{sec:gw170817}
In this section, we discuss which model parameters would be suitable to interpret the observed optical and infrared light curves in GW170817. We focus only on the light curves observed from the direction close to the polar axis since the GW data analysis of GW170817 infers that the event is observed from $\theta\lesssim28^\circ$ (Abbott et al. 2017a).

For our setup of ejecta profile, we find that the post-merger ejecta is required to explain the brightness in {\it griz}-band. Indeed, we confirm that the {\it griz}-band emission only with the dynamical ejecta is too faint to interpret the observational results as long as we assume $M_{\rm d}\le0.01\,M_\odot$ (see Figure~\ref{fig:mag_dyn} in the appendix). The results obtained only with post-merger ejecta suggest that the mass of $\approx 0.03\,M\odot$ would be suitable to interpret the brightness in {\it griz}-band (see Figure~\ref{fig:mag_pm} in the appendix). On the other hand, a smaller mass is preferred for the multi-components model due to the effect of the diffusion of photons preferentially in the polar direction. Indeed, the fiducial model ({\tt HMNS\_YH}), in which post-merger ejecta is lanthanide-free with mass $0.03\,M_\odot$, overproduces the {\it griz}-band brightness for the early phase  ($\lesssim\,2\,{\rm days}$). This suggests that the post-merger ejecta mass less than $0.03\,M_\odot$ is preferred for our setup if it is lanthanide-free ($X_{\rm pm,lan}\ll10^{-3}$).

For interpreting the observed brightness in {\it griz}-band, larger mass would be allowed for lanthanide-rich post-merger ejecta than lanthanide-free case because {\it griz}-band emission is fainter for such models. However, too massive lanthanide-rich post-merger ejecta would also be unsuitable because it overproduces the {\it JHK}-band brightness for the late phase ($t\gtrsim7\,{\rm days}$). Indeed, even in the absence of the dynamical ejecta, the model with highly lanthanide-rich post-merger ejecta ({\tt PM\_YL}, $X_{\rm pm,lan}\approx0.14$) overproduces the {\it JHK}-band brightness for $t\gtrsim7{\rm days}$, while the {\it griz}-band brightness is underproduced. This indicates that the lanthanide fraction of the post-merger ejecta is likely to $\ll0.1$ for interpreting the observed light curves in GW170817.

A tighter upper limit to the dynamical ejecta mass is suggested for multi-components models than a model only with the dynamical ejecta. The result only with the dynamical ejecta ({\tt DYN0.01}) shows that $M_{\rm d}\lesssim0.01\,M_\odot$ is preferred not to overproduce the observed {\it JHK}-band light curves for $t\lesssim3\,{\rm days}$ (see Figure~\ref{fig:mag_dyn} in the appendix). On the other hand, Figure~\ref{fig:mag_dc2} shows that the model with $M_{\rm d}\lesssim0.003\,M_\odot$ is consistent with the observed {\it JHK}-band light curves particularly for the early phase ($t\lesssim4\,{\rm days}$) if the post-merger ejecta mass is $\approx0.03\,M_\odot$. This shows that the enhancement of the infrared brightness due to the heating effect to the dynamical ejecta by the post-merger ejecta is important in the presence of multi-components. 

\begin{figure*}
 	 \includegraphics[width=.5\linewidth]{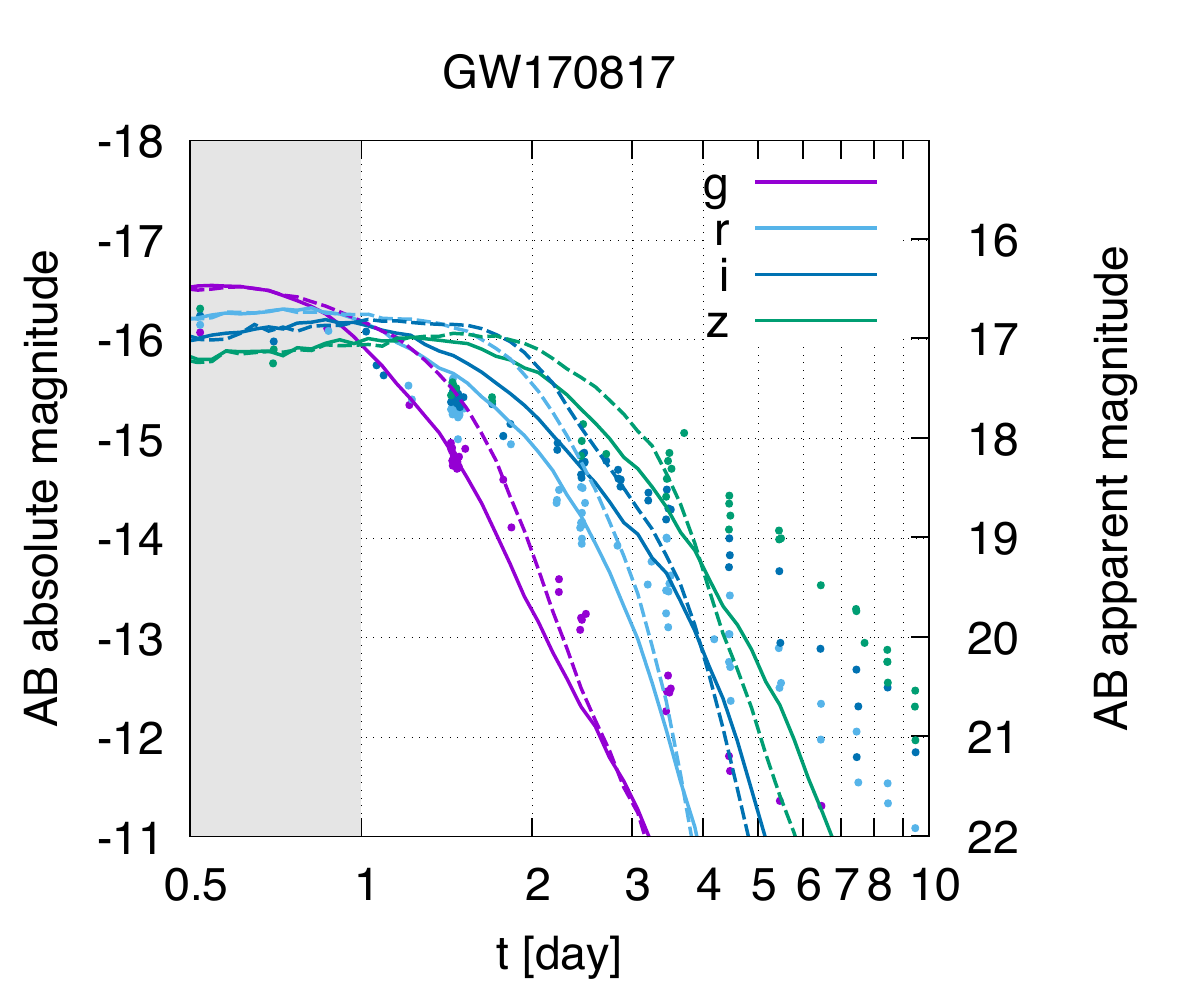}
 	 \includegraphics[width=.5\linewidth]{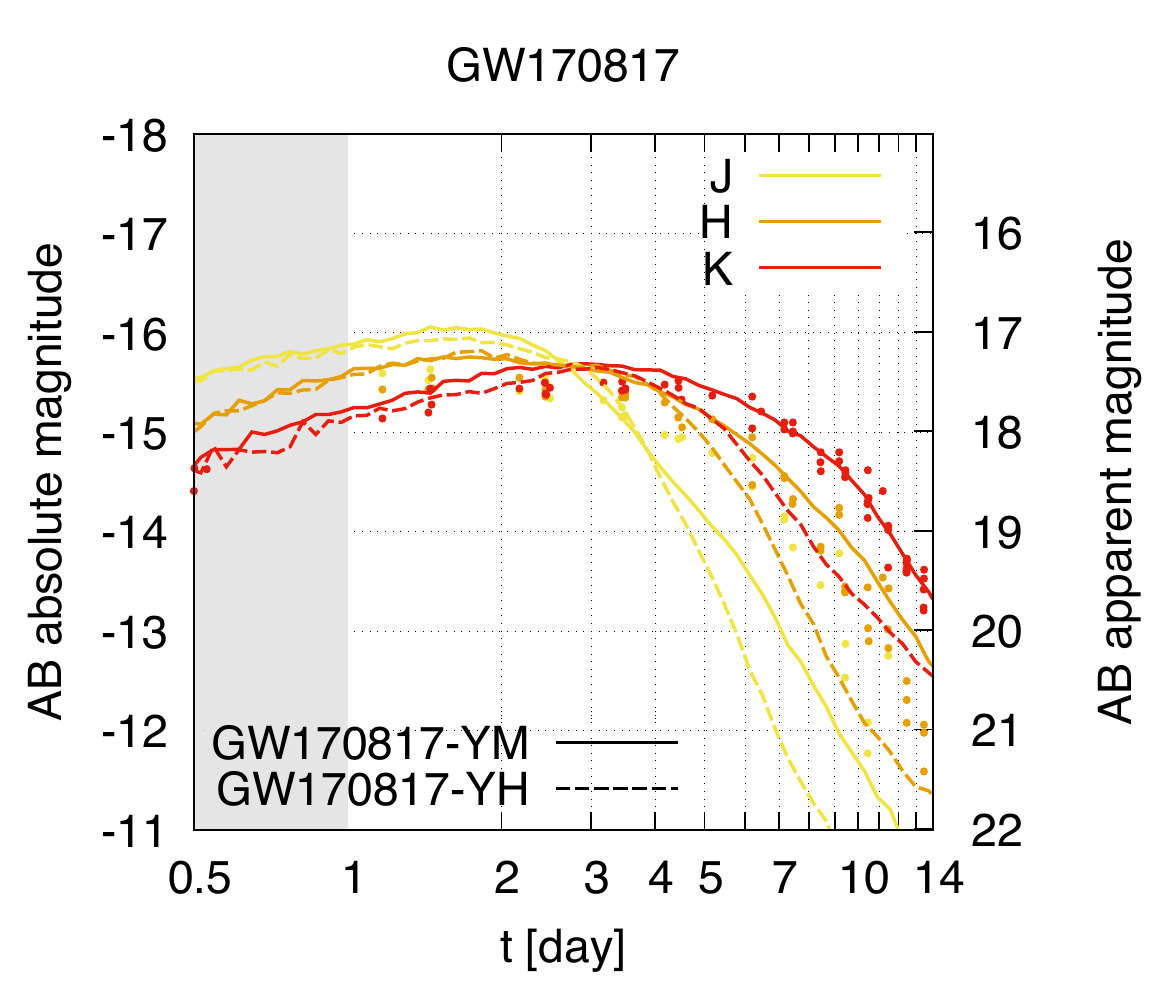}\\
 	 \caption{The {\it grizJHK}-band light curves observed from the polar direction for the models which approximately reproduce the observed peak brightness of GW170817. The solid and dashed curves denote the models with mildly lanthanide-rich ({\tt GW170817\_YM}, $X_{\rm pm,lan}\approx0.025$) and lanthanide-free  ({\tt GW170817\_YH}, $X_{\rm pm,lan}\ll10^{-3}$) post-merger ejecta, respectively.  light curves observed from $0^\circ\le\theta<20^\circ$ and $20^\circ\le\theta<28^\circ$ are shown for the model with mildly lanthanide-rich ({\tt GW170817\_YM}) and lanthanide-free ({\tt GW170817\_YH}) post-merger ejecta, respectively. For a reference, we also plot the data points of GW170817~\citep{Villar:2017wcc}.}\label{fig:mag_can}
\end{figure*}

Motivated by the discussion above, we select two models, {\tt GW170817\_YM} and {\tt GW170817\_YH}, which reproduce the peak brightness observed in GW170817. Figure~\ref{fig:mag_can} shows the light curves with mildly lanthanide-rich post-merger ejecta ({\tt GW170817\_YM}, $X_{\rm pm,lan}\approx0.025$) and with lanthanide-free post-merger ejecta ({\tt GW170817\_YH}, $X_{\rm pm,lan}\ll10^{-3}$) observed from $0^\circ\le\theta<20^\circ$ and $20^\circ\le\theta<28^\circ$, respectively. For both models, the mass of dynamical and post-merger ejecta are $0.003\,M_\odot$ and  $0.02\,M_\odot$, respectively. Both models shown in Figure~\ref{fig:mag_can} reproduce the peak brightness observed in GW170817. In particular,  {\tt GW170817\_YM} reproduces the observed brightness of the optical light curves for the early phase ($t\lesssim2\,{\rm days}$) even for the case that the post-merger ejecta is mildly lanthanide-rich. This indicates that the post-merger ejecta is not necessarily required to be completely lanthanide-free to interpret the observed peak brightness in the optical bands~\citep{Kasen:2017sxr,Perego:2017wtu}. Moreover, the {\it HK}-band light curves consistent with the observation are obtained for the model with mildly lanthanide-rich post-merger ejecta, while the model with lanthanide-free post-merger ejecta underproduces those observed light curves for the late phase ($t\approx5\,{\rm days}$). Thus, the model with mildly lanthanide-rich post-merger ejecta is preferred to interpret the observation for our setup.

\begin{figure}
 	 \includegraphics[width=1.\linewidth]{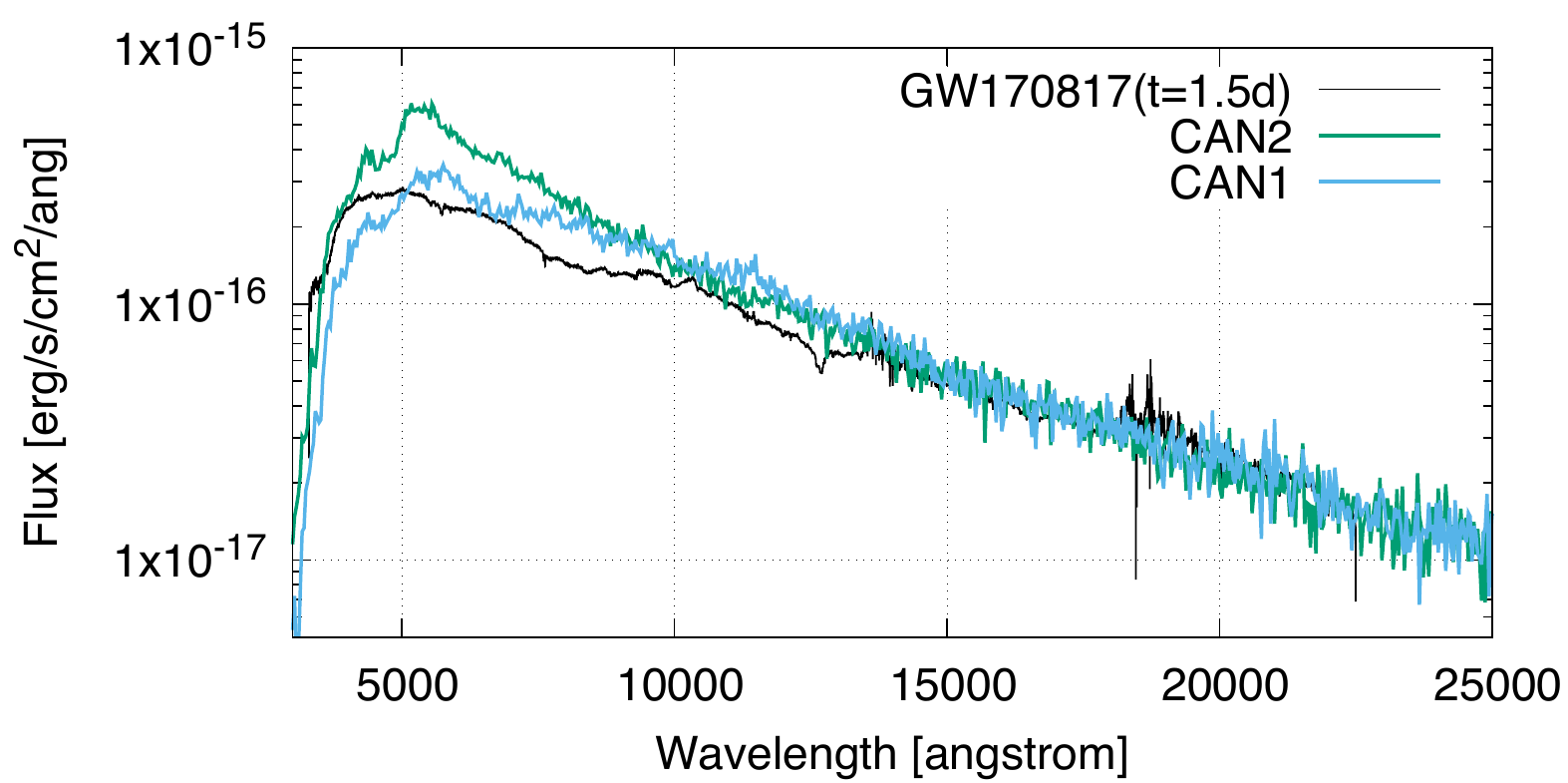}\\
 	 \includegraphics[width=1.\linewidth]{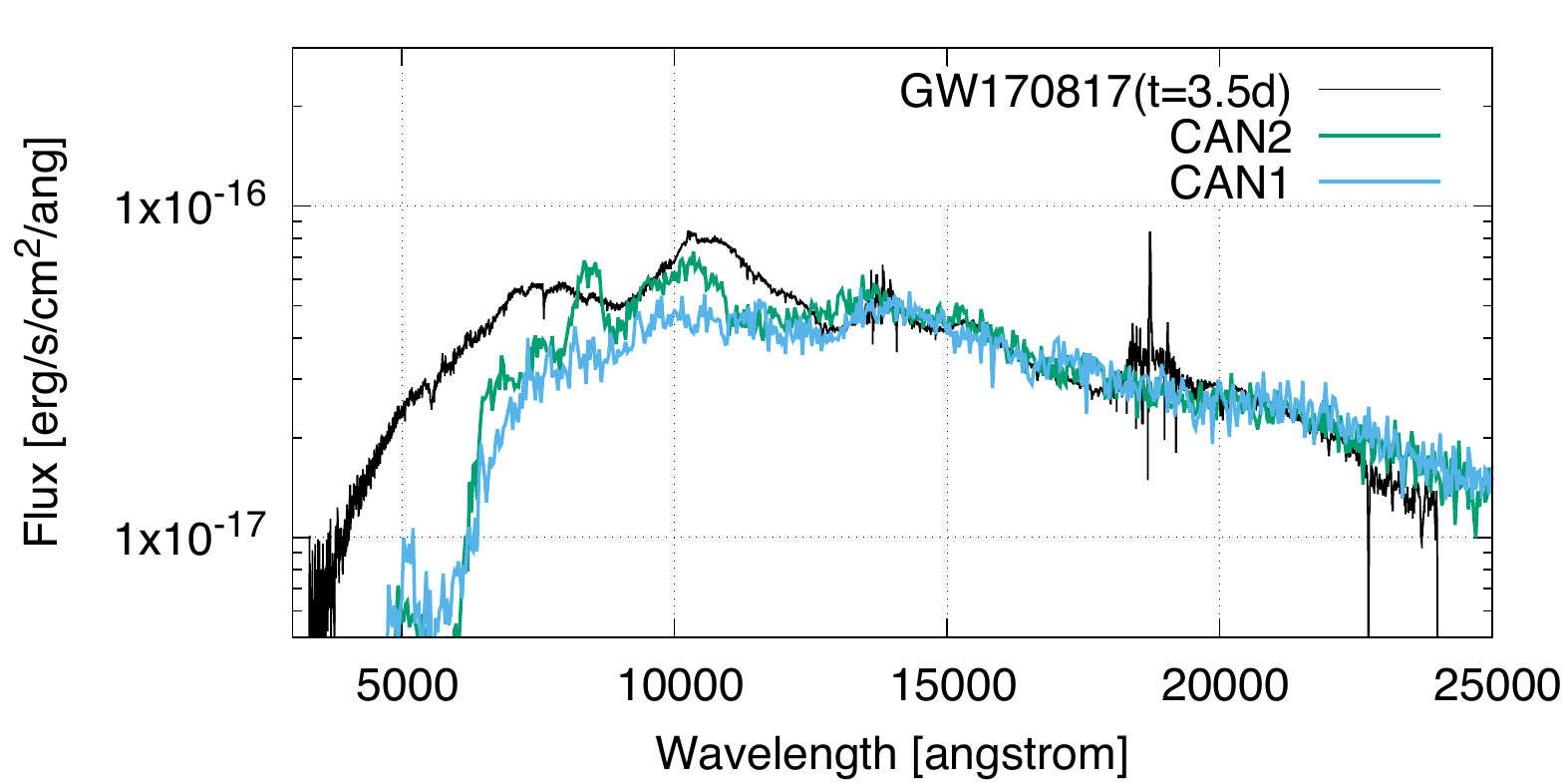}
 	 \caption{The spectra at $t\approx1.5\,{\rm days}$ (top panel) and $t\approx3.5\,{\rm days}$ (bottom panel) observed from the polar direction for the models which approximately reproduce the observed peak brightness of GW170817. The blue and green curves denote the models with mildly lanthanide-rich ({\tt GW170817\_YM}) and lanthanide-free  ({\tt GW170817\_YH}) post-merger ejecta, respectively. The black curves denotes the spectra of GW170817 taken with VLT/X-Shooter~\citep{Pian:2017gtc,Smartt:2017fuw}. Spectra observed from $0^\circ\le\theta<20^\circ$ and $20^\circ\le\theta<28^\circ$ are shown for the model with mildly lanthanide-rich ({\tt GW170817\_YM}) and lanthanide-free ({\tt GW170817\_YH}) post-merger ejecta, respectively.}\label{fig:comp_spec}
\end{figure}

The lanthanide-rich post-merger ejecta could also be well-suited for interpreting the featureless spectrum found in the early part of the observation. Figure~\ref{fig:comp_spec} shows the spectra of {\tt GW170817\_YM} and {\tt GW170817\_YH} at $t\approx1.5\,{\rm days}$ and $t\approx3.5\,{\rm days}$ together with the observed data. The spectrum with mildly lanthanide-rich post-merger ejecta ({\tt GW170817\_YM}) exhibits much less feature than that with lanthanide-free one ({\tt GW170817\_YH}). We find that this is also the case for the later phase ($t\gtrsim3.5\,{\rm days}$). This is a consequence of the huge number of spectral line mixing, and thus, this indicates that high velocity ejecta is not necessarily required to reproduce the observed featureless spectrum unless the line emitting region is lanthanide-free. We also note that this shows the importance of employing the complete atomic line list for the light curve prediction.

We note that, however, there is a drawback in these models. Neither models reproduce the {\it grizJ}-band light curves for $t\ge2$--$4\,{\rm days}$ due to too steep decline in the model light curves. Both models in Figure~\ref{fig:mag_can} exhibit too fast and exponential-like decline feature in the {\it griz}-band light curves compared to the observation (see also the spectrum for $\lesssim 7000 \AA$ at 3.5 day in Figure~\ref{fig:comp_spec}). On the other hand, the observed {\it griz}-band light curves of GW170817 exhibits approximately power-law like decline for the late phase. Actually, it is natural to have exponential-like decline feature by employing the power-like heating rate because the spectrum declines exponentially in the high-frequency part if the spectrum is approximately the black-body. Indeed, the similar drawback is also found in the model proposed in~\cite{Waxman:2017sqv}, though the ejecta configuration, treatment of radiative transfer, and the energy deposition rate are different from our model (see Figure 10 in the reference). The observed power-law like decline of the optical light curves for the late phase may indicate that additional ejecta components of which diffusion timescales are different from the post-merger ejecta models may be needed to explain the {\it griz}-band light curves below $\gtrsim-14\,{\rm mag}$ or $m_{\rm app}\gtrsim 19\,{\rm mag}$, where $m_{\rm app}$ denotes the apparent magnitude. For example, such power-law like feature for the late phase are described by the superposition of light curves from three ejecta components with different diffusion timescales for the model proposed in~\cite{Villar:2017wcc} (see Figure 4 in the reference). 

Alternatively, this issue may be due to our assumption for the heating rate. The model of the radioactive heating rate which we employ for the post-merger ejecta is derived based on a nucleosynthesis calculation for the dynamical ejecta~\citep{Wanajo:2014wha}, but it may differ from that for the post-merger ejecta in reality due to the difference in ejecta composition, entropy and velocity~\citep{Lippuner:2015gwa,Wanajo:2017cyq,Wanajo:2018wra}. In addition, the LTE assumption, which we assume to determine the ejecta temperature, would not be valid in such a late phase because the density becomes low. Thus, we note that the behavior of the light curves for the late phase as well as the ejecta mass suitable for interpreting the observational results could change from the current results if more realistic heating rate model is employed or the non-LTE effect is taken into account.

\section{Kilonova light curves in various NS mergers}\label{sec:variety}
In the previous section, we present  models that could capture the features for the light curve of GW170817. These models are considered to be typical models for the case that a HMNS is formed after the merger of binary neutron stars. In this section, we show the predicted light curves for a variety of NS mergers.
\subsection{Prompt collapse cases}
\begin{figure*}
 	 \includegraphics[width=.5\linewidth]{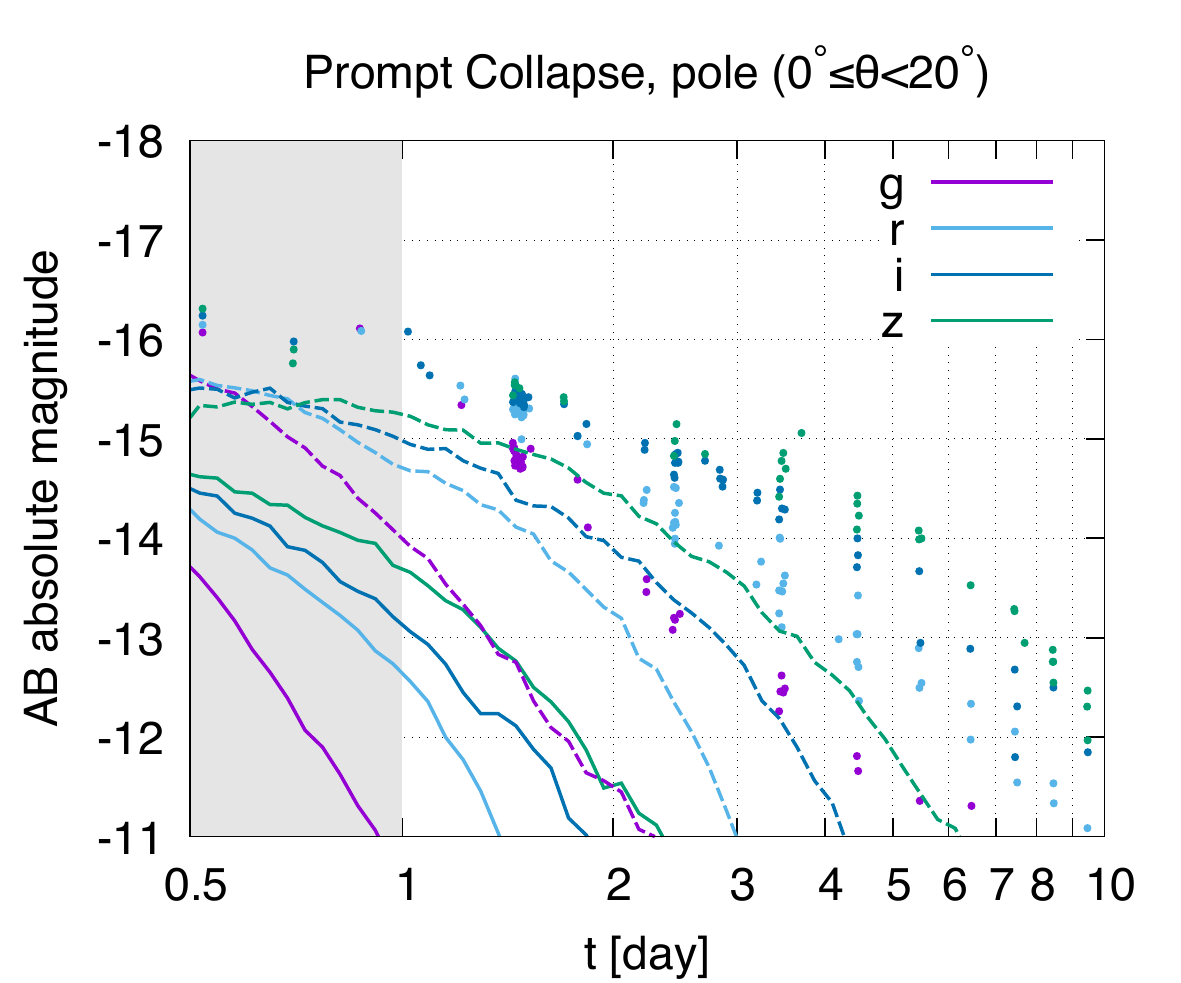}
 	 \includegraphics[width=.5\linewidth]{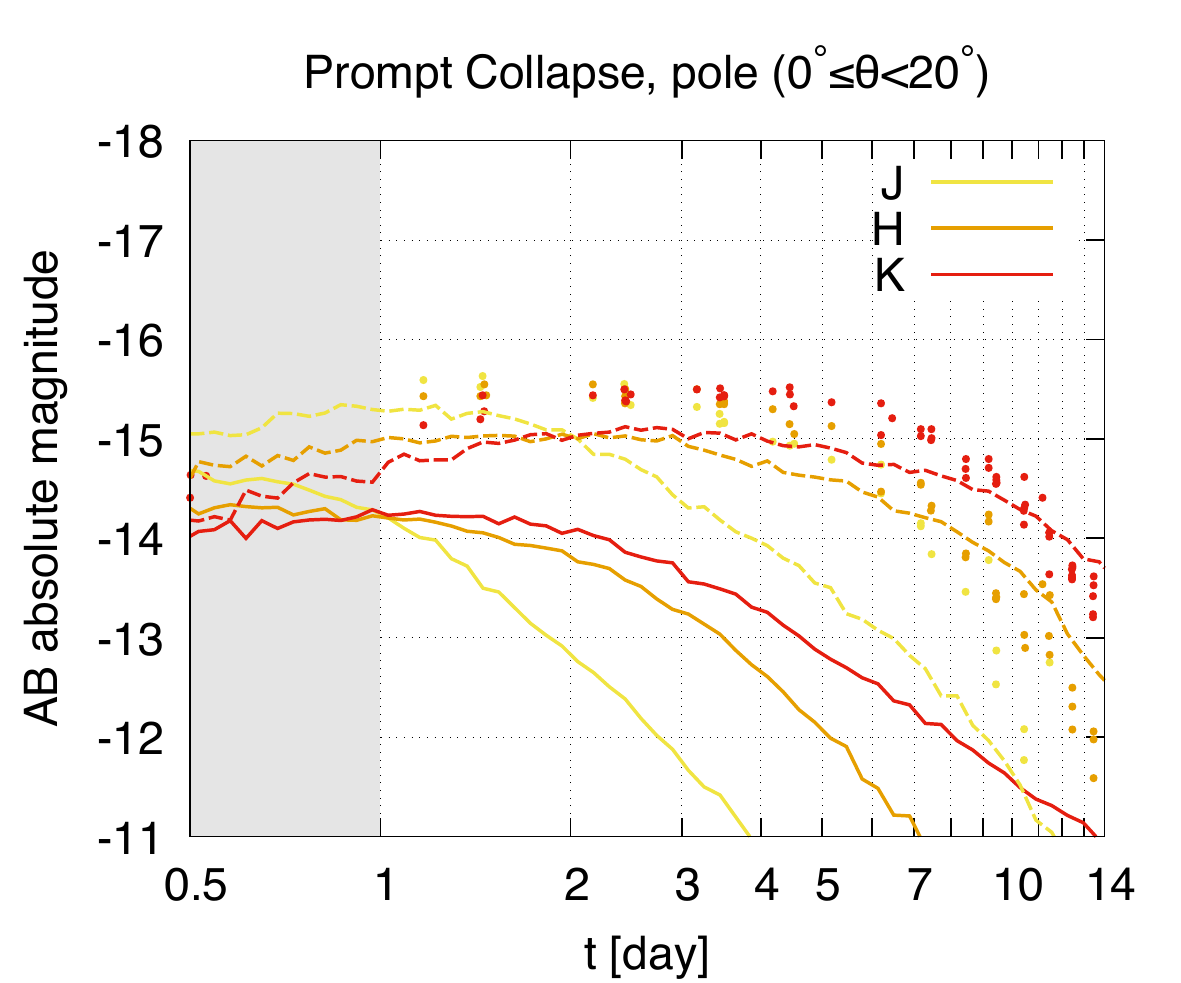}\\
 	 \includegraphics[width=.5\linewidth]{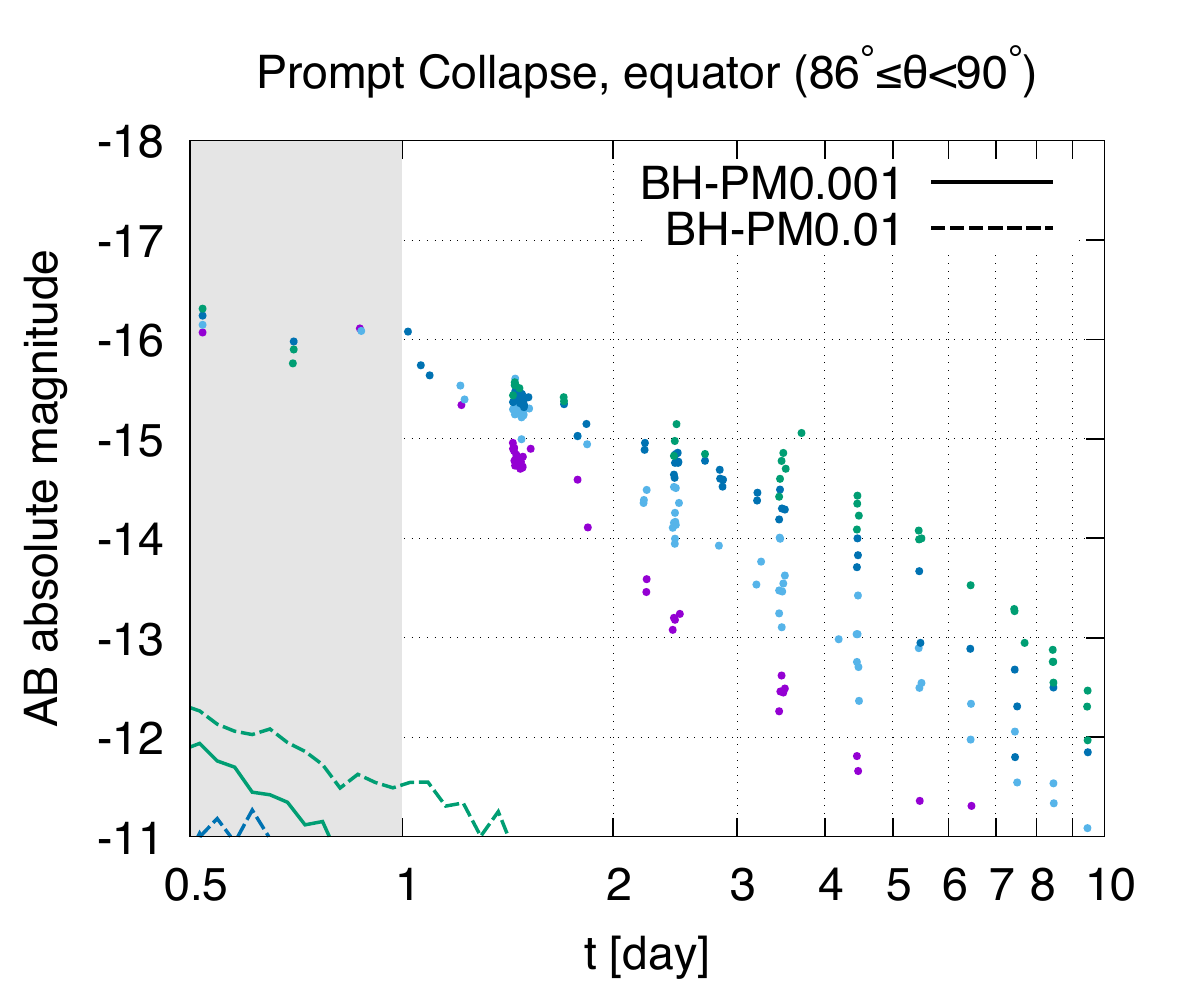}
 	 \includegraphics[width=.5\linewidth]{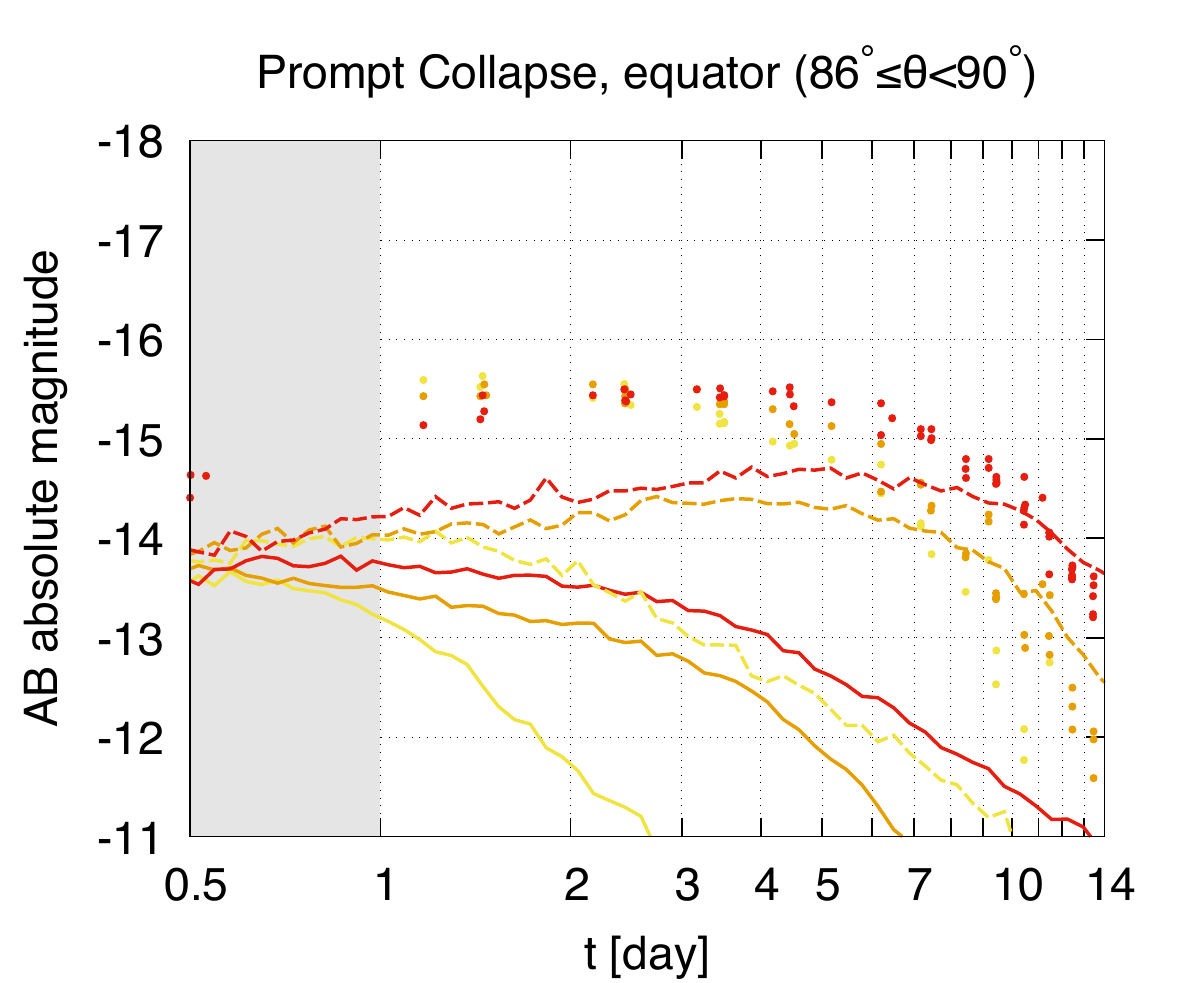}
 	 \caption{The {\it grizJHK}-band light curves for prompt collapse ejecta models. The solid and dashed curves denote the small post-merger ejecta mass model ({\tt BH\_PM0.001}; $M_{\rm pm}=0.001\,M_\odot$) and relatively large post-merger ejecta mass model ({\tt BH\_PM0.01}; $M_{\rm pm}=0.01\,M_\odot$), respectively. For a reference, we also plot the data points of GW170817~\citep{Villar:2017wcc}.}
	 \label{fig:mag_pc}
\end{figure*}

Figure~\ref{fig:mag_pc} shows the {\it grizJHK}-band light curves for prompt collapse ejecta models in which a BH is supposed to be immediately formed after the onset of NS-NS merger. The dynamical ejecta mass is set to be $0.001\,M_\odot$ for both {\tt BH\_PM0.001} and {\tt BH\_PM0.01}, and the post-merger ejecta mass to be $0.001\,M_\odot$  and $0.01\,M_\odot$ for {\tt BH\_PM0.001} and {\tt BH\_PM0.01}, respectively. These values are employed motivated by the numerical results that the ejecta mass is suppressed for the prompt collapse case~\citep{Hotokezaka:2012ze}. Kilonovae for these models are fainter than those in the HMNS formation case shown in Section~\ref{sec:multi} due to the small ejecta mass. Indeed, the peak brightness of the {\it griz}-band light curves is always fainter than $-16\,{\rm mag}$. However, for the model with the post-merger ejecta mass of $0.01\,M_\odot$, the {\it riz}-band emission remains to be brighter than $-13\,{\rm mag}$ (which corresponds to $m_{\rm app}=20\,{\rm mag}$ for the distance of $40\,{\rm Mpc}$) until $2$--$3.5\,{\rm days}$ after the merger due to relatively larger mass and longer diffusion timescale. This indicates that the {\it riz}-band emission for $t\lesssim3\,{\rm days}$ would be observed by $1$-m class telescopes, such as ZTF~\citep{2017NatAs...1E..71B}\footnote{We assume the limiting magnitude to be $m_{\rm app}=20\,{\rm mag}$.}, even for the prompt collapse cases if the event occurs as close as in GW170817 and is observed close to face-on. On the other hand, for the model of which post-merger ejecta mass is $0.001\,M_\odot$, the {\it griz}-band emission becomes fainter than $-11\,{\rm mag}$ within $2\,{\rm days}$. The optical emission observed from the equatorial direction is completely blocked by the dynamical ejecta even for the case that the dynamical ejecta is not massive $0.001\,M_\odot$. Thus, if the post-merger ejecta is much smaller than $0.01\,M_\odot$ or the event is observed from the equatorial direction, the emission would not be brighter than $m_{\rm app}=22\,{\rm mag}$ in any optical bands after $\approx2\,{\rm day}$ unless the event occurs within $\sim 40$ Mpc.

While then, the {\it HK}-band emission brighter than $-11\,{\rm mag}$ could last even for a week for both models. In particular, the {\it HK}-band emission is as bright as in that observed in GW170817 for $\approx7$--$10$ days. This is also the case for the emission observed from the equatorial direction. Thus, the HK-band emission would be brighter than $m_{\rm app}=22$ mag and could be observed by 4-m class telescopes, such as VISTA~\citep{Dalton2006} even for the prompt collapse cases if the distance to the event is the same as in GW170817 ($\sim 40$ Mpc).

\subsection{SMNS cases (accelerated ejecta)}

\begin{figure*}
 	 \includegraphics[width=.5\linewidth]{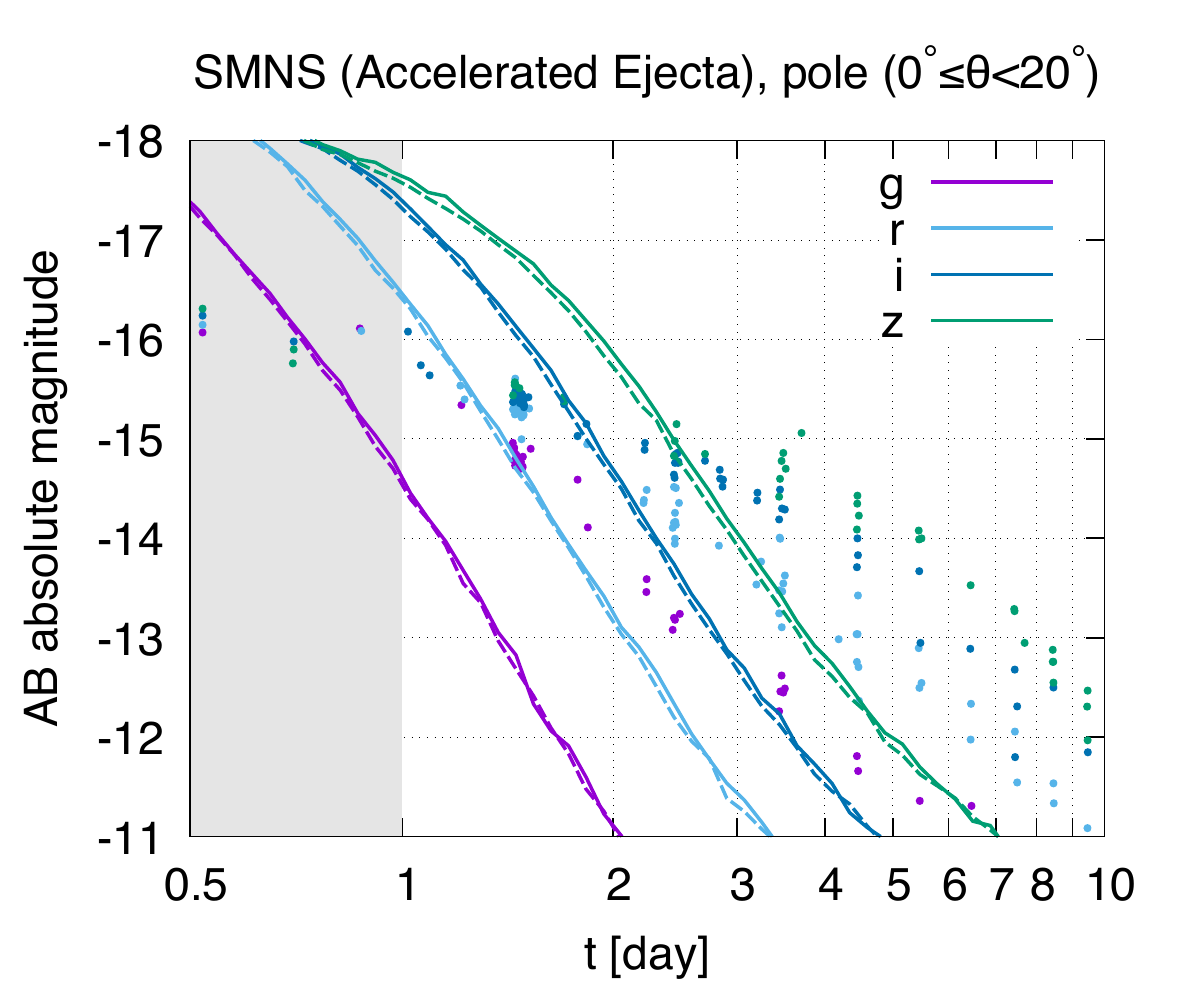}
 	 \includegraphics[width=.5\linewidth]{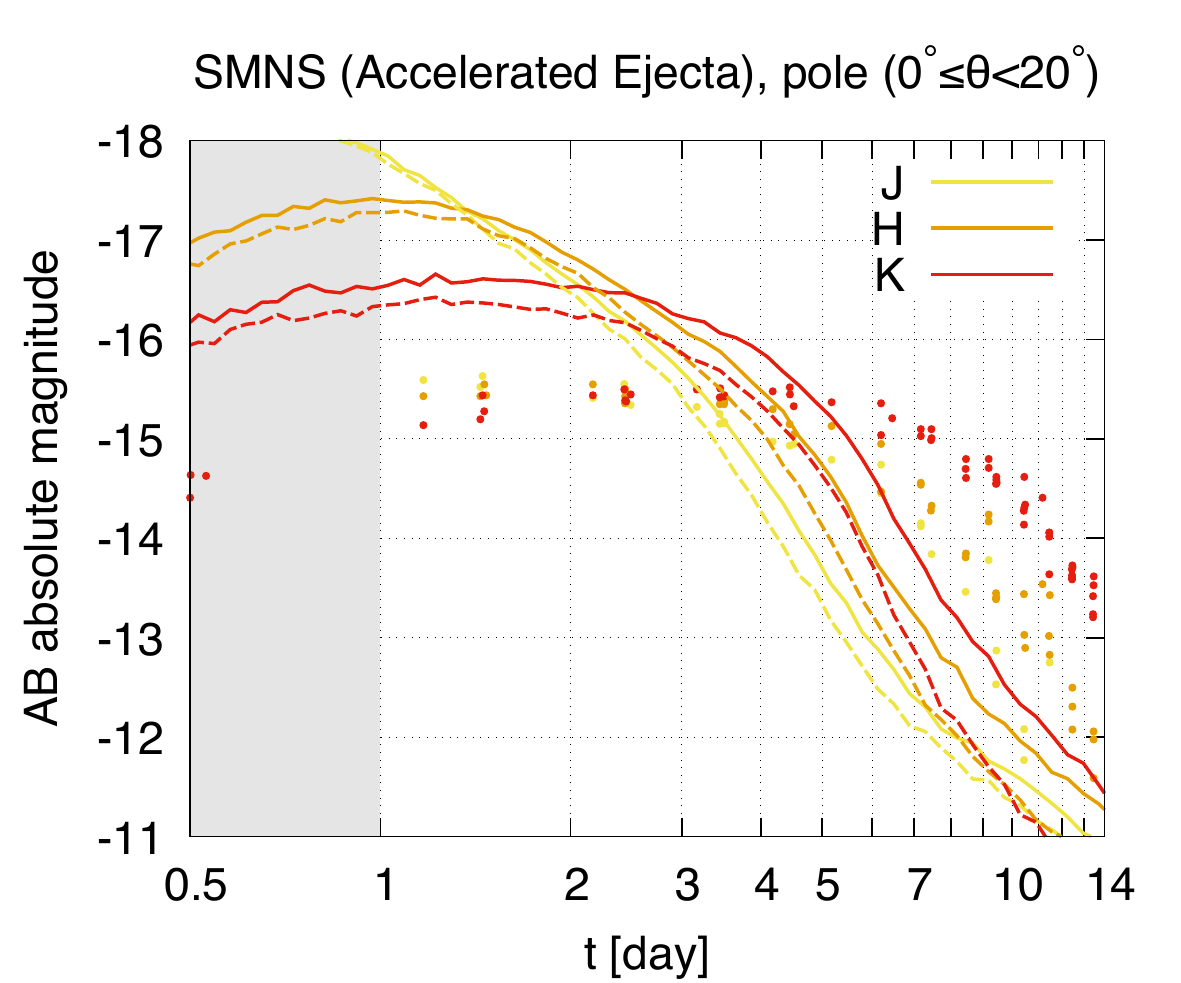}\\
 	 \includegraphics[width=.5\linewidth]{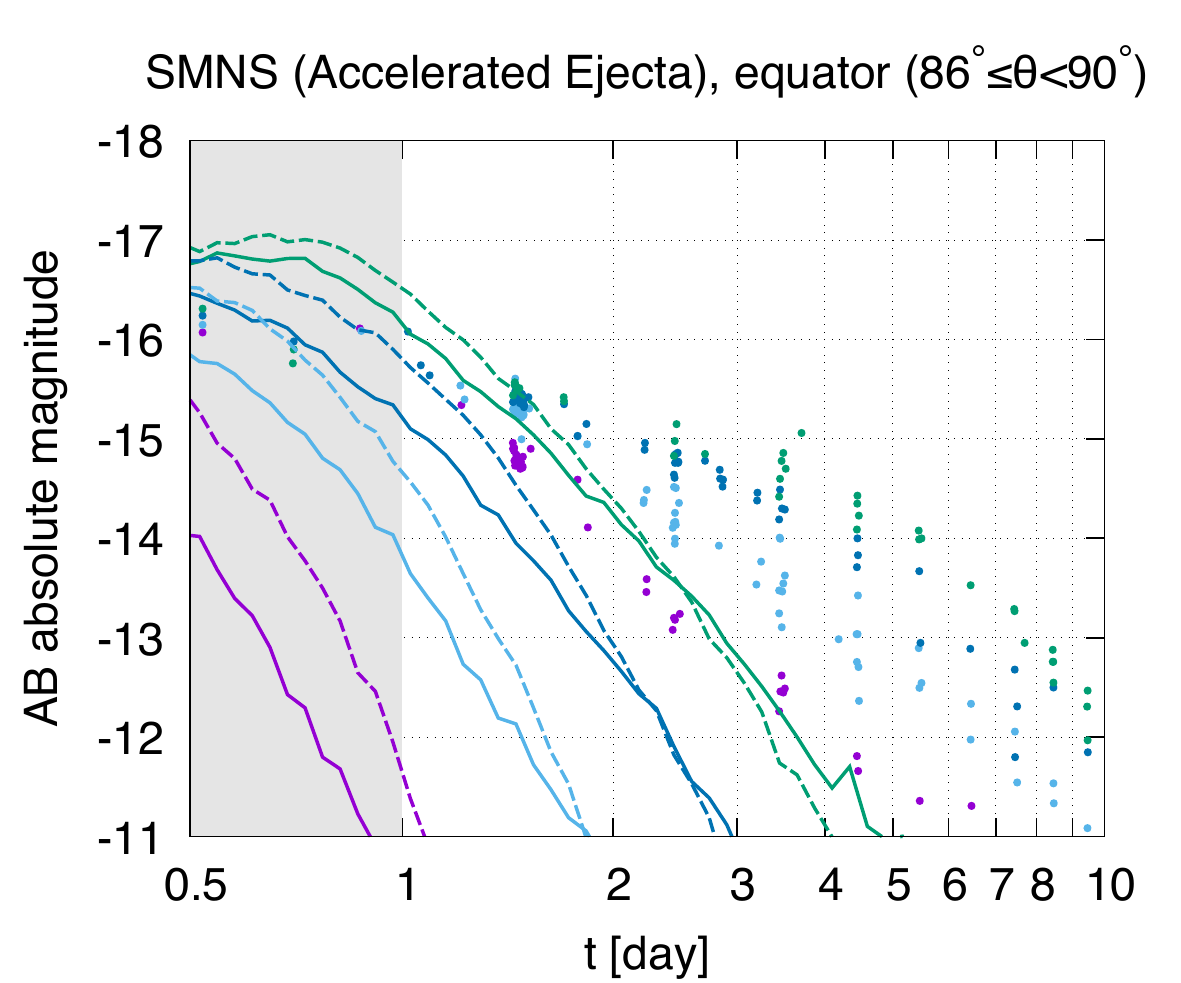}
 	 \includegraphics[width=.5\linewidth]{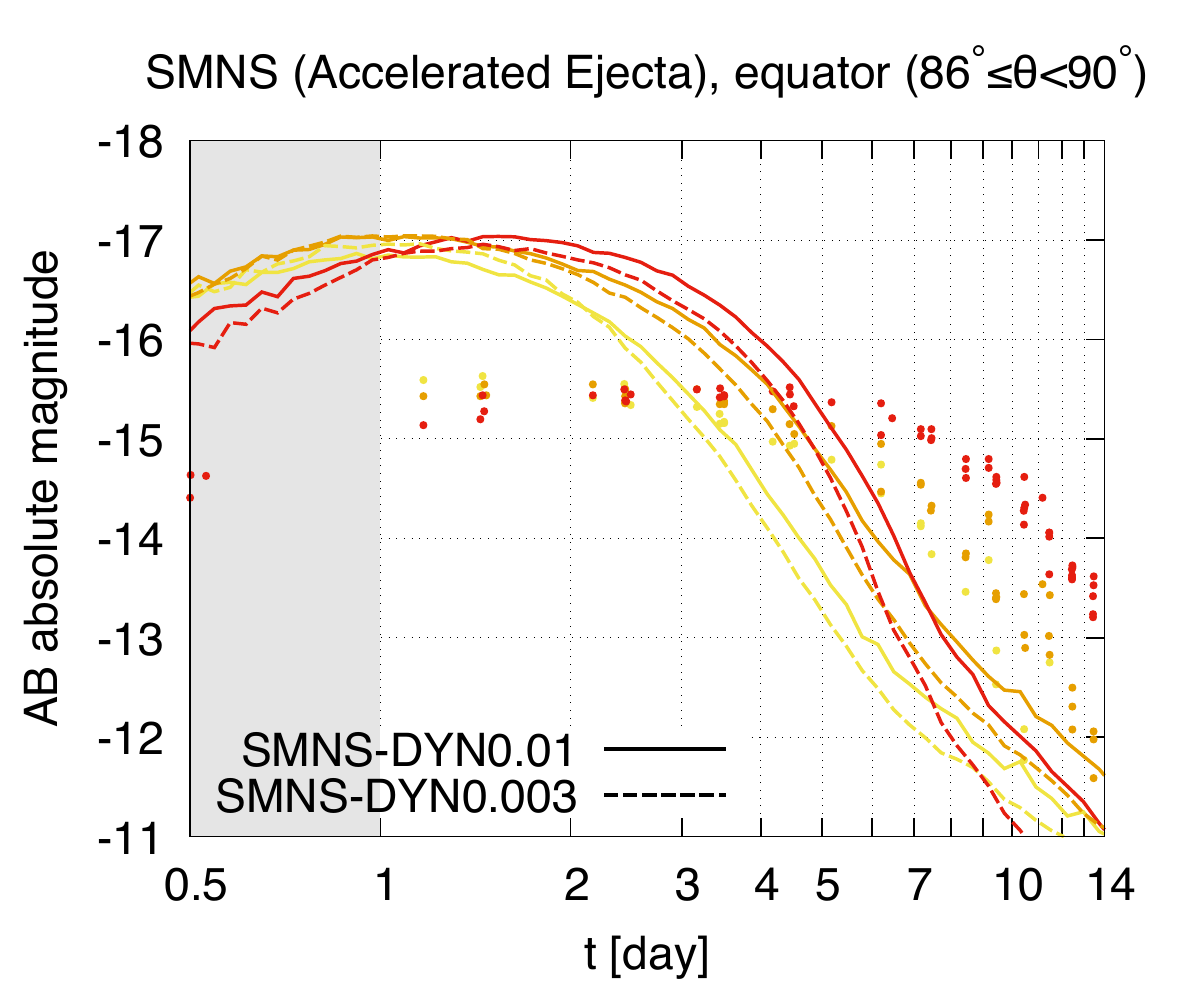}
 	 \caption{The {\it grizJHK}-band light curves for the SMNS models with highly accelerated ejecta. The solid and dashed curves denote the models with $0.01\,M_\odot$ dynamical ejecta ({\tt SMNS\_DYN0.01}) and with $0.003\,M_\odot$ dynamical ejecta ({\tt SMNS\_DYN0.003}), respectively.  For a reference, we also plot the data points of GW170817~\citep{Villar:2017wcc}.}\label{fig:mag_dc3}
\end{figure*}

Figure~\ref{fig:mag_dc3} shows the {\it grizJHK} light curves for the SMNS models with highly accelerated ejecta. For the SMNS models, bright peak luminosity for the early phase is realized due to the short diffusion timescale associated with the high velocity of the post-merger ejecta. For both SMNS models ({\tt SMNS\_DYN0.01} and {\tt SMNS\_DYN0.003}), the {\it riz}-band emission in the polar region is brighter than that observed in GW170817 by $1$--$2\,{\rm mag}$ for $t\lesssim1$--$3\,{\rm days}$. On the other hand, the brightness of those light curves becomes fainter after $1$--$3\,{\rm days}$, and particularly, the {\it g}-band emission is always fainter than that in GW170817 after $\approx0.7\,{\rm day}$. In addition, the {\it griz}-band light emission after $t\gtrsim2\,{\rm days}$ is always fainter than that of the fiducial model though the post-merger ejecta masses of {\tt SMNS\_DYN0.01} and {\tt SMNS\_DYN0.003} are larger than the fiducial case. These imply that the rapid follow-up observation is crucial to find this type of kilonovae and to determine the ejecta properties from the light curves if ejecta is highly accelerated~\citep{Arcavi:2018mzm,Matsumoto:2018mra}.

The same features are also found in the {\it JHK}-band light curves. The {\it JHK}-band emission is brighter for $\lesssim 3\,{\rm days}$, but it declines much faster than that observed in GW170817 or that of the fiducial model. The {\it JHK}-band emission is brighter for the model with larger dynamical ejecta mass ({\tt SMNS\_DYN0.01}) particularly for the late phase ($t\lesssim4\,{\rm days}$), but yet, it is fainter than that observed in GW170817. Indeed, we find that the post-merger ejecta more massive than $0.1\,M_\odot$ is needed for the {\it JHK}-band emission to be as bright as that observed in GW170817 for the late phase ($t\ge6\,{\rm days}$) in the presence of ejecta acceleration. Thus, these results indicate that faint emission observed for $t\gtrsim5$ days does not always imply small ejecta mass, and the data points in early phase $(t\gtrsim3\,{\rm days})$ are crucial to estimate the ejecta properties from the observation.

The clear difference among the SMNS models and the fiducial model is also found in the optical light curves observed from the equatorial direction. The SMNS models with highly accelerated ejecta exhibit brighter {\it iz}-band emission than that of the fiducial model, and it is as bright as that observed in GW170817 for $t\approx1$--$2\,{\rm days}$. This reflects the fact that optical photons emitted from the high-velocity part of post-merger ejecta are less absorbed by the dynamical ejecta because the column density measured from the equatorial direction is small. 

\subsection{BH-NS cases (NS tidal disruption cases)}

\begin{figure*}
 	 \includegraphics[width=.5\linewidth]{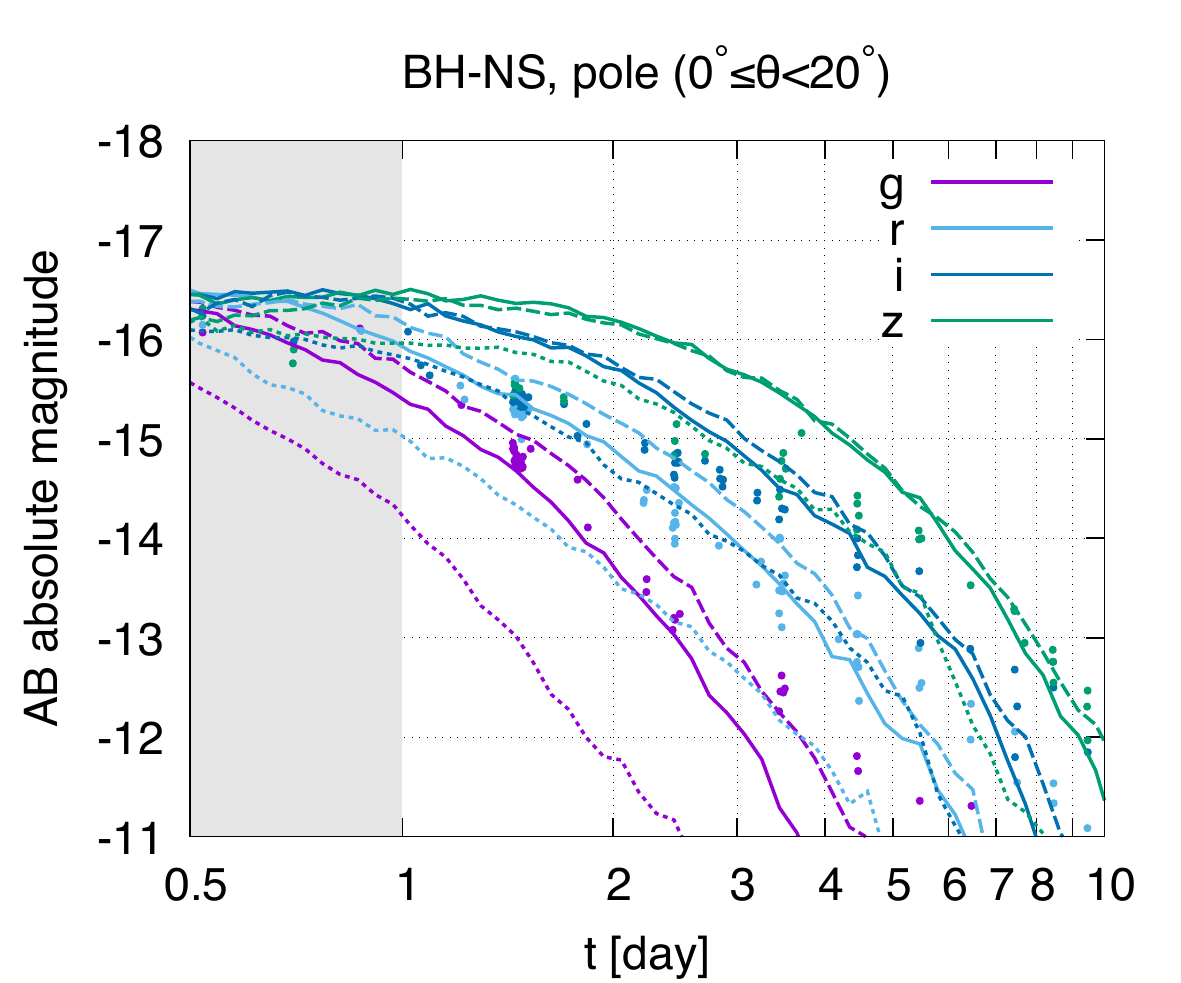}
 	 \includegraphics[width=.5\linewidth]{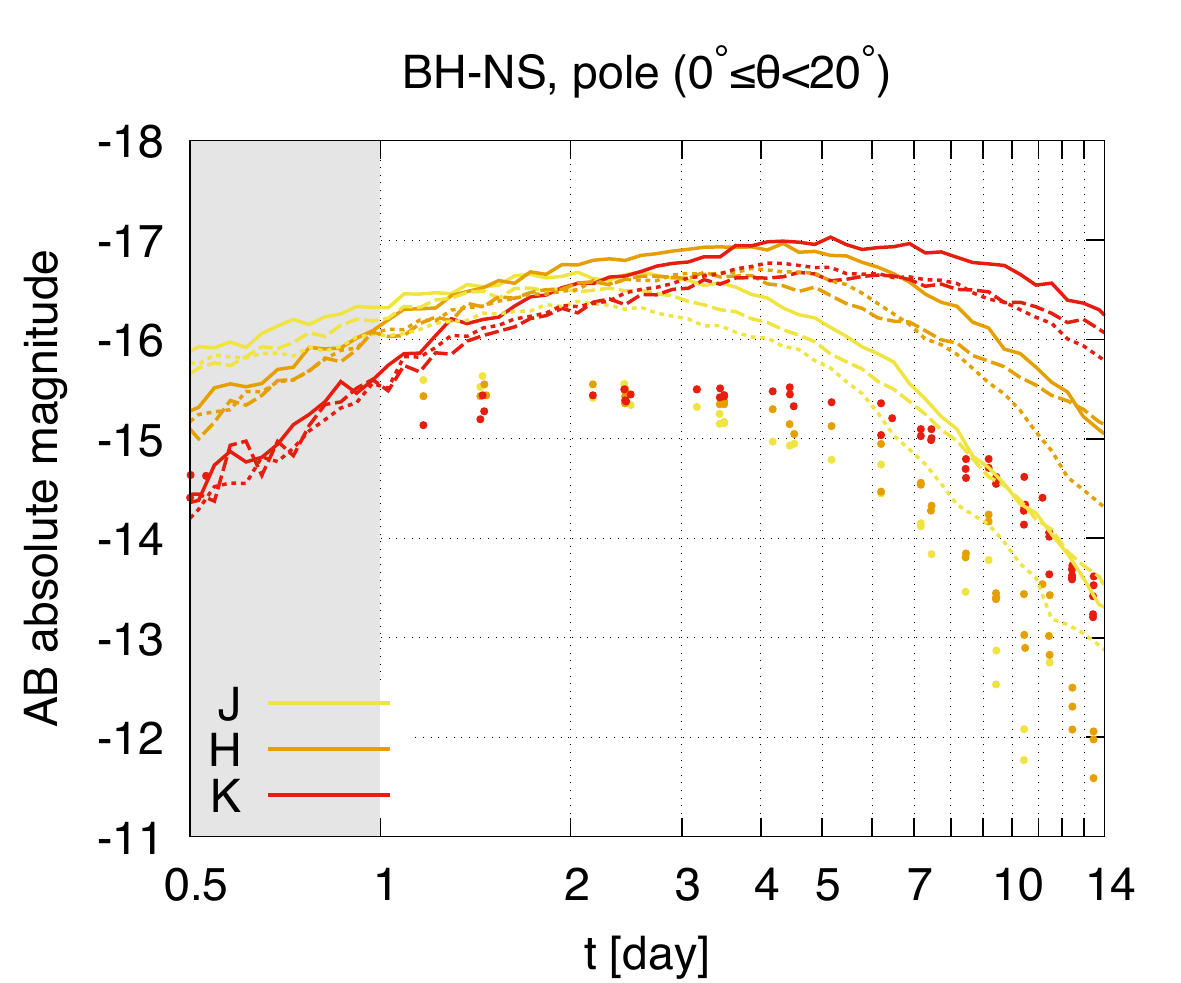}\\
 	 \includegraphics[width=.5\linewidth]{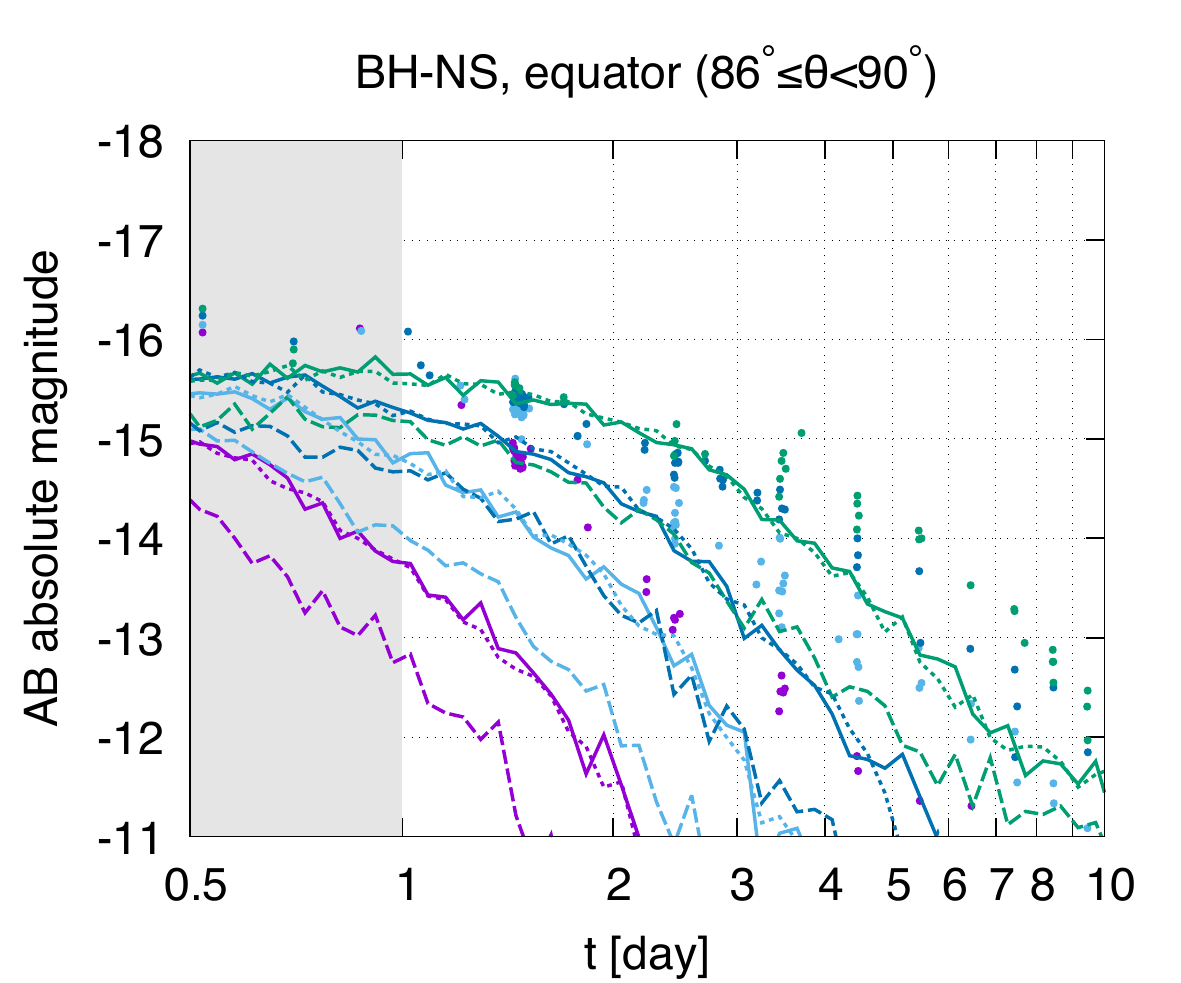}
 	 \includegraphics[width=.5\linewidth]{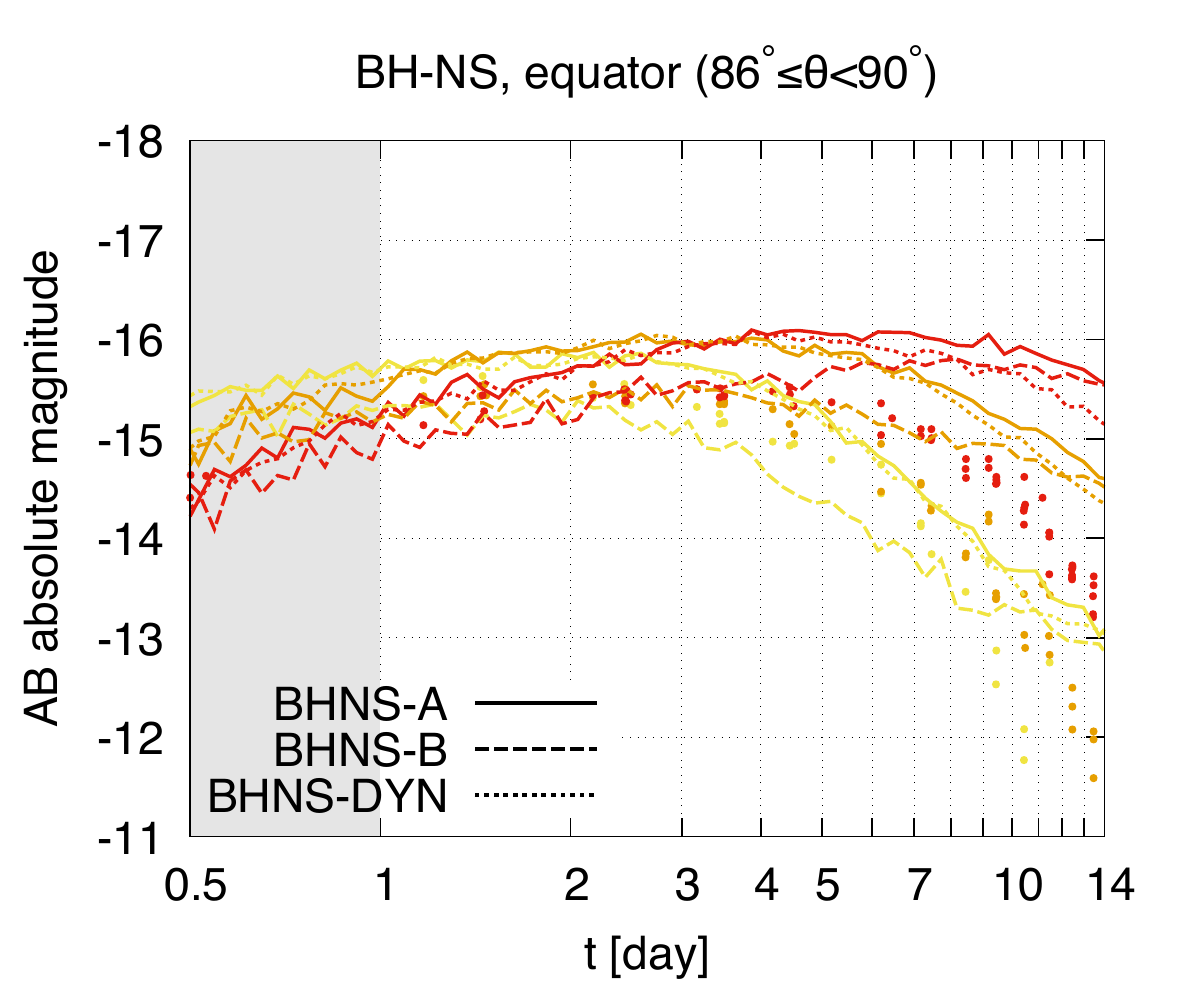}
 	 \caption{The {\it grizJHK}-band light curves for the BH-NS ejecta models. The solid curves denote the result of the BH-NS ejecta model with $0.02\,M_\odot$ post-merger ejecta and $0.02\,M_\odot$ dynamical ejecta ({\tt BHNS\_A}). Dashed curves denote the results of the BH-NS ejecta models ({\tt BHNS\_B}) with more massive post-merger ejecta ($0.04\,M_\odot$) and less massive dynamical ejecta  ($0.01\,M_\odot$). Dotted curves denote the results for the model only with dynamical ejecta  ({\tt BHNS\_DYN}). For a reference, we also plot the data points of GW170817~\citep{Villar:2017wcc}.}
	 \label{fig:mag_bhns}
\end{figure*}

Figure~\ref{fig:mag_bhns} shows {\it grizJHK} light curves for the BH-NS ejecta models for which the NS is supposed to be tidally disrupted. As is expected from the presence of massive and highly lanthanide-rich dynamical ejecta, the BH-NS ejecta models show bright emission in particular in the {\it JHK}-band. The {\it izJHK}-band emission observed from the polar direction is much brighter than that in GW170817, while the {\it griz}-band emission is as bright as that in GW170817. In particular, the BH-NS ejecta model with $0.02\,M_\odot$ post-merger ejecta coincidently reproduces the {\it gr}-band data points of GW170817. This implies that kilonovae of BH-NS mergers could be as bright as in GW170817 in the optical wavelengths, but at the same time,  emission in the {\it JHK}-band would be much brighter. Focusing on the models with the same total ejecta mass ({\tt HMNS\_YL} and {\tt BHNS\_A}), we found that the kilonova of the BH-NS model is always brighter in all the {\it grizJHK} bands than that of the NS-NS ejecta model with lanthanide-rich post-merger ejecta. This is due to higher thermalization efficiency of the dynamical ejecta resulting from its higher density profile, while the absence of dynamical ejecta in the polar region for the BH-NS case would also be the reason for difference in the optical light curves. This indicates that the ejecta mass could be overestimated if the kilonova model for NS-NS is employed for a kilonova from BH-NS.

The optical light curves of {\tt BHNS\_A} and {\tt BHNS\_B} are dominated by the emission from the post-merger ejecta. Indeed,  the model with larger post-merger ejecta mass ({\tt BHNS\_B}) and the model only with dynamical ejecta ({\tt BHNS\_DYN}) show slightly brighter and significantly fainter optical emission than that of  {\tt BHNS\_A}, respectively. This implies that the presence of the post-merger ejecta has a large impact on the optical light curves also for BH-NS cases, and the difference in the post-merger ejecta mass would be reflected in the observed optical light curves particularly in the {\it gr}-band. On the other hand, the {\tt BHNS\_A} and {\tt BHNS\_B} show approximately the same brightness of {\it iz}-band light curves in spite of the fact that the masses of dynamical and post-merger ejecta for those models are different. {\tt BHNS\_B} and {\tt BHNS\_DYN} also show approximately the same brightness of {\it K}-band light curves. These imply that the light curve only in the specific band filter particularly in the infrared bands would be degenerate with respect to the ejecta mass. Thus, the simultaneous observation in the several bands are crucial to estimate the ejecta mass.

As in other models, the {\it griz}-band emission observed from the equatorial direction is suppressed due to the presence of dynamical ejecta. However, it is still as bright as that of the NS-NS ejecta model with low-$Y_e$ post-merger ejecta observed from the polar direction, and particularly, the {\it z}-band  emission is as bright as in GW170817. We note that the emission observed from the equatorial direction is dominated by the emission from the dynamical ejecta. Indeed, the BH-NS ejecta models with the same mass of dynamical ejecta ({\tt BHNS\_A} and {\tt BHNS\_DYN}) exhibit approximately the same brightness of light curves. On the other hand, the model with smaller dynamical ejecta mass ({\tt BHNS\_B}) shows fainter emission. Thus, both optical and infrared emission for BH-NS mergers could be as bright as that of NS-NS mergers even from the equatorial direction if dynamical ejecta with sufficiently large mass (say $\gtrsim0.02\,M_\odot$) is launched. 

Before closing this subsection, we note that the ejecta mass for BH-NS mergers could have a large variety depending on the binary parameters. For example, a massive post-merger ejecta ($\gtrsim0.01$--$0.05\,M_\odot$) associated with only small amount of dynamical ejecta ($\lesssim0.001\,M_\odot$) would be launched for a BH-NS merger if the mass ratio of the BH to the NS is close to unity~\citep{Foucart:2019bxj}. For such a case, the kilonova light curves would be similar to those for {\tt PM\_YL}: The peak brightness observed from the polar direction would be fainter than that for BH-NS mergers with substantial amount of dynamical ejecta ({\tt BHNS\_A} and {\tt BHNS\_B}) due to the absence of the radiative transfer effect in the multiple ejecta components, while the peak brightness observed from the equatorial direction would be similar (see the results of {\tt PM\_YL} in Figures~\ref{fig:peak_comp_pol} and~\ref{fig:peak_comp_eq}). It should be also noted that eject mass for BH-NS mergers could be quite small 
($\lesssim 0.001\,M_\odot$) if NS is only weakly disrupted before the merger. For such a case, the light curves are likely to be similar to those for NS-NS mergers collapsing promptly to a BH.

\subsection{Comparison among various models}
\begin{figure*}
 	 \includegraphics[width=1\linewidth]{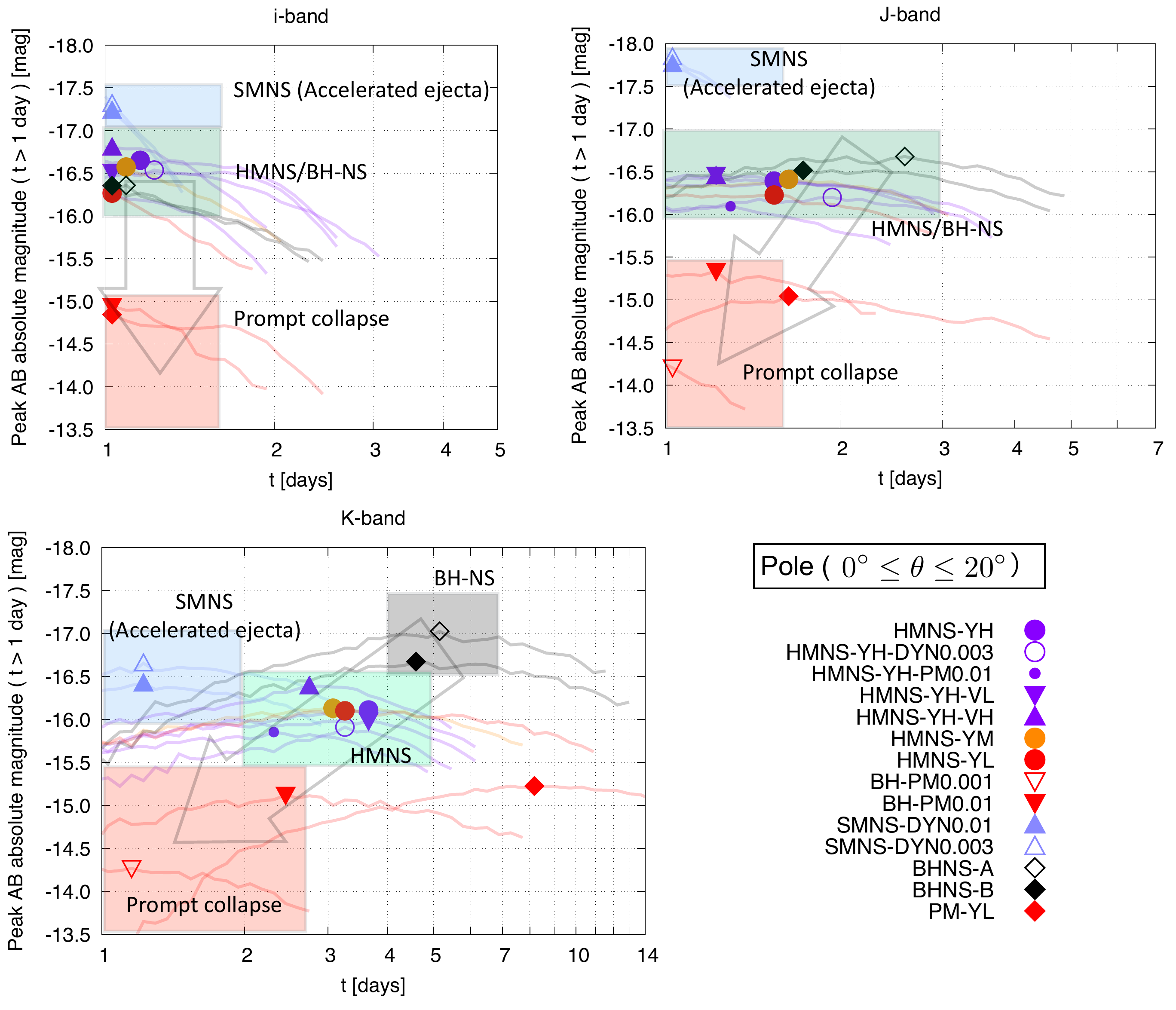}
 	 \caption{Comparison of the peak magnitude (the brightest magnitude for $t\ge1\,{\rm day}$) in the {\it iJK}-band observed from the polar direction ($0^\circ\le\theta\le20^\circ$) among various kilonova models. Each point in the plot shows the time of peak and its magnitude for each kilonova model. The light curves of which brightness is within $\approx1\,{\rm mag}$ and $\approx0.5\,{\rm mag}$ from the peak magnitude are shown in the plots for the {\it i}-band and the {\it JK}-band, respectively. The blue, green, red, and black regions denote the regions in which the peak brightness and time of peak approximately cluster for the SMNS, HMNS, prompt collapse, and BH-NS models (NS tidal disruption cases), respectively. We note that the ejecta mass for BH-NS mergers could have a large variety depending on the binary parameters, and the peak brightness and time of peak would become faint and shifted toward the early phase, respectively, (toward the direction of arrows in the figure) for small amount of ejecta.}
	 \label{fig:peak_comp_pol}
\end{figure*}

\begin{figure*}
 	 \includegraphics[width=1\linewidth]{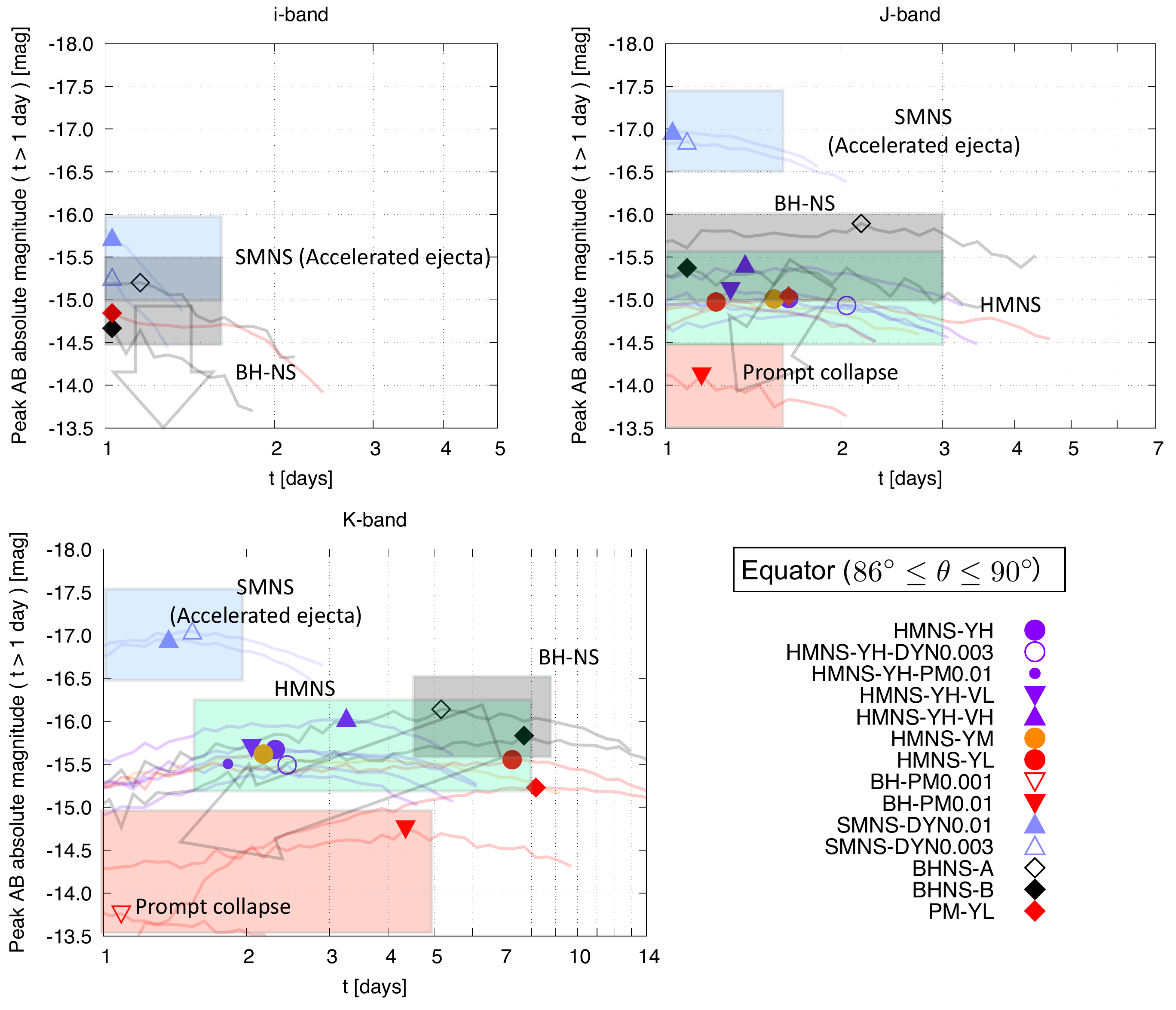}
 	 \caption{The same as Figure~\ref{fig:peak_comp_pol} but for the equatorial direction ($86^\circ\le\theta\le90^\circ$).}
	 \label{fig:peak_comp_eq}
\end{figure*}

As a summary for the variety in kilonova light curves, Figures~\ref{fig:peak_comp_pol} and~\ref{fig:peak_comp_eq} compare the peak magnitude and time of peak in the {\it iJK}-band observed from the polar direction ($0^\circ\le\theta\le20^\circ$) and equatorial direction ($86^\circ\le\theta\le90^\circ$), respectively, among various kilonova models. Here, we note that the peak magnitude is practically defined as the brightest magnitude for $t\ge1\,{\rm day}$ due to the limitation in our calculation. 

The peak magnitude in the {\it K}-band shows a rich variety in both brightness and time of peak among the models. In particular, the models of the same category (SMNS, HMNS, and Prompt collapse) are approximately clustered in the same region of peak brightness and time of peak; for the polar emission, the HMNS models (the models referred to as {\tt HMNS\_}...) are in the region of peak brightness $\approx-16.5$--$15.5\,{\rm mag}$ and time of peak $\approx2$--$5\,{\rm days}$, the SMNS models ({\tt SMNS\_DYN0.01} and {\tt SMNS\_DYN0.003}) in $\approx-16.5\,{\rm mag}$ and $\lesssim2\,{\rm days}$, the prompt-collapse models ({\tt BH\_PM0.001} and {\tt BH\_PM0.01}) in $\gtrsim-15\,{\rm mag}$ and $\lesssim2.5\,{\rm days}$. The peak magnitude and time of peak observed from the equatorial direction also show approximately the same qualitative feature, while the brightness at the peak is typically fainter by $\approx0.5\,{\rm mag}$ than that observed from the polar direction. This indicates that we may be able to infer the type of the central engine for kilonovae by observing the peak of the light curves in the infrared bands.

%, and the BH-NS models (NS tidal disruption cases; {\tt BHNS\_A} and {\tt BHNS\_B}) in $\lesssim-16.5\,{\rm mag}$ and $\gtrsim4\,{\rm days}$

The peak magnitudes in the {\it iJ}-band for the polar emission agree with each other within $\approx0.5\,{\rm mag}$ among the various models except the prompt collapse ({\tt BH\_PM0.001} and {\tt BH\_PM0.01}) and SMNS (accelerated ejecta) models ({\tt SMNS\_DYN0.01} and {\tt SMNS\_DYN0.003}). The results of the prompt collapse ({\tt BH\_PM0.001} and {\tt BH\_PM0.01}) and SMNS (accelerated ejecta) models ({\tt SMNS\_DYN0.01} and {\tt SMNS\_DYN0.003}) show brighter and fainter peak brightness in the {\it i} or {\it J}-band than the other models by more than $1\,{\rm mag}$, respectively. The peak magnitudes in the {\it J}-band for the equatorial emission also show approximately the same qualitative feature. This indicates that we may be able to distinguish the prompt collapse and SMNS cases by the other cases from the observation of the peak brightness in the near-infrared bands.

We note that the peak magnitude and time of peak for a BH-NS merger would have diversity reflecting the large variety of ejecta mass depending on the binary parameters. For example, the peak emission would be faint and the light curves would decline fast as is the case for the prompt collapse models if NS tidal disruption only weakly occurs and only small amount of ejecta is launched. Thus, we should note that a kilonova for a BH-NS merger could also exhibit peak brightness and time of peak similar to those of the prompt collapse or HMNS cases.
 
\section{Summary}\label{sec:summary}
We performed radiative transfer simulations for kilonova light curves for a variety of ejecta profiles suggested by latest numerical simulations. The radiative transfer simulations were performed employing a new line list obtained by systematic atomic structure calculations for all the r-process elements~\citep{Tanaka:2019iqp}. 

We demonstrated the strong effect of the radiative transfer of photons in the multiple ejecta components on the resulting light curve of kilonovae and clarified the dependence of the light curves on the ejecta parameters. We showed that the brightness of the optical light curves observed in the polar direction could be enhanced by $50$--$100\%$ in the presence of optically thick dynamical ejecta concentrated near the equatorial plane due to the preferential diffusion to the polar direction. This indicates that the ejecta mass could be overestimated by a factor of $\sim2$ if one fails to take into account this effect. We found that such enhancement of the optical emission particularly for the early phase of $t\lesssim2\,{\rm days}$ depends only weakly on the dynamical ejecta mass as long as $M_{\rm d}\ge0.001\,M_\odot$. In addition, significant angular dependence of the optical emission was found in the presence of dynamical ejecta with $M_{\rm d}\ge0.001\,M_\odot$. Indeed, we found that the optical emission observed from the equatorial direction would be fainter by more than $3\,{\rm mag}$ than that observed from the polar direction for such a case. Since numerical-relativity simulations with a variety of binary parameters show that the dynamical ejecta with its mass larger than $0.001\,M_\odot$ are often driven~\citep{Hotokezaka:2012ze,Bauswein:2013yna,Sekiguchi:2016bjd,Radice:2016dwd,Dietrich:2016hky,Bovard:2017mvn}, taking the effect of this enhancement in the optical bands due to the preferential diffusion as well as the angular dependence of the light curves into account would be indispensable for estimating the ejecta mass.

We showed that the infrared brightness is also enhanced due to the emission from the post-merger ejecta reprocessed in the dynamical ejecta. In addition, we showed that the infrared emission is also enhanced by emission from the post-merger ejecta particularly for the presence of lanthanide in it. These imply that the infrared light curves depend not only on the mass and velocity of the dynamical ejecta but also significantly on the mass, velocity, and the lanthanide fraction of the post-merger ejecta, and that the infrared light curves have strong degeneracy with respect to the ejecta parameters. Thus, the observation of the light curves in a wide range of wavelengths and considering the effect of the radiative transfer of photons in the multiple ejecta components is crucial to extract the physical information of the ejecta.

Based on the parameter dependence of the light curves studied in this work, we searched for the ejecta parameters that can interpret the optical and infrared light curves of the EM counterpart observed in GW170817. We showed that the lanthanide-free post-merger ejecta ($X_{\rm pm,lan}\ll10^{-3}$) is not necessarily required to interpret the peak brightness in the optical bands observed in GW170817~\citep{Kasen:2017sxr,Perego:2017wtu}. Indeed, we showed that mildly lanthanide-rich post-merger ejecta ($X_{\rm pm,lan}\approx0.025$) reproduces the peak brightness in the optical bands due to the preferential diffusion to the polar region, and it reproduces the observed infrared light curves for the late phase of $\lesssim7\,{\rm days}$ for our setup. Furthermore, we found that the spectra become much featureless by employing lanthanide-rich post-merger ejecta than by employing lanthanide-free one, which is consistent with the observation. This suggests that featureless spectra observed in GW170817 might be explained by mildly lanthanide-rich ejecta with $\lesssim 0.1\,c$ while lanthanide-free ejecta with high velocity $\lesssim 0.3\,c$ is often claimed to be necessary~\citep[e.g.,][]{Kilpatrick:2017mhz,McCully:2017lgx,Nicholl:2017ahq,Shappee:2017zly,Tanaka:2019iqp}. On the other hand, we found that it is not easy to reproduce the optical brightness for the late phase ($t\gtrsim2\,{\rm days}$) in our setups of ejecta model. This suggests that additional ejecta components may be needed to interpret the observed optical brightness of the light curves for the late phase. Alternatively, the different function for the heating rate as well as taking the non-LTE effect into account could also modify the results, and thus, the further investigation is required to achieve conclusive statements.

We found that the optical emission of kilonova for the prompt collapse of NS-NS mergers to a BH would be much fainter than that observed in GW170817 due to small ejecta masses. However, if the event occurs in the same distance as in GW170817 and the post-merger ejecta is as massive as $0.01\,M_\odot$, the {\it riz}-band emission observed from the polar direction may still be brighter than $m_{\rm app}=20\,{\rm mag}$ for $t\lesssim3\,{\rm days}$. In particular, for such a case, the {\it JHK}-band emission would be as bright as in GW170817. %Thus, the rapid follow-up observation in the {\it riz}-band is crucial for finding the EM counterparts, and once the candidates are found, further observation in the {\it JHK}-band would help us to confirm that it is a kilonova.

For the case that the ejecta is accelerated by an energy injection from the merger remnant, we found that the optical and infrared emission could be brighter by $1$--$2\,{\rm mag}$ than that observed in GW170817 for $t\lesssim 4\,{\rm days}$ and for $t\lesssim 7\,{\rm days}$, respectively, due to the short diffusion timescale associated with the high velocity of the post-merger ejecta. On the other hand, we also found that it declines much faster than that observed in GW170817. We also showed that the suppression of the optical emission observed from the equatorial direction is less significant for the case that post-merger ejecta is highly accelerated than that in the fiducial case. This indicates that the presence of high-velocity ejecta component may be confirmed from the optical light curves for the case that the edge-on observation is suggested by the GW data analysis.

We also studied kilonova light curves for BH-NS mergers that result in NS tidal disruption. We found that the optical emission could be as bright as or even brighter than that observed in GW170817 if substantial mass ($\gtrsim0.02\,M_\odot$) of post-merger ejecta is ejected, although we should note that the ejecta mass could have a large variety depending on the binary parameters. For such high ejecta mass cases, we showed that the infrared emission would be brighter by $1$--$2\,{\rm mag}$ than that observed in GW170817 due to the large amount of lanthanide-rich ejecta, and thus, we may be able to distinguish the NS-NS from the BH-NS merger. On the other hand, the similar light curves could be realized in a specific band even if the ejecta mass is different. This indicates that the simultaneous observation in the multiple wavelengths is quite crucial for the ejecta mass estimation. We also note that eject mass for BH-NS mergers could be quite small ($\lesssim 0.001\,M_\odot$) if NS is only weakly disrupted before the merger. For such a case, the light curves are likely to be similar to those for NS-NS mergers collapsing promptly to a BH. We also note that ejecta would not launched for BH-NS mergers if NS tidal disruption does not occur before the merger. For such a case, the kilonova would not associate with the gravitational wave event.

We showed that the difference in the ejecta properties would be imprinted in the differences in the peak brightness and time of peak of kilonovae (Figures~\ref{fig:peak_comp_pol}and~\ref{fig:peak_comp_eq}). This indicates that we may be able to distinguish the different merger evolution by observations of the peak in the light curves. For this purpose, it is crucial to obtain multi-color data from optical to infrared since there is a degeneracy in the peak brightness and peak time in some specific filters whose wavelengths are close to each other. We note that more systematic studies varying the ejecta parameters as well as the assumptions for calculations, such as the model for the heating rates, would be necessary to quantitatively understand how well we can distinguish models, while such work is beyond the scope of this paper.

\begin{acknowledgments}
We thank S. Fujibayashi, K. Hotokezaka, and S. Wanajo for valuable discussions. We thank the Yukawa Institute for Theoretical Physics for support in the framework of International Molecule-type Workshop (YITP-T-18-06), where a part of this work has been done. Numerical computation was performed on Cray XC40 at Yukawa Institute for Theoretical Physics, Kyoto University. This work was supported by Grant-in-Aid for Scientific Research (JP16H02183, JP16H06342, JP17H01131, JP15K05077, JP17K05447, JP17H06361, JP15H02075, JP17H06363, 18H05859) of JSPS and by a post-K computer project (Priority issue No. 9) of Japanese MEXT.
\end{acknowledgments}

\appendix
\section{light curves of each ejecta component}\label{sec:app}
In this section, we show the results and their basic properties of the light curves obtained by the radiative transfer simulations for each ejecta component. 

\subsection{Post-merger ejecta}

\begin{figure*}
 	 \includegraphics[width=.5\linewidth]{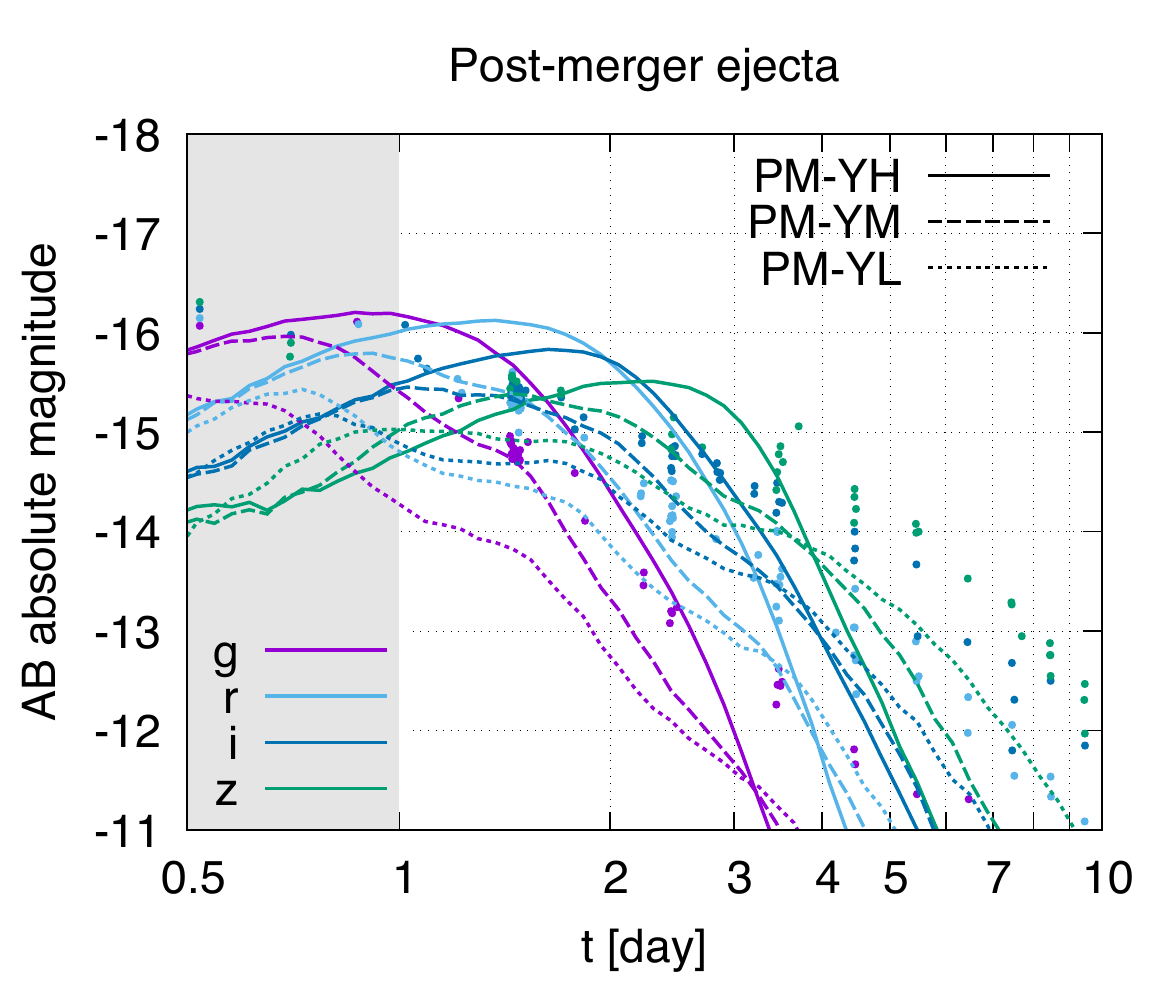}
 	 \includegraphics[width=.5\linewidth]{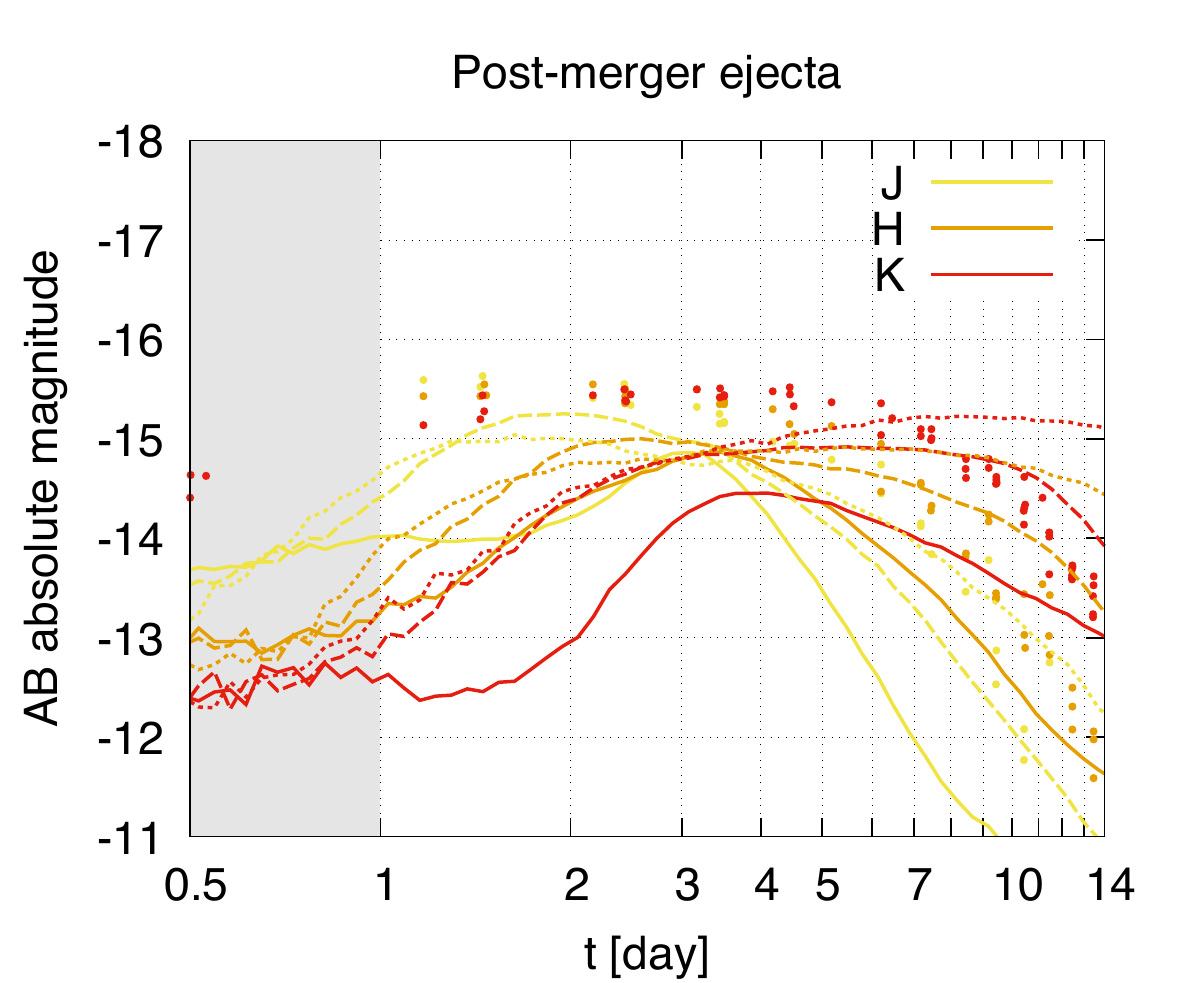}
 	 \caption{The {\it grizJHK}-band light curves for the models only with the post-merger ejecta. The solid, dashed, and dotted curves denote the light curves for the models with flat $Y_e$ distributions in $0.3$--$0.4$ ({\tt PM\_YH}, $X_{\rm pm,lan}\approx0.025$), $0.2$--$0.4$ ({\tt PM\_YM} $X_{\rm pm,lan}\ll10^{-3}$), and $0.1$--$0.3$ ({\tt PM\_YL} $X_{\rm pm,lan}\approx0.14$), respectively. For a reference, we also plot the data points of GW170817~\citep{Villar:2017wcc}.}
	 \label{fig:mag_pm}
\end{figure*}

Figure~\ref{fig:mag_pm} shows the {\it grizJHK}-band light curves of the post-merger ejecta models with different $Y_e$ distributions. The dependence on the $Y_e$ distribution is basically the same as is found in the section~\ref{sec:multi}. The {\it griz}-band emission becomes faint, and the slopes of decline become shallow as the lanthanide fraction of the post-merger ejecta increases (i.e., as the fraction of low-$Y_e$ components increases).

The {\it JHK}-band emission for $t\gtrsim 3.5\,{\rm days}$ becomes bright monotonically as the lanthanide fraction of the post-merger ejecta increases, while the brightness of the {\it JHK}-band light curves for $t\lesssim 3.5\,{\rm days}$ saturates for the case that the lanthanide fraction is higher than that of {\tt PM\_YM}. In particular, the {\it HK}-band emission for the early phase ($t\approx1\,{\rm day}$) is fainter than those peak magnitudes by $\approx 1\,{\rm mag}$. This is because the time of peak magnitudes typically delays as the lanthanide fraction of the post-merger ejecta increases due to longer diffusion timescale. 

For $t\le2$--$3\,{\rm days}$, the {\it griz}-band light curves with lanthanide-free post-merger ejecta ({\tt PM\_YH}) and with mildly lanthanide-rich post-merger ejecta ({\tt PM\_YM}) are broadly consistent with the observed data points in GW170817. On the other hand, we find that those with highly lanthanide-rich post-merger ejecta ({\tt PM\_YL}) are fainter than the data in GW170817. 

\subsection{Dynamical ejecta}

\begin{figure*}
 	 \includegraphics[width=.5\linewidth]{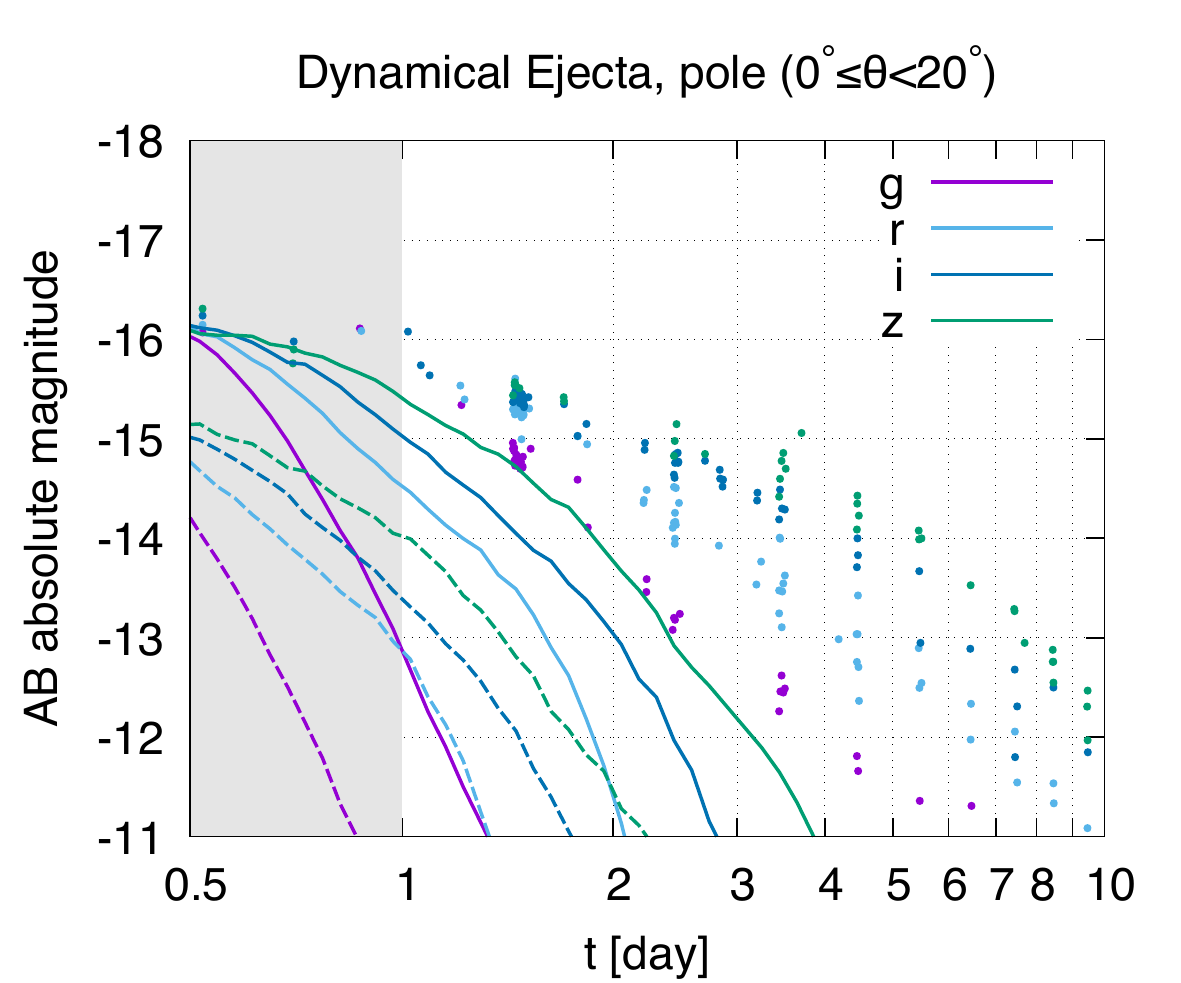}
 	 \includegraphics[width=.5\linewidth]{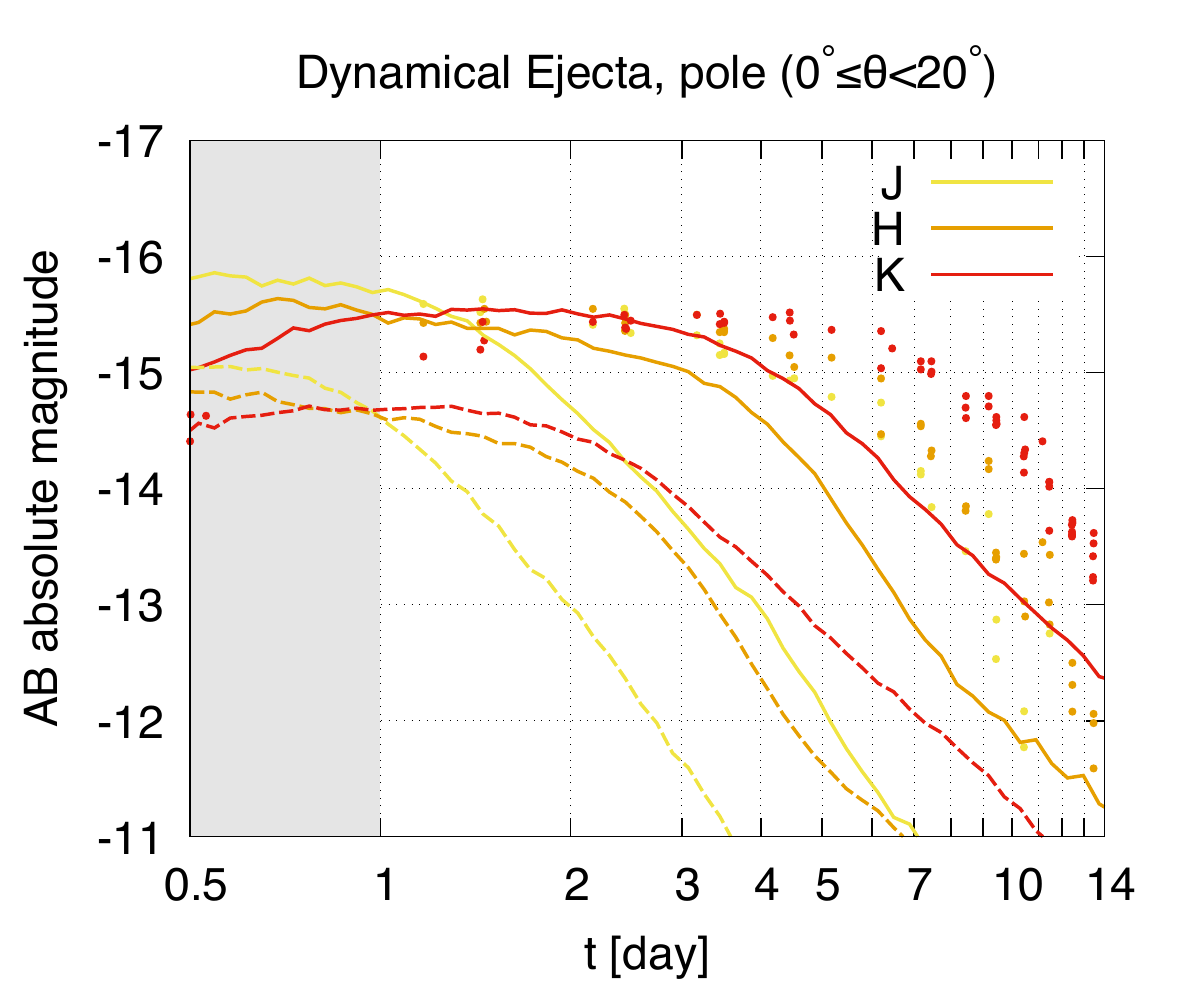}\\
 	 \includegraphics[width=.5\linewidth]{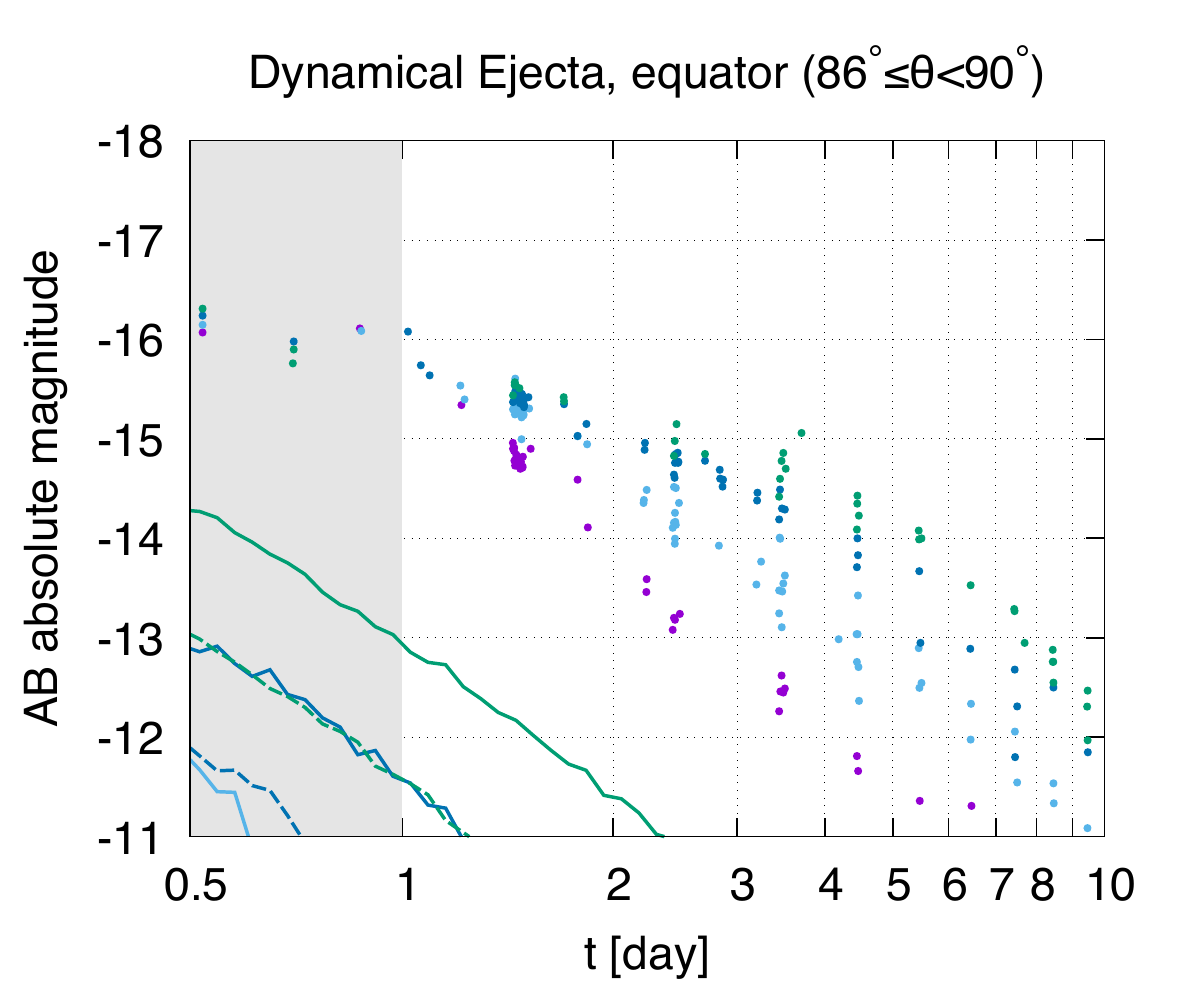}
 	 \includegraphics[width=.5\linewidth]{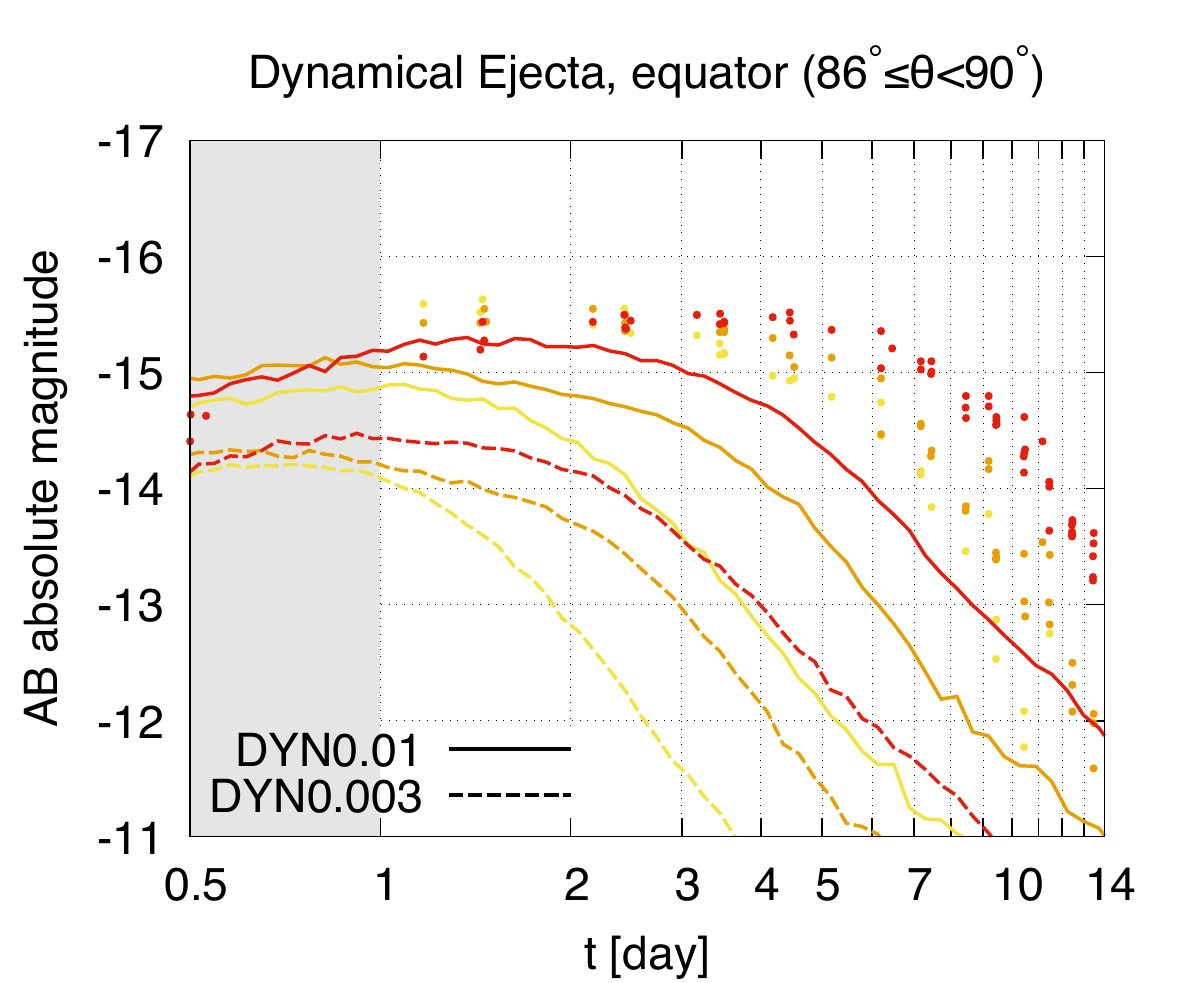}
 	 \caption{The {\it grizJHK}-band light curves for the models only with dynamical ejecta. The solid and dashed curves denote the light curves obtained by the models of which dynamical ejecta mass are $0.01\,M_\odot$ ({\tt DYN0.01}) and $0.003\,M_\odot$ ({\tt DYN0.003}), respectively. For a reference, we also plot the data points of GW170817~\citep{Villar:2017wcc}.}
	 \label{fig:mag_dyn}
\end{figure*}

Figure~\ref{fig:mag_dyn} shows the results for the models only with dynamical ejecta of different masses. As is expected, a kilonova is brighter for larger mass. In addition, significant angular dependence is present in the optical light curves due to the non-spherical morphology, and their brightness is strongly suppressed when observed from the equatorial direction.

The {\it JHK}-band emission is brighter than that in the {\it griz}-band due to high lanthanide fraction of the dynamical ejecta. The {\it JHK}-band light curves exhibit approximately the same brightness as in the peak ($t\approx 1\,{\rm days}$), and are sustained for a few days. The angular dependence of the {\it JHK}-band light curves is much weaker than the {\it griz}-band light curves. Indeed, the {\it JHK}-band  emission observed from the equatorial direction is fainter than that observed from the polar direction only by $\approx 0.5\,{\rm mag}$.

\section{Light curves observed from $20^\circ\le\theta<28^\circ$}\label{sec:misaligned}
\begin{figure*}
 	 \includegraphics[width=.5\linewidth]{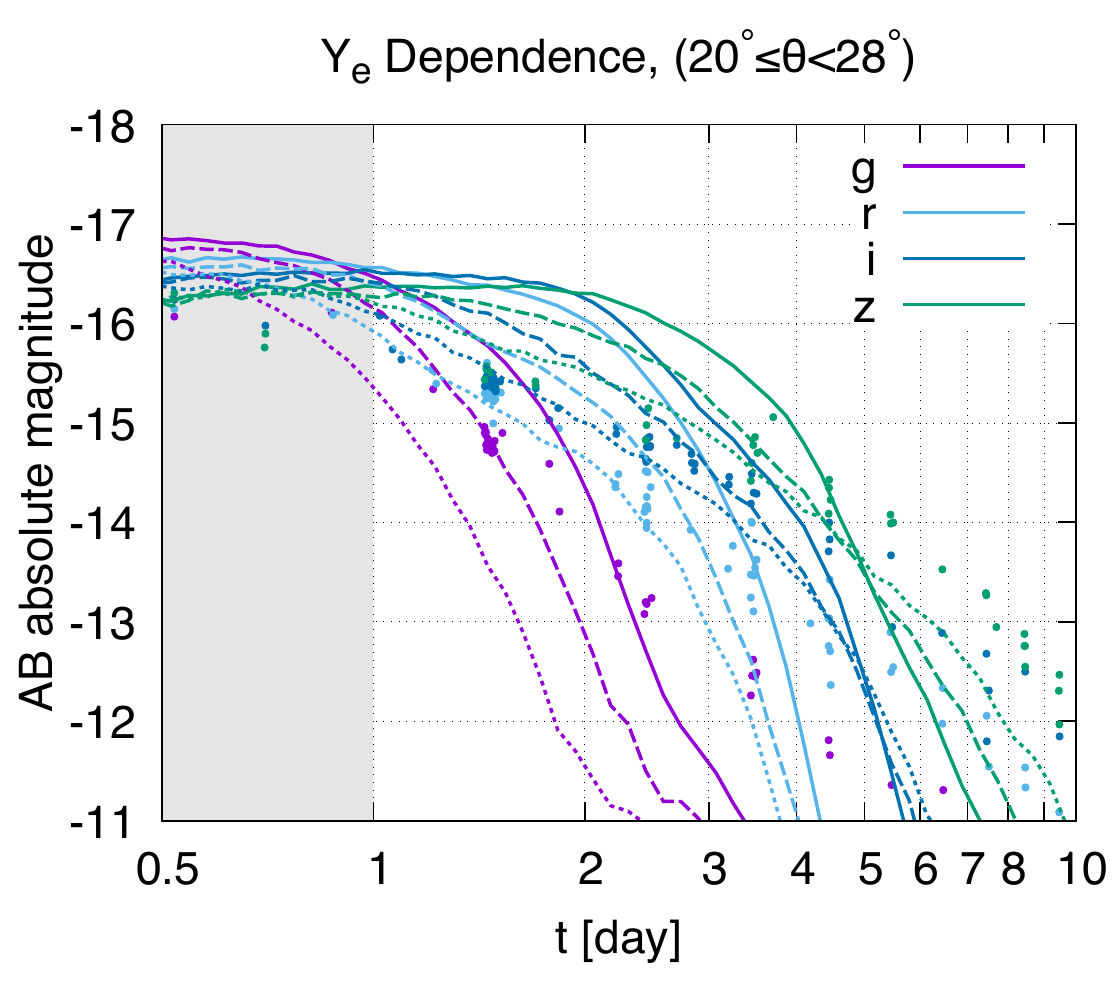}
 	 \includegraphics[width=.5\linewidth]{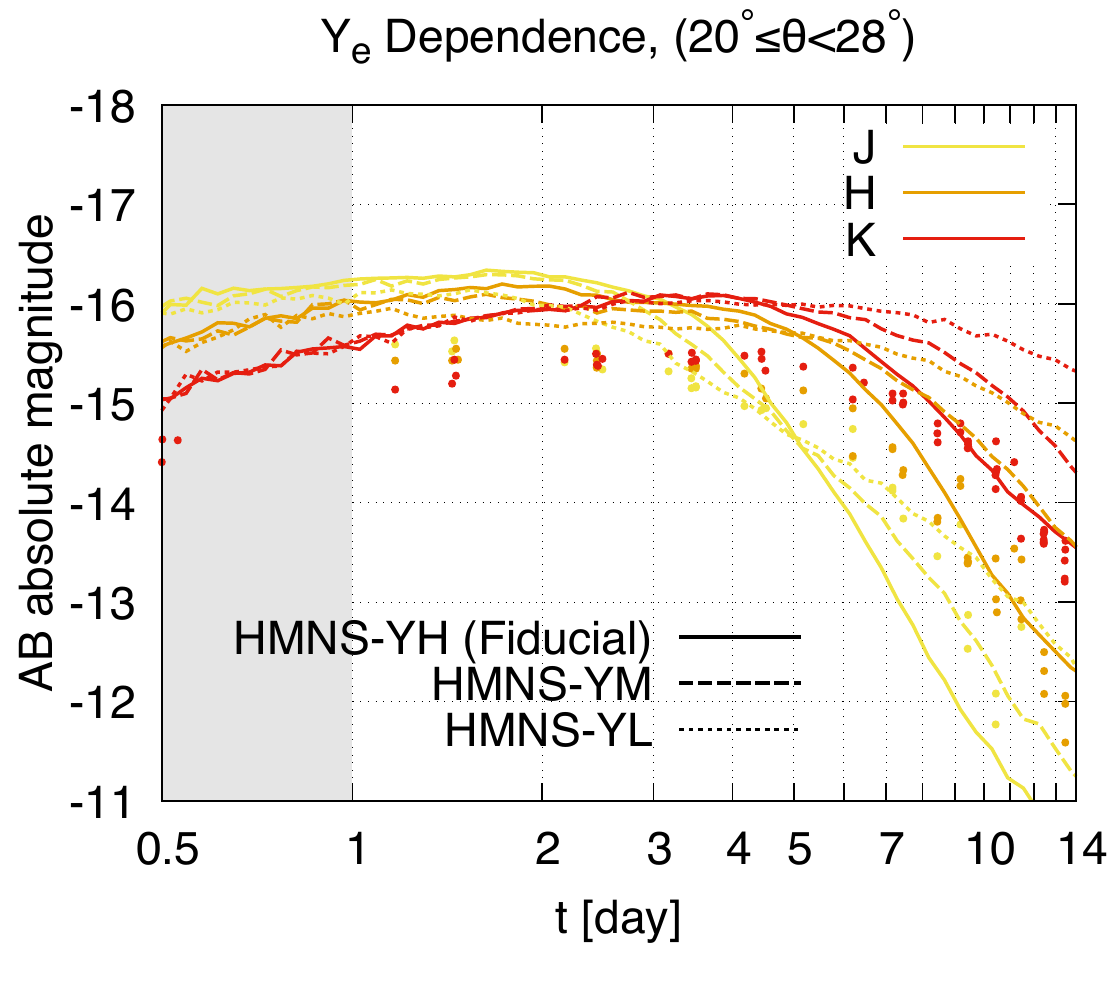}\\
 	 \includegraphics[width=.5\linewidth]{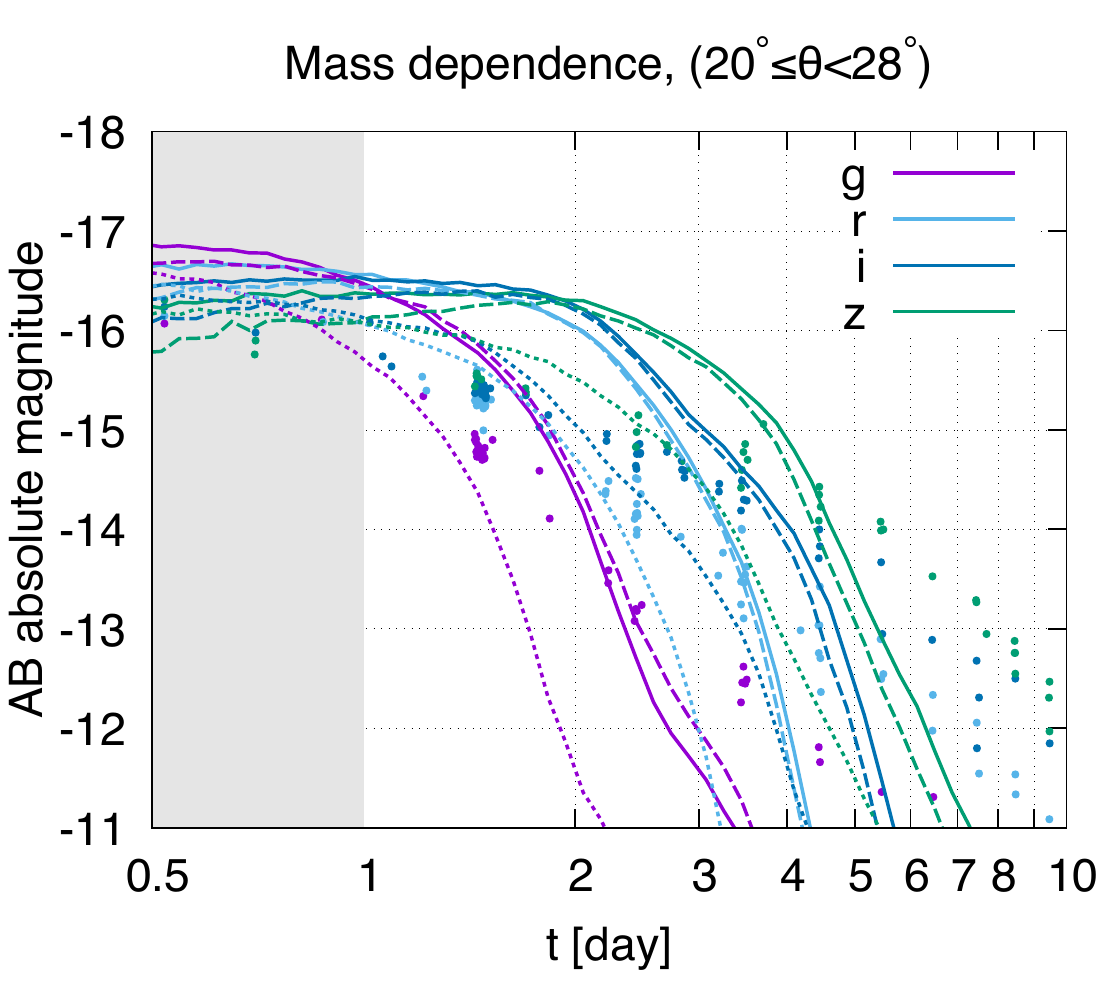}
 	 \includegraphics[width=.5\linewidth]{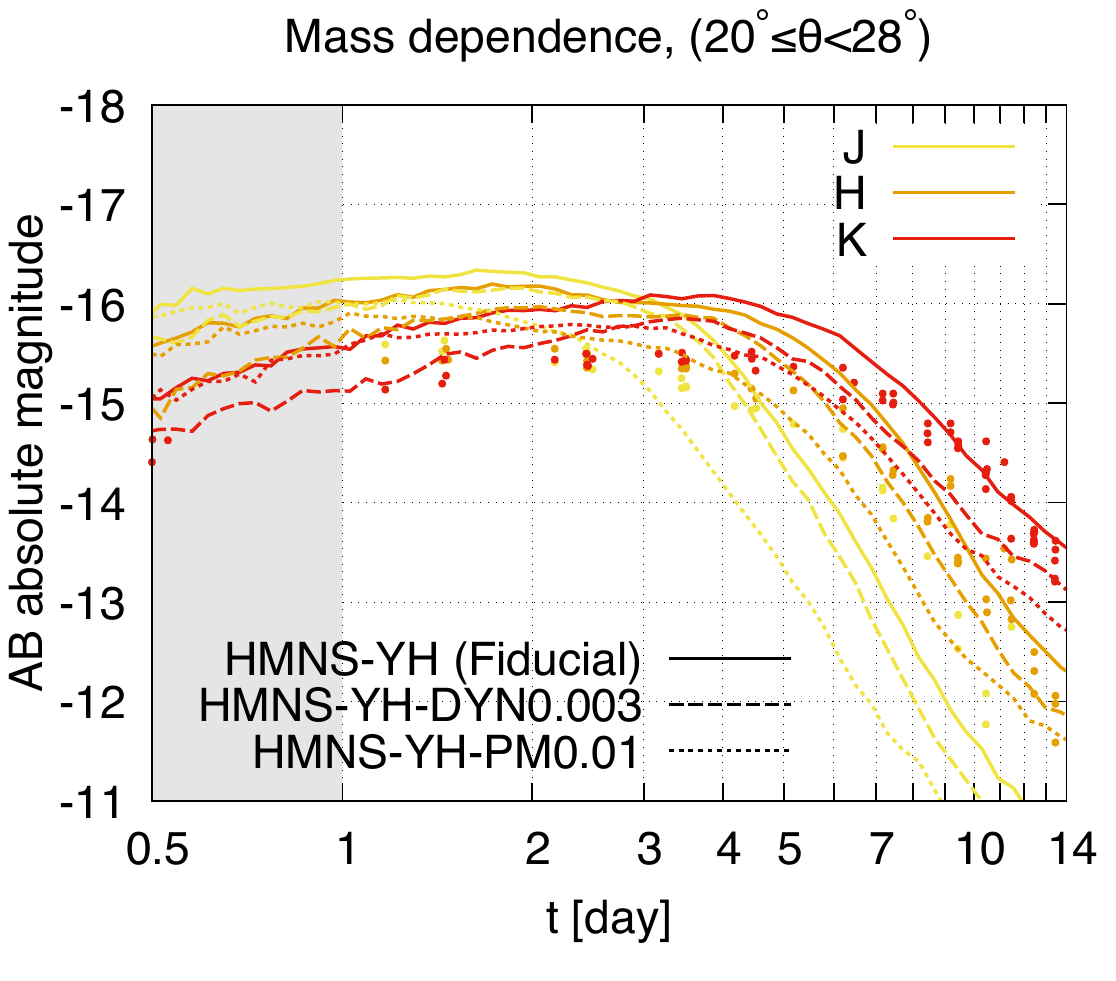}\\
 	 \includegraphics[width=.5\linewidth]{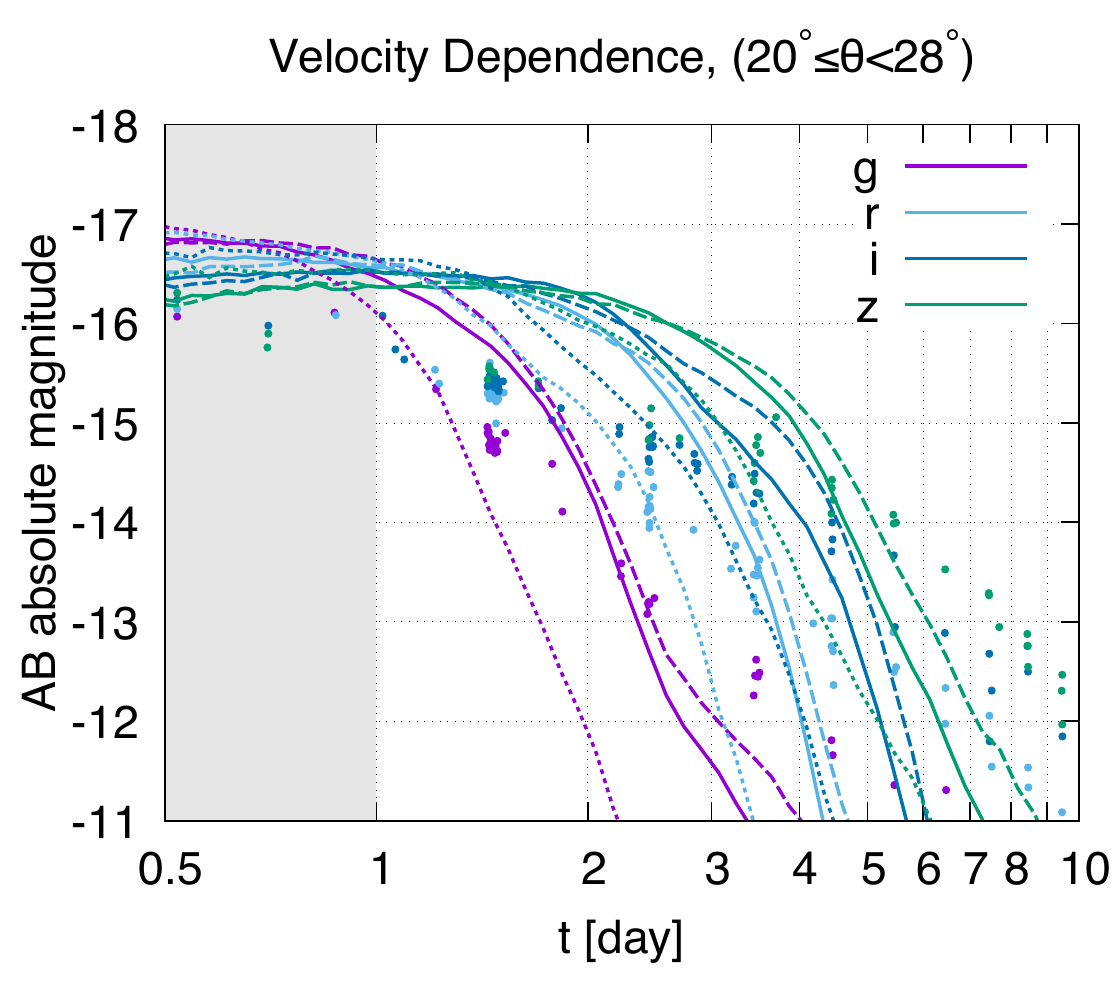}
 	 \includegraphics[width=.5\linewidth]{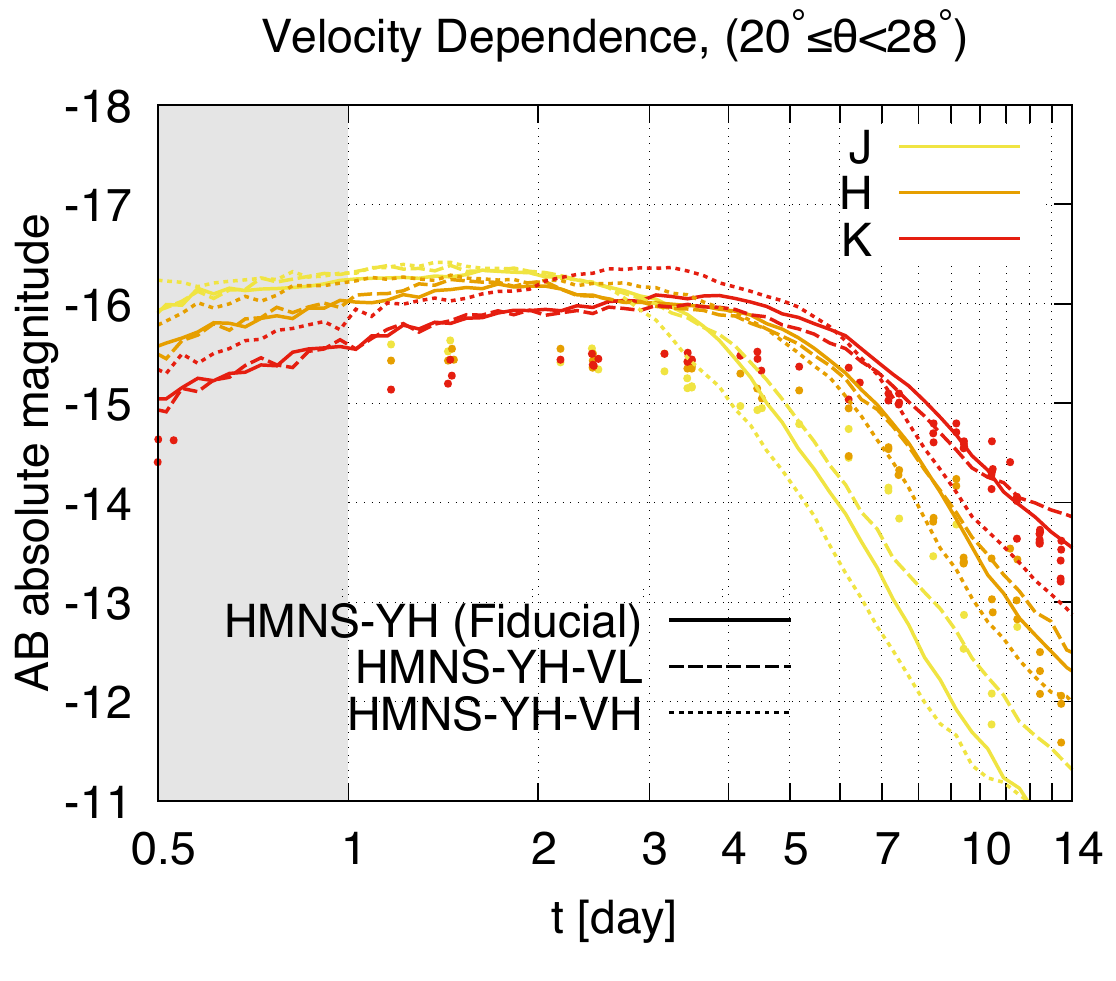}\\
 	 \caption{The {\it grizJHK}-band light curves observed from $20^\circ\le\theta<28^\circ$ for the models shown in Section~\ref{sec:multi}. For a reference, we also plot the data points of GW170817~\citep{Villar:2017wcc}.}
	 \label{fig:mag_misaligned}
\end{figure*}

The GW data analysis infers that GW170817 is observed from $\theta\lesssim28^\circ$ (Abbott et al. 2017a), while the light curves observed from $0^\circ\le\theta<20^\circ$ are only shown in Section~\ref{sec:multi}. Thus, here, we plot the {\it grizJHK}-band light curves observed from $20^\circ\le\theta<28^\circ$ as complementary information to enable the readers to make a comparison to the observed data (see Figure~\ref{fig:mag_misaligned}). 

\section{Contribution of fission fragments to the heating rate}\label{sec:fission}

Contribution of fission fragments to the heating rate is highly uncertain~\citep[e.g.,][]{Wanajo:2018wra}. To check how such uncertainty affects the results in this paper, we compare the light curves with and without considering the contribution of fission fragments to the heating rate for several models studied in this paper.

\begin{figure*}
 	 \includegraphics[width=.5\linewidth]{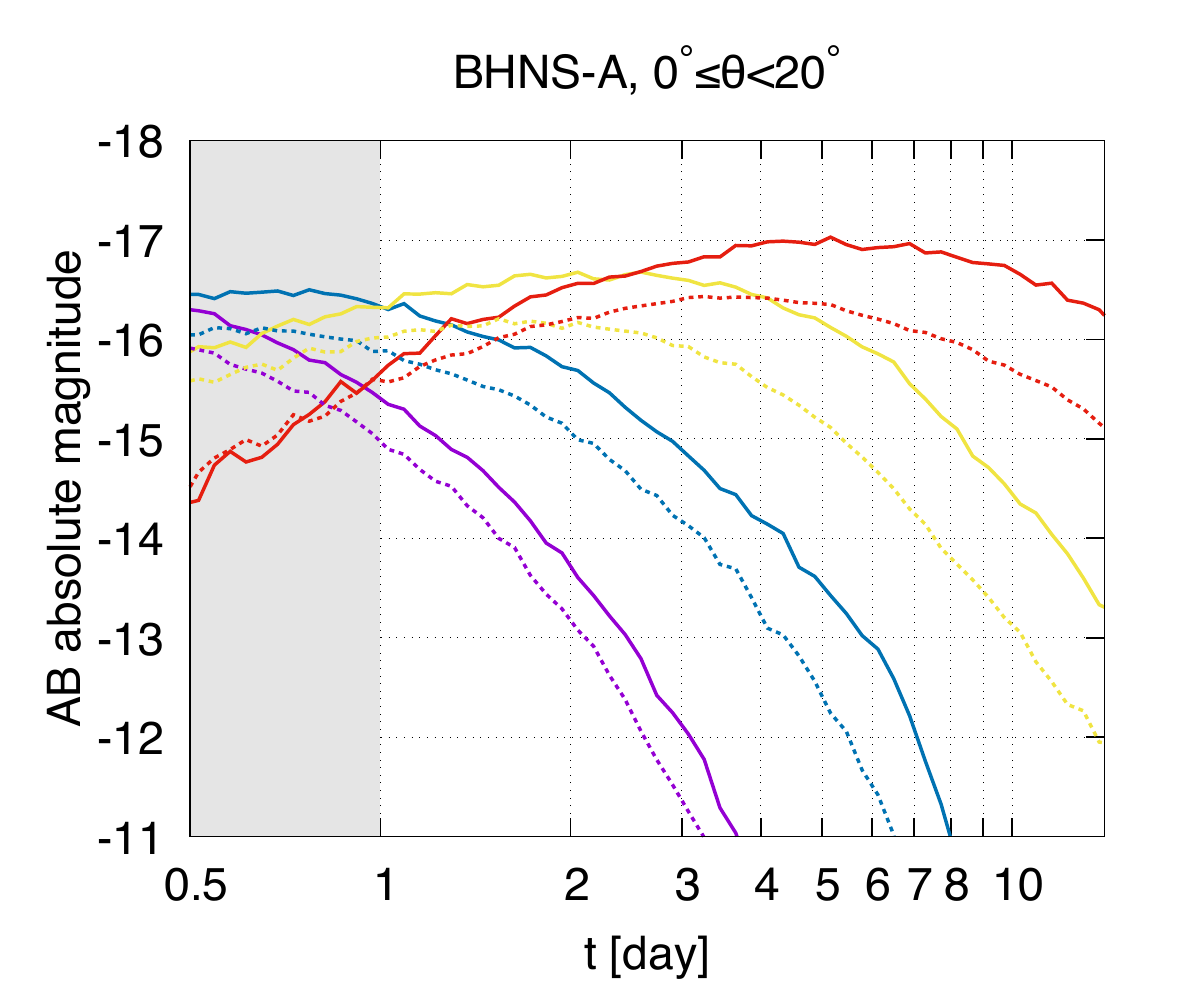}
 	 \includegraphics[width=.5\linewidth]{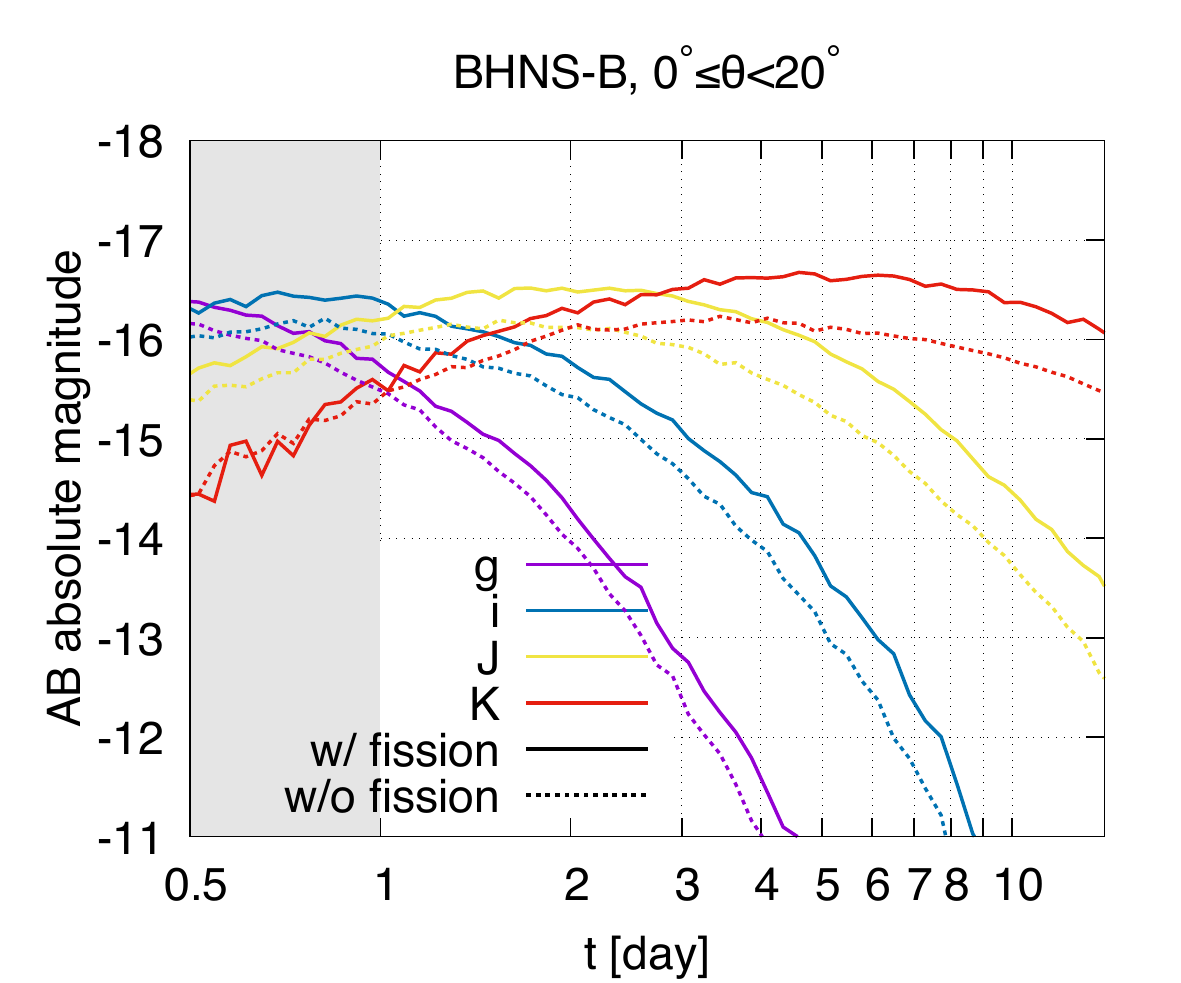}
 	 \caption{Comparison of the {\it giJK}-band light curves with (solid curves) and without (dotted curves) the contribution of fission to the heating rate for the BH-NS ejecta models ({\tt BHNS\_A} and {\tt BHNS\_B}).}
	 \label{fig:mag_fis34}
\end{figure*}

Figure~\ref{fig:mag_fis34} compares the {\it giJK}-band light curves with (solid curves) and without (dotted curves) the contribution of fission fragments to the heating rate for the BH-NS ejecta models ({\tt BHNS\_A} and {\tt BHNS\_B}). The kilonova emission for the models without fission fragments is fainter than that with fission fragments by more than $\approx0.5\,{\rm mag}$. The difference is more significant for the model with larger dynamical ejecta ({\tt BHNS\_A}) than the model with less dynamical ejecta ({\tt BHNS\_B}), and this reflects that uncertainty in fission fragments has a large impact on the emission particularly from the dynamical ejecta.

\begin{figure*}
 	 \includegraphics[width=.5\linewidth]{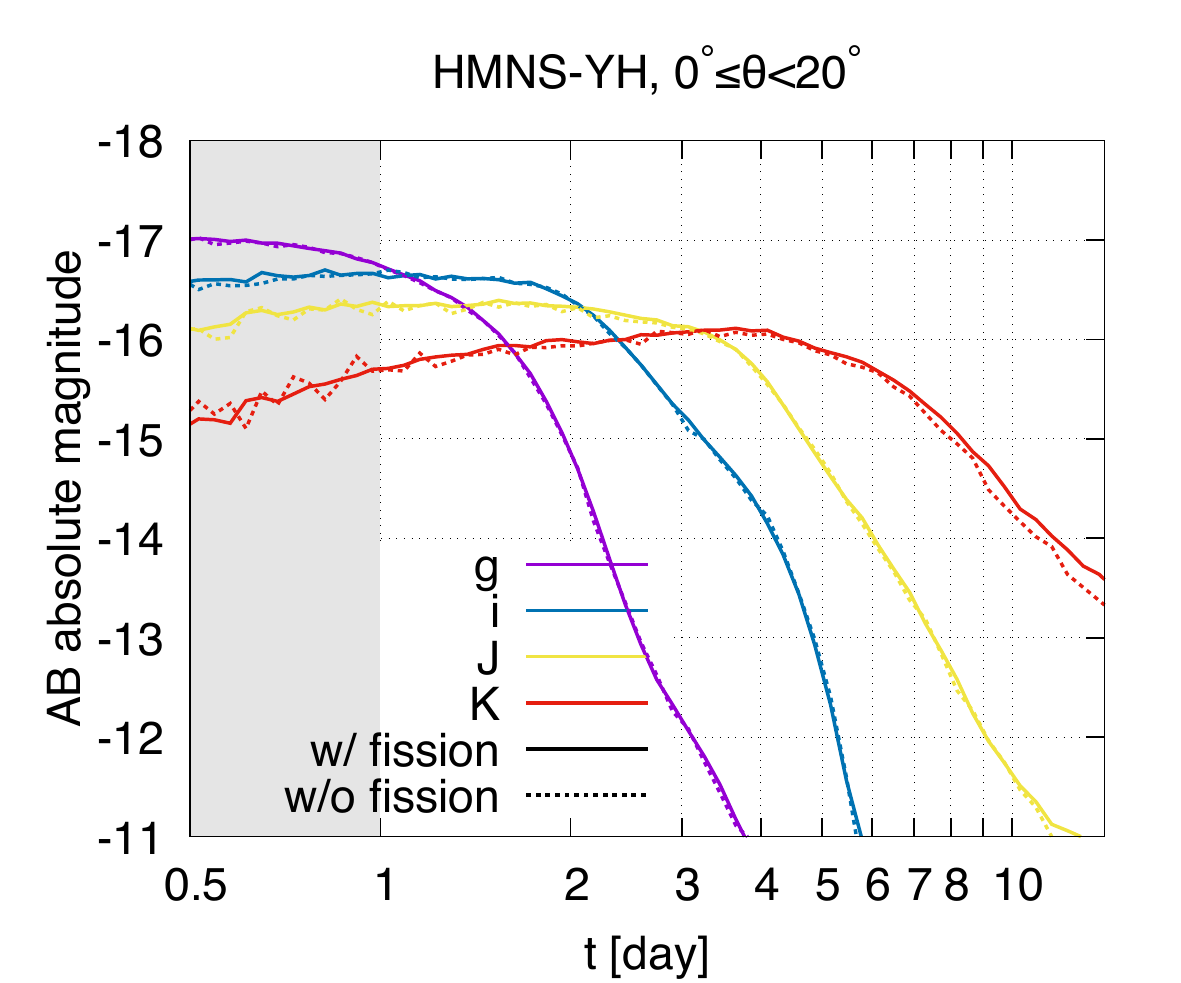}
 	 \includegraphics[width=.5\linewidth]{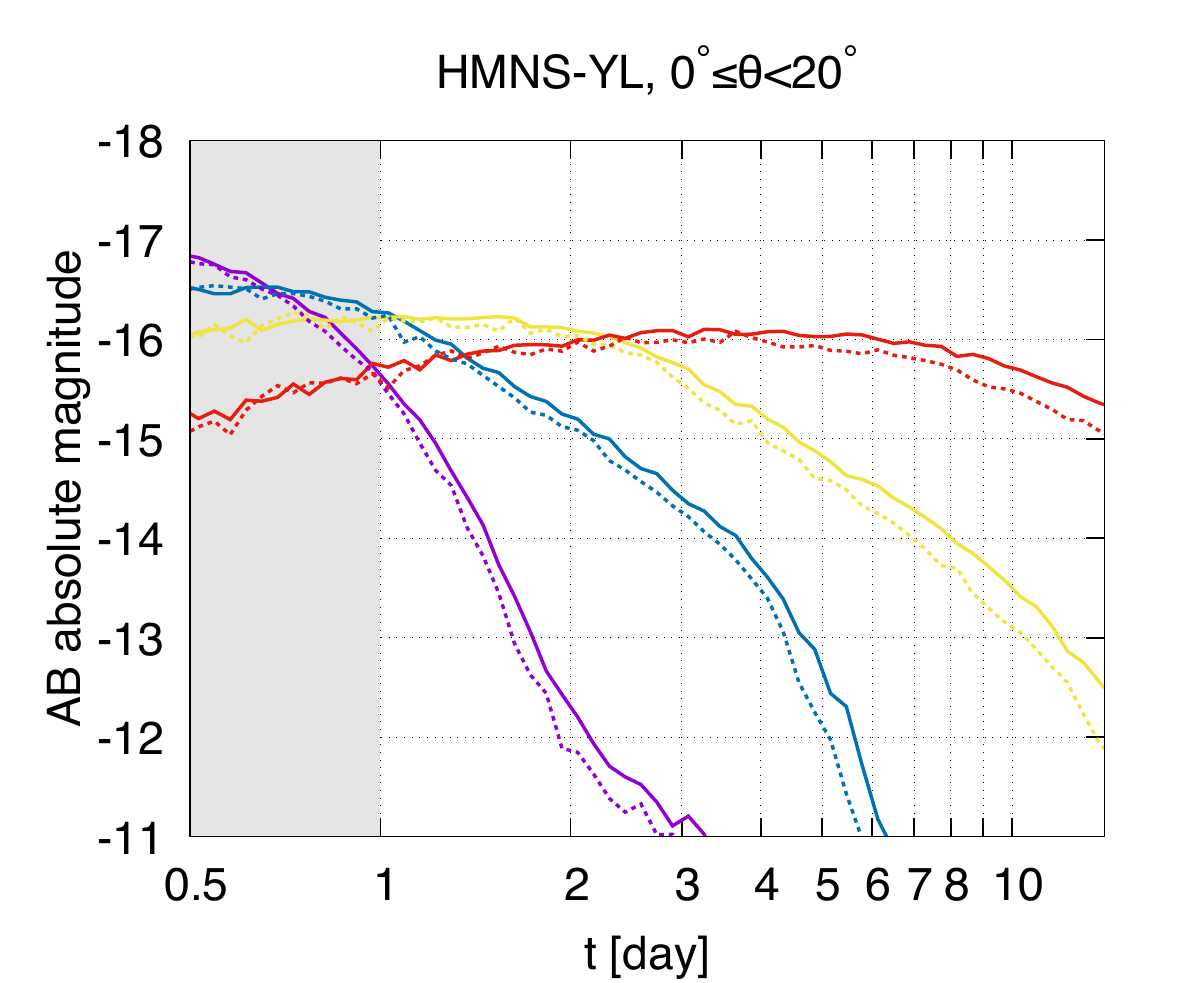}
 	 \caption{Comparison of the {\it giJK}-band light curves with (solid curves) and without (dotted curves) the contribution of fission to the heating rate for the HMNS ejecta models ({\tt HMNS\_YH} and {\tt HMNS\_YL}).}
	 \label{fig:mag_fis12}
\end{figure*}

Figure~\ref{fig:mag_fis12} is the same as Figure~\ref{fig:mag_fis34} but for the HMNS ejecta models ({\tt HMNS\_YH} and {\tt HMNS\_YL}). In contrast to the BH-NS ejecta models, the difference in the light curves between the models with and without fission fragments are always smaller than $\approx 0.5\,{\rm mag}$ for the HMNS ejecta models, and is approximately absent for the model with the lanthanide-poor post-merger ejecta ({\tt HMNS\_YH}). The reason why the difference in the HMNS ejecta models is less significant than the BH-NS models is that the average $Y_e$ value of the dynamical ejecta is set to be higher values for the HMNS ejecta models than the BH-NS ejecta models, and the contribution of fission fragments to the heating rate is less significant. Indeed, the electron fraction of the dynamical ejecta for the HMNS models is widely distributed from $0.1$ to $0.3$ while that of the BH-NS models is set to be $\alt0.1$, and the heating is dominated by the contribution from $\beta$-decays or $\alpha$-decays for the former cases. Thus, uncertainty in the contribution of fission fragments to the heating rate has a largest impact on the light curves particularly for the case that $Y_e$ value is as low as in the cases of BH-NS mergers (or the prompt collapse cases of NS-NS mergers).

\bibliographystyle{apj}
\bibliography{ref}

\end{document}